\documentclass[
 reprint,
 amsmath,amssymb,
 aps,
 noeprint,
 longbibliography
]{revtex4-2}

\usepackage{graphicx}
\usepackage[dvipsnames]{xcolor}
\usepackage[colorlinks,linkcolor=black,citecolor=black,urlcolor=black,filecolor=black]{hyperref}
\usepackage{amsmath, amssymb, amsthm}
\usepackage{wasysym}
\usepackage{lipsum}
\usepackage[ruled]{algorithm2e}
\usepackage{outlines}
\usepackage{enumerate}
\usepackage{xcolor}
\usepackage[export]{adjustbox}
\usepackage{tikz}
\usetikzlibrary{decorations.pathreplacing}
\usetikzlibrary{arrows,arrows.meta,calc,shapes.geometric,shapes.misc}
\tikzset{
	>=stealth',
	help lines/.style={dashed, thick},
	important line/.style={thick},
	connection/.style={thick, dotted},
}
\DeclareMathAlphabet{\mymathbb}{U}{BOONDOX-ds}{m}{n}
\usepackage{dsfont}
\usepackage{braket}
\usepackage{hyperref}
\hypersetup{
    pdfstartview={FitH}, 
    colorlinks=true, 
    linkcolor=blue, 
    citecolor=blue, 
    filecolor=magenta,
    urlcolor=blue
}

\usepackage{wrapfig}
\newtheorem{theorem}{Theorem}
\newcommand{\setsize}{b}
\newcommand{\restrb}{(r,\setsize)}
\newcommand{\nbranch}{{n_{b}}}
\newcommand{\tsa}{\tau^{}_{\textnormal{SA}}}
\newcommand{\tpt}{\tau^{}_{\textnormal{PT}}}

\newcommand{\tqmc}{\tau^{}_{\textnormal{QMC}}}
\newcommand{\gsa}{\Delta_{\textnormal{SA}}}
\newcommand{\gpt}{\Delta_{\textnormal{PT}}}
\newcommand{\gqaa}{\Delta_{\textnormal{QAA}}}
\newcommand{\gqmc}{\Delta_{\textnormal{QMC}}} 
\newcommand{\hqaa}{H_{\textnormal{QAA}}}
\newcommand{\tr}{\textnormal{Tr}}
\newcommand{\barket}[1]{\ket{S_{#1}}}
\newcommand{\barbra}[1]{\bra{S_{#1}}}
\newcommand{\ketE}{\ket{\mathcal{E}}} 
\newcommand{\ketEi}[1]{\ket{\mathcal{E}_{#1}}} 
\newcommand{\kettildeEi}[1]{\ket{\tilde{\mathcal{E}}_{#1}}} 
\newcommand{\ketG}{\ket{\mathcal{G}}}
\newcommand{\braE}{\bra{\mathcal{E}}} 
\newcommand{\braG}{\bra{\mathcal{G}}}
\newcommand{\kettildeG}{\ket{\tilde{\mathcal{G}}}}
\newcommand{\mee}{m_{ee}} 
\newcommand{\mge}{m_{ge}} 
\newcommand{\mgg}{m_{gg}}

\newcommand{\ezero}{\bar{E}} 
\newcommand{\eone}{\Delta \bar{E}} 
\newcommand{\eg}[1]{\braE H_{\mathrm{eff}}(#1)\ketG} 
\newcommand{\hcost}{H_{\textnormal{cost}}}

\newcommand{\hse}{H_{se}}
\newcommand{\loc}{(\Omega/\delta)_{\star}}
\newcommand{\ecrit}{E_{\star}}
\newcommand{\locinv}{(\delta/\Omega)_{\star}}

\newcommand{\hdrive}{H_q}
\newcommand{\hbulk}{H_\textnormal{bulk}}
\newcommand{\hryd}{H_{\text{Ryd}}}
\newcommand{\htb}{H_{tb}}
\newcommand{\hlaplace}{H_{\ell}}
\newcommand{\bottleneck}{t_{\alpha-1}}
\newcommand{\heff}{H_{\rm{eff}}}
\newcommand{\gqaanaive}{\tilde{\Delta}_{\rm{QAA}}}
\newcommand{\glaplace}[1]{\Delta_{\ell,{#1}}}
\newcommand{\gbulk}{\Delta_{\text{bulk}}}
\newcommand{\gqaaest}{\Delta^{\text{est.}}_{\textnormal{QAA}}}

\begin{document}

\title{Quantum speedup for combinatorial optimization with flat energy landscapes}

\author{M. Cain$^{1}$, S. Chattopadhyay$^{1}$,  J.-G. Liu$^{1,2}$, R. Samajdar$^{3,4}$,  H. Pichler$^{5,6}$, M. D. Lukin$^{1}$}

\date{\today}

\affiliation{\mbox{$^1$Department of Physics, Harvard University, Cambridge, MA 02138, USA}\\
\mbox{$^2$Advanced Materials Thrust, Hong Kong University of Science and Technology (Guangzhou), Guangdong 511453, China}\\ 
\mbox{$^3$Department of Physics, Princeton University, Princeton, NJ 08544, USA}\\
\mbox{$^4$Princeton Center for Theoretical Science, Princeton University, Princeton, NJ 08544, USA}\\
\mbox{$^{5}$Institute for Theoretical Physics, University of Innsbruck, Innsbruck A-6020, Austria}\\
\mbox{$^{6}$Institute for Quantum Optics and Quantum Information, Austrian Academy of Sciences, Innsbruck A-6020, Austria}}

\begin{abstract}
    Designing quantum algorithms with a speedup over their classical analogs is a central challenge in quantum information science.
    Motivated by recent experimental observations of a superlinear quantum speedup in solving the Maximum Independent Set problem on certain unit-disk graph instances [Ebadi \textit{et al.}, \href{https://www.science.org/doi/10.1126/science.abo6587}{Science \textbf{376}, 6598 (2022)}], we develop a theoretical framework to analyze the relative performance of the optimized quantum adiabatic algorithm and a broad class of classical Markov chain Monte Carlo algorithms.
    We outline conditions for the optimized adiabatic algorithm to achieve a quadratic speedup on hard problem instances featuring flat low-energy landscapes and provide example instances with either a quantum speedup or slowdown.
    We then introduce an additional local Hamiltonian with no sign problem to the optimized adiabatic algorithm to achieve a quadratic speedup over a wide class of classical simulated annealing, parallel tempering, and quantum Monte Carlo algorithms in solving these hard problem instances. 
    Finally, we use this framework to analyze the experimental observations.
\end{abstract}
\maketitle
 
\newcommand{\nocontentsline}[3]{}
\newcommand{\tocless}[2]{\bgroup\let\addcontentsline=\nocontentsline#1{#2}\egroup}

\tocless\section{Introduction}
Combinatorial optimization problems have wide-ranging applications in science and technology~\cite{arora_barak_2016}. 
They are foundational to modern computer science because they encompass NP-hard problems which cannot be solved efficiently by known algorithms.
A central challenge in quantum information science is to understand when quantum algorithms can outperform their classical counterparts in solving such NP-hard combinatorial optimization problems~\cite{Montanaro2016, albash_lidar}.  
The most general classical combinatorial optimization algorithms seek to minimize a cost function over a set of bit strings. 
This includes broad classes of Markov chain Monte Carlo algorithms such as simulated annealing (SA) and parallel tempering~\cite{Earl_2005}, which simulate cooling to low-temperature states of a classical Hamiltonian encoding the cost function. 

Quantum adiabatic algorithms (QAAs)~\cite{farhi_2001} can be viewed as quantum analogs of such general-purpose classical solvers. 
QAA prepares low-energy states of a classical cost Hamiltonian~\cite{Lucas_2014} by adiabatic evolution. 
The relative performance of QAA and SA is not generically well understood beyond numerical studies~\cite{young_2010, Guidetti_2011, Hen_2011}, and theoretical examples of quantum speedup are either restricted to specifically constructed problem instances~\cite{farhi_sa_2002} or require unphysical Hamiltonians~\cite{Roland_2002, Muthukrishnan_2016, Crosson_2016}.
However, unlike other quantum algorithms that are known to generically achieve a quadratic speedup over SA~\cite{szegedy_marked, somma2008annealing, Montanaro_2015}, QAA can be studied experimentally on existing quantum devices. 
Although early experimental implementations of QAA lacked the many-body coherence believed to be necessary for quantum speedup~\cite{Johnson2011, dwave_how_quantum, R_nnow_2014, Boixo_2014, katzgraber_2015, Boixo2016}, a recent study using a programmable Rydberg atom array~\cite{Ebadi_2022} observed a superlinear speedup over SA in solving certain hard instances of the NP-hard Maximum Independent Set problem on unit-disk graphs. 

Motivated by these experimental results, in this work we develop a theoretical framework to analyze the relative performance of optimized QAA and several classical Markov chain Monte Carlo algorithms. 
Specifically, we focus on problem instances with flat energy landscapes comprised of many suboptimal configurations of the same cost, over which algorithms must search to find the optimal solution. 
We show that the QAA's performance is determined by (de)localization of the low-energy eigenstates of the adiabatic Hamiltonian in configuration space: when the low-energy eigenstates are delocalized, and the quantum evolution is optimized to maintain adiabaticity, QAA achieves a quadratic speedup over a wide class of SA and parallel tempering algorithms. 
To illustrate these concepts, we provide examples of problem instances that feature either a quantum speedup or slowdown depending on the localization of the low-energy eigenstates. 

Having developed this framework, we then use it to introduce a modification of QAA that achieves a quadratic speedup over SA and parallel tempering on certain hard Maximum Independent Set problem instances. 
Importantly, our algorithm only uses local Hamiltonians with no sign problem, meaning that all the off-diagonal matrix elements are non-positive.
While QAA Hamiltonians without a sign problem are typically amenable to simulation with quantum Monte Carlo (QMC) -- and many prior speedups over SA in this setting have indeed been recovered by QMC~\cite{Crosson_2016, denchev_dwave_2016} -- we nevertheless show that our algorithm maintains a quadratic speedup over a wide class of path-integral QMC algorithms. 
Finally, we apply these techniques to interpret the experimental observations reported in Ref.~\cite{Ebadi_2022}. 
We identify instances with better-than-classical performance due to either delocalization or favorable localization of the low-energy eigenstates. 
Instances with worse-than-classical performance can be explained by unfavorable localization of the eigenstates, as introduced by Ref.~\cite{altshuler_2010}. 

Before proceeding, we note that state-of-the-art classical heuristic algorithms specialized to the Maximum Independent Set problem can outperform SA (e.g.,~\cite{redumis}). 
These algorithms accelerate the computation by exploiting the problem-specific graph structure. 
In contrast, SA is a general-purpose solver that only uses the energy of a configuration in decision-making to prepare the Gibbs distribution of the cost Hamiltonian. 
Similarly, QAA only takes in the cost Hamiltonian as an input, and prepares its ground state by adiabatic evolution.
We will restrict our analysis to the case where the QAA evolution is slow enough to maintain adiabaticity, and the SA evolution is long enough to equilibrate to the Gibbs distribution. 
Running these algorithms at short, \textit{diabatic} timescales and exploring shortcuts to adiabaticity is of independent interest~\cite{Gu_ry_Odelin_2019, Crosson_2021, King2023, Schiffer2023}. \\ 

\begin{figure}[t!]
    \centering
    \includegraphics[width=.5\textwidth]{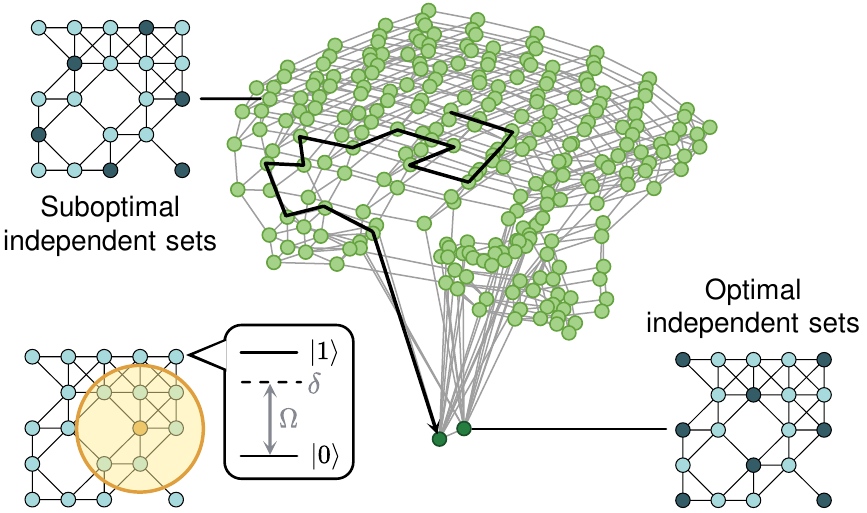}
    \caption{
    Flat energy landscapes in combinatorial optimization. 
    The goal of the Maximum Independent Set problem is to find the largest independent sets of a graph (e.g., the dark blue vertices, bottom right) among many suboptimal independent sets (top left). 
    The dynamics of SA on this problem can be visualized by a configuration graph (center), where vertices represent individual independent sets and edges link sets connected by an SA update. 
    SA algorithms randomly walk (black lines) between suboptimal independent sets of the same size (light green vertices) until finding an optimal independent set (dark green vertices).
    We study QAA's performance on unit-disk graphs (bottom left), where vertices are connected within a unit radius (yellow circle). 
    Each vertex is associated with a qubit with a time-dependent drive $\Omega(t)$ and detuning $\delta(t)$. 
    } 
    \label{fig:speedup_1}
\end{figure}

\tocless\subsection{Maximum Independent Set}
Throughout this work, we focus on the Maximum Independent Set problem, a paradigmatic NP-hard optimization problem that involves finding the largest independent set of a graph.
An independent set is a subset of vertices where no two vertices are connected by an edge.
The largest independent set for a graph $G=(V, E)$  with $n$ vertices is a  configuration $\ket{z}\in \{\ket{0}, \ket{1}\}^n$ minimizing $\hcost(z) = \bra{z}\hcost\ket{z}$ for $\delta > 0$, where 
\begin{align}
    \hcost = -\delta\sum_{u\in V}n_u +  U\sum_{(u, v)\in E} n_u n_v \label{eq:hcost}
\end{align}
is the classical cost Hamiltonian.
Here, $n_u$\,$\equiv$\,$\ket{1_u}\bra{1_u}$, and $\ket{1_u}$ ($\ket{0_u}$) denotes that vertex $u$ is present (absent) in the independent set. 
\mbox{$U\gg |\delta|$} penalizes edges that violate the independent set constraint. 
We focus primarily on unit-disk graphs, where edges connect vertices within a unit radius on a two-dimensional plane. 
These graphs naturally model problems with geometrically local connectivity, such as wireless communication networks~\cite{unit_disk_review}.

The Maximum Independent Set problem on unit-disk graphs can be naturally encoded in Rydberg atom arrays as follows~\cite{pichler_2018}. 
Every vertex is associated with an atomic qubit placed on a square grid at position $r_u$ (Fig.~\ref{fig:speedup_1}). 
The full system is described by the many-body Hamiltonian \mbox{$H = \hryd-\hdrive$}, where 
\begin{align}
    \hdrive &= 
    \Omega\sum_{u\in V}\ket{1_u}\bra{0_u}+\mathrm{h.c.},\label{eq:hdrive}\\
    \hryd &= -\delta\sum_{u\in V} n_u + \sum_{u, v} V_{uv} n_u n_v,\label{eq:hryd}
\end{align}
and $\Omega(t)>0$ and $\delta(t)$ are time-dependent energies controlled by a coherent laser drive.  
The distance-dependent Rydberg blockade interaction energy \mbox{$V_{uv}$\,$\sim$\,$ 1/|r_u-r_v|^6$} makes simultaneous excitation of two atoms in the Rydberg state $\ket{1_u 1_v}$ within a certain radius energetically unfavorable, mimicking $U$ in Eq.~\eqref{eq:hcost}. 
In practice, the blockade radius is chosen to encompass nearest and next-nearest neighbors on the grid. \\

\tocless\section{Simulated annealing runtime on hard instances}
We first characterize fundamentally hard graph instances for SA to find the largest independent set.
SA stochastically samples spin configurations from the thermal Gibbs distribution $\pi$ of $\hcost$ at a low temperature $1/\beta$. 
We consider any Metropolis-Hastings SA algorithm~\cite{Metropolis, Hastings} in which the probability $P_{z, z'}$ to update $\ket{z}$ to $\ket{z'}$ satisfies the detailed balance condition,
\begin{align}
     P_{z, z'} \pi_z = P_{z', z} \pi_{z'} \qquad \pi_z =e^{-\beta \hcost(z)}/\mathcal{Z}_{\beta}, \label{eq:detailed_balance}
\end{align}
where $\pi_z$ is the Gibbs population of $\ket{z}$ and $\mathcal{Z}_\beta$ is the partition function. 
We allow the update rule to be arbitrarily non-local.

Within this general setting, we find that flat energy landscapes, defined as many suboptimal independent sets of the same size with few larger independent sets, form a fundamental obstacle for SA to find the solution~\cite{Znidari_2005, Ebadi_2022}.
Figure~\ref{fig:speedup_1} visualizes the flat energy landscape of an example unit-disk graph as a \textit{configuration graph}, where vertices represent independent sets and edges represent SA updates (here, spin-exchange and spin-flip operations).
This instance has many suboptimal independent sets of size $\alpha-1$ and few optimal largest independent sets of size $\alpha$.
The SA dynamics, governed by Eq.~\eqref{eq:detailed_balance}, are dominated by a random walk among the suboptimal, equal-energy configurations, reminiscent of unstructured search for the optimal solutions. 
Therefore, we expect the SA runtime to go like the inverse rate \mbox{$\simeq D_{\alpha-1}/D_\alpha$} of randomly choosing an optimal independent set, where $D_{\setsize}$ is the number of independent sets of size $\setsize$.

We now formalize this intuition and describe a lower bound on the SA runtime $\tsa(\varepsilon)$. 
$\tsa(\varepsilon)$ is a proxy for the time needed for SA to find an optimal solution. 
In particular, it given by the SA \textit{mixing time}: the number of proposed updates, normalized by $n$, needed to prepare the Gibbs distribution with total variation distance \mbox{$\varepsilon < 1/2$} starting from any initial configuration~\footnote{The error \mbox{$\varepsilon<1/2$} is the total variation distance (equal to half the $l_1$ distance) between the Gibbs distribution and the distribution prepared by SA.}. 
As the temperature $1/\beta\to 0$, the Gibbs distribution approaches the uniform mixture of optimal configurations. 
Thus, if the time for SA to equilibrate amongst the optimal configurations is small compared to the time to find an optimal configuration, we expect $\tsa(\varepsilon)$ to represent the time to find a solution. 
We confirm that this is the case in Appendix~\ref{subsec:finite_lambda}, because the optimal configurations are well-connected under spin-exchange updates. 

We prove the lower bound on $\tsa(\varepsilon)$ in Appendix~\ref{subsec:sa_runtime} by relating $\tsa(\varepsilon)$ to the inverse spectral gap $\gsa^{-1}$ of the SA Markov chain transition matrix \mbox{$P=(P_{z,z'})$}~\cite{markov_chain_mixing}. 
We then use the Cheeger inequality~\cite{cheeger_ref_diaconis} to relate $\gsa$ to the flow of population in the Gibbs distribution from independent sets of size $\leq\setsize - 1$ to size $\geq \setsize$ during a single SA update. 
This flow is proportional to $D_{\setsize} / D_{\setsize-1}$, which gives us
\begin{equation}
    \tsa(\varepsilon) \geq  \frac{\ln\big(\frac{1}{2\varepsilon}\big)}{2nk} \max \frac{D_{\setsize-1}}{D_\setsize},\label{eq:tsa_lower_bound}
\end{equation}
where $k$ is the maximum number of spins altered during a proposed update~\footnote{
The SA and QMC runtime lower bounds in Eqs.~\eqref{eq:tsa_lower_bound} and~\eqref{eq:tqmc_bound} assume that the independence polynomial of the graph instance is unimodal, i.e., $D_0\leq D_1\leq\dots\leq D_{\setsize^\star}\geq \dots \geq D_{\alpha-1}\geq D_\alpha$ for some $\setsize^\star$. 
In Appendix~\ref{sec:runtime_system_size}, we study 24,000 unit disk graph instances with system sizes of up to 720 vertices, and find that every instance has a unimodal independence polynomial, which may be of independent interest~\cite{Levit2005TheIP}. 
If the independence polynomial is not unimodal, then the same bounds apply, but the maximum must be taken over all $\setsize > \setsize^\star$, as in Appendix~\ref{sec:mcmc_runtime}}.
We numerically find in Appendix~\ref{sec:runtime_system_size} that $\max_\setsize (D_{\setsize-1}/D_{\setsize})$, and therefore the SA runtime, grows exponentially in $\sqrt{n}$.
Moreover, we demonstrate that a similar bound holds for a wide class of parallel tempering algorithms in Appendix~\ref{subsec:pt_runtime}. 
As our proofs are framed in terms of a generic discrete cost function, they also apply to combinatorial optimization problems beyond Maximum Independent Set. 

\begin{figure}[tb]
    \centering
    \includegraphics[width=.5\textwidth]{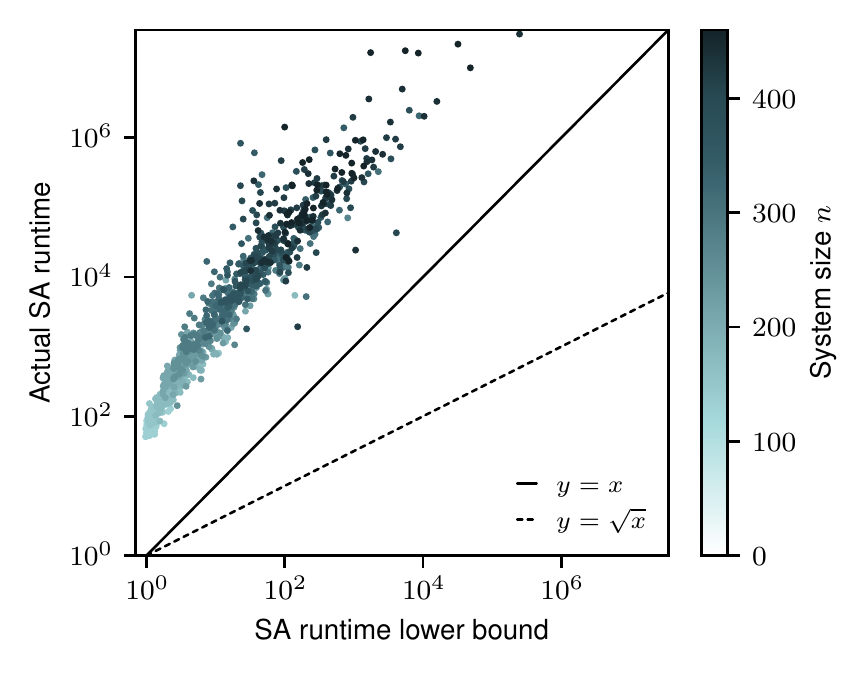}
    \caption{
    Flat energy landscapes determine SA runtime.
    The actual SA runtime to find an optimal solution with probability $3/4$ is linearly related to the analytic SA runtime lower bound in Eq.~\eqref{eq:tsa_lower_bound}, confirming that SA runtime is dominated by overcoming flat energy landscapes.
    }
    \label{fig:speedup_2}
\end{figure}

Figure~\ref{fig:speedup_2} shows the time for an optimized SA algorithm~\cite{Ebadi_2022} to find an optimal solution with probability $3/4$ against Eq.~\eqref{eq:tsa_lower_bound}, which we compute via a tensor-network algorithm~\cite{liu_tensor_network_2022}. 
We plot the data for the top $5\%$ hardest unit-disk graphs maximizing Eq.~\eqref{eq:tsa_lower_bound} within each system size (\mbox{$n=39$\,--\,$460$}, see Appendix~\ref{sec:runtime_system_size}), omitting a small fraction ($0.9\%$) of instances for which the SA runtime is too long to collect sufficient statistics. 
The strong linear relationship in Fig.~\ref{fig:speedup_2} confirms that the SA runtime is dominated by unstructured search over flat energy landscapes. 
Furthermore, it indicates that $\tsa(\varepsilon)$ is representative of the time to find an optimal solution. \\

\begin{figure*}[th!]
    \centering
    \includegraphics[width=\textwidth]{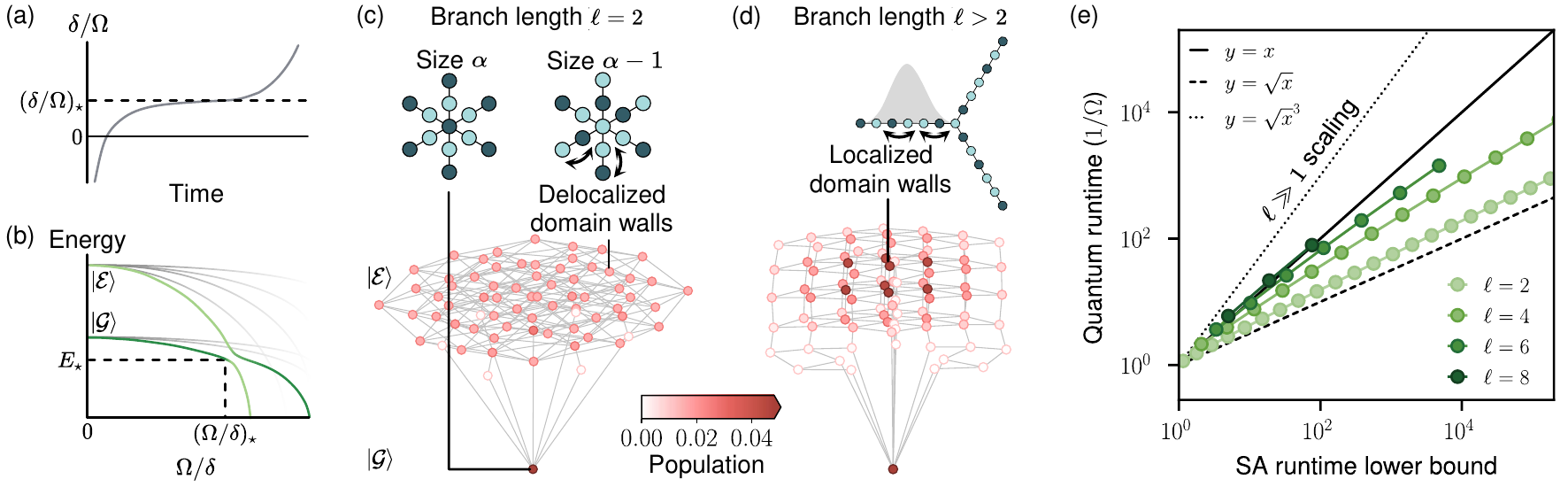}
    \caption{
    Eigenstate localization determines QAA runtime.
    (a)~The optimized QAA runtime is proportional to $\gqaa^{-1}$ when the system Hamiltonian changes slowly at $\locinv$, the location of the avoided level crossing. 
    (b)~$\gqaa$ can be computed perturbatively when $\loc\ll 1$ from Eq.~\eqref{eq:off_diag_expansion}, which describes the coupling under the Hamiltonian between the estimated eigenstates $\ketG, \ketE$ involved in the avoided level crossing.
    (c)~The star graph with $\nbranch$ branches of even length $\ell$ has a unique optimal independent set of size $\alpha$ with the central vertex in the independent set (top left). 
    It has approximately $(\ell/2+1)^\nbranch$ suboptimal independent sets of size $\alpha-1$ with the central vertex absent (top right), corresponding to all possible locations of a domain wall on each branch.
    When $\ell=2$, the two possible domain wall locations on each branch are equally energetically favored, causing $\ketE$ to delocalize over all domain wall locations. 
    (d)~When $\ell > 2$, $\ketE$ localizes around configurations with the domain walls near the center of each branch. 
    (e)~QAA has a quadratic speedup in runtime over SA as a function of $\nbranch$ for the delocalized case of $\ell = 2$. 
    As $\ell$ increases, $\ketE$ localizes away from $\ketG$, causing SA to outperform QAA when $\ell \gg 1$.
    }
    \label{fig:speedup_3}
\end{figure*}

\tocless\section{Instance-by-instance performance of QAA}
We now establish conditions for which QAA outperforms SA on such hard instances. 
QAA prepares the ground state of $\hcost$ by adiabatic evolution under
\begin{equation}
    \hqaa = \hcost-\hdrive,\label{eq:hqaa}
\end{equation} 
where the energies $\Omega(t), \delta(t)$ [Eqs.~\eqref{eq:hcost} and~\eqref{eq:hdrive}] vary in time as shown in Fig.~\ref{fig:speedup_3}(a).
In particular, we assume that $\Omega(t), \delta(t)$ are optimized to minimize the evolution time while maintaining adiabaticity near the minimum energy gap $\gqaa$ between the ground and first-excited states of the dominant avoided level crossing, so the runtime of QAA goes as $\gqaa^{-1}$ (specifically, \mbox{$|dH/dt| \propto \gqaa^2$} at the avoided crossing location $\loc$, see Refs.~\cite{Roland_2002, jarret_2019} and Sec.~\ref{sec:outlook} for further discussion).
We will show that $\gqaa$ is controlled by the properties of two states, $\ketG$ and $\ketE$, which approximate the ground and first-excited states at this avoided crossing, as shown in Fig.~\ref{fig:speedup_3}(b).
We analyze three qualitatively distinct behaviors for $\ketG, \ketE$, which we term \textit{delocalized, favorably localized}, and \textit{unfavorably localized}. 
The former two result in a speedup over SA, while the latter causes a slowdown. 

As argued in Appendix~\ref{subsec:late_crossing}, generically, the avoided level crossing occurs near the end of the ramp, at \mbox{$\loc\ll 1$}. 
We will show later that $\ketG$ and $\ketE$ can be computed at leading order in $\Omega/\delta$ as non-negative superpositions of optimal and suboptimal independent sets, respectively:
\begin{align}
    \ketG &= \sum_{z: \hcost(z) = -\delta\alpha} \sqrt{\mathcal{G}_z}\ket{z},\nonumber \\ 
    \ketE &= \sum_{z: \hcost(z) = -\delta(\alpha-1)} \sqrt{\mathcal{E}_z}\ket{z}.
\end{align}
In the examples we consider, $\ketE$ is a superposition of independent sets of size \mbox{$\alpha-1$}, though our arguments can be generalized when $\ketE$ is a superposition of smaller independent sets (see Appendix~\ref{subsec:late_crossing}). 
We can estimate $\gqaa$ in powers of $\loc$ as the coupling between $\ketG$ and $\ketE$~\cite{cohen_tannoudji, Amin_2008, altshuler_2010, choi_first_order, Choi2020},
\begin{align}
    \label{eq:off_diag_expansion} \gqaanaive&=
    2\left|\sum_{l=0}^{\infty} \braE  \Big(\hdrive\frac{Q}{\ecrit - \hcost }\Big)^l \hdrive \ketG\right|,
\end{align}
where \mbox{$Q=\mathds{1}-\ketE\braE-\ketG\braG$}.
In Appendix~\ref{subsec:resolvent_derivation}, we derive a bounded proportionality factor relating  $\gqaa$ and  $\gqaanaive$. 
We note that these results provide a perturbative approach to \textit{exactly} compute $\gqaa$ and are thus of broader utility and interest beyond the specifics of the problem considered here.

Per Eq.~\eqref{eq:off_diag_expansion}, $\gqaanaive$ is determined by the distribution of wavefunction amplitudes in $\ketG$ and $\ketE$. 
At each order $l$ in $\loc$, factors of $\hdrive$ generate $l+1$ spin flips to connect pairs of configurations in $\ketG$ and $\ketE$. 
The leading-order coupling between two configurations $\ket{z}$ and $\ket{z'}$ within Hamming distance $l+1$ goes like $ \sqrt{\mathcal{G}_z \mathcal{E}_{z'}}\loc^l$. 
This coupling is enhanced for sets with larger amplitude but is suppressed exponentially in $l$. 
This intuition leads us to distinguish between problem instances where $\mathcal{G}_z$ and $\mathcal{E}_{z}$ are localized on comparatively few sets, and those where they are distributed more evenly among all sets.
We refer to instances where $\ketG$ and $\ketE$ localize on sets sufficiently far apart in Hamming distance such that QAA suffers a slowdown relative to SA ($\gqaanaive\ll D_{\alpha}/D_{\alpha-1}$) as \textit{unfavorably localized}~\cite{altshuler_2010}. 
By contrast, on \textit{favorably localized} instances, $\ketG$ and $\ketE$ localize at small Hamming distances, such that QAA has speedup over SA ($\gqaanaive\gg D_{\alpha}/D_{\alpha-1}$). 
Several previous notable instances where QAA has an exponential speedup~\cite{farhi_sa_2002} or slowdown~\cite{Schiffer2023} fall into these two categories.

QAA also outperforms SA on \textit{delocalized} instances, where the amplitudes $\sqrt{\mathcal{G}_z}$ and $\sqrt{\mathcal{E}_z}$ are close to uniform. 
Suppose that $\ketG = \barket{\alpha}$ and $\ketE = \barket{\alpha-1}$, where
\begin{equation}
    \barket{\setsize} = \frac{1}{\sqrt{D_{\setsize}}}\sum_{z: \hcost(z) = -\delta\setsize}\ket{z}. \label{eq:equal_superposition}
\end{equation} 
The lowest-order ($l=0$) contribution to Eq.~\eqref{eq:off_diag_expansion} is then 
\begin{align}
    \gqaanaive &=  2|\barbra{\alpha-1} H_q \barket{\alpha}|=\frac{2}{\sqrt{D_{\alpha-1 }D_{\alpha}}}\sum_{z: \hcost(z)=-\delta\alpha}\hspace*{-0.4cm}\Omega \alpha \nonumber\\
    &=2\Omega\alpha\sqrt{\frac{D_\alpha}{D_{\alpha-1}}}\label{eq:delocalized_coupling}.
\end{align}
Due to coherent enhancement in the coupling, here, the QAA runtime $\gqaanaive^{-1}$ is quadratically smaller than the SA runtime [Eq.~\eqref{eq:tsa_lower_bound}] up to polynomial factors in $n$. 
This is reminiscent of the adiabatic version of Grover's search~\cite{Roland_2002}, which has a similar quadratic speedup over randomly guessing in $\{\ket{0}, \ket{1}\}^n$ for optimal solutions. 
However, we emphasize that the runtimes of QAA and SA in Eqs.~\eqref{eq:delocalized_coupling} and~\eqref{eq:tsa_lower_bound}, respectively, are asymptotically faster than Grover's search, because they search only among near-optimal configurations for the largest independent set. \\

\tocless\subsection{Determining eigenstate localization}
Given a problem instance, we can determine $\ketG$ and $\ketE$ by performing second-order perturbation theory in the degenerate manifolds of $\hcost$. 
For simplicity, we take the energy penalty on independent set violations $U\to \infty$, so that each degenerate manifold contains independent sets of the same size. 
The perturbed eigenstates (energy shifts) are the eigenvectors (eigenvalues) of the matrix
\begin{align}\label{eq:second_order_ham}
    H^{(2)}
    =-\frac{\Omega^2}{\delta}\bigg(&\hse+\sum_{u\in V}\Big[n_u -(\mathds{1}-n_u)\hspace*{-0.2cm}\prod_{(u,v)\in E}\hspace*{-0.2cm}(\mathds{1}-n_v)\Big]\bigg),
\end{align}
where $\hse$ is the spin-exchange Hamiltonian, 
\begin{align}
    \hse = \sum_{(u, v)\in E} \sigma^+_u\sigma^-_v +  \sigma^-_u\sigma^+_v,\label{eq:spin_exchange}
\end{align}
$\sigma_u^+ = \ket{1_u}\bra{0_u},$ and $\sigma_u^- = \ket{0_u}\bra{1_u}$. 
$\ketG$ is the ground state of $H^{(2)}$ in the $\hcost = -\delta\alpha$ manifold, and $\ketE$ is the ground state of the excited manifold whose energy first intersects $\ketG$ at a finite $\loc$. 
As $H^{(2)}$ has no sign problem, $\ketG$ and $\ketE$ have non-negative amplitudes.

We find that first term in Eq.~\eqref{eq:second_order_ham}, $-(\Omega^2/\delta)\hse$, primarily determines the (de)localization of $\ketG$ and $\ketE$. 
This is because the second term is uniform within a manifold, and the third term (which counts the number of vertices that can be added to the independent set) is small for near-optimal independent sets. 
In particular, the expectation value of the third term is at most $-(\Omega^2/\delta)(\alpha-\setsize)$ for an independent set of size $\setsize$, and is zero when no vertices can be added to a set without removing existing vertices.
In order to minimize  $-(\Omega^2/\delta)\hse$, $\ketG$ and $\ketE$ will thus have larger overlap with independent sets that have more neighboring independent sets connected by spin exchanges in the configuration graph. 
In contrast, if all configurations in the $\hcost = -\delta \setsize$ manifold have the same degree (number of neighbors), the ground state in that manifold is the delocalized superposition $\barket{\setsize}$.
This follows from viewing $\hse$ as the adjacency matrix of the configuration graph within that manifold, and noting that the principal eigenvector of the adjacency matrix of a graph with regular degree is uniform~\cite{chung_1997}. \\

\tocless\subsection{Delocalization--localization crossover for a family of star graphs}
To concretely illustrate these concepts, we explore a family of star graphs, where $\ketE$ can be tuned from delocalized to unfavorably localized. 
A star graph contains $\nbranch$ branches of even length $\ell$ connected by a central vertex. 
We will compare the QAA and SA runtimes at fixed $\ell$ as $\nbranch$ grows.
The unique largest independent set includes the central vertex plus alternating vertices on each branch (Fig.~\ref{fig:speedup_3}(c), top left). 
All but a vanishing fraction of the suboptimal independent sets of size $\alpha-1$ have the central vertex absent and alternating antiferromagnetic order on the branches, each of which has a single domain wall located in one of $\ell/2+1$ possible positions (Fig.~\ref{fig:speedup_3}(c), top right). 
The SA runtime is thus exponential in $\nbranch$,
\begin{align}
    \tsa(\varepsilon)  \geq \frac{\ln\big(\frac{1}{2\varepsilon}\big)}{2nk}\frac{D_{\alpha-1}}{D_\alpha} \geq \frac{\ln\big(\frac{1}{2\varepsilon}\big)}{2nk}(\ell/2+1)^{\nbranch}.
\end{align}

To compute the QAA runtime from Eq.~\eqref{eq:off_diag_expansion}, we first calculate $\ketG$ and $\ketE$. 
$\ketG$ is the unique largest independent set, and $\ketE$ is the ground state of $H^{(2)}$ in the \mbox{$\hcost = -\delta(\alpha-1)$} manifold. 
By the reasoning above, on each branch, $\ketE$ is well-approximated by the ground state of $-(\Omega^2/\delta)\hse$, which acts as a one-dimensional hopping Hamiltonian, with open boundary conditions, for each domain wall. 
Therefore, $\ketE$ is given by 
\begin{align}\label{eq:wavefunction_e}
    \braket{x_1 x_2 \dots x_{\nbranch}|\mathcal{E}} \simeq  \prod_{i=1}^\nbranch \frac{1}{\sqrt{\ell/4+1}}\sin\Big(\frac{\pi x_i}{\ell/2+2}\Big),
\end{align}
where $\ket{x_i}, x_i\in \{1, 2, \dots, \ell/2+1\}$ is the state with the domain wall on the $i$th branch located between sites \mbox{$2x_i-2$} and \mbox{$2x_i-1$} (see Fig.~\ref{fig:speedup_3}(d), top and Appendix~\ref{subsec:star_graph_level_crossing_params}). 
$\gqaanaive$ can be computed to leading order in $\Omega/\delta$ from Eq.~\eqref{eq:off_diag_expansion} by connecting $\ketG$ to the set in $\ketE$ with all domain walls adjacent to the central vertex ($x_i=1$) by flipping the central vertex, 
\begin{align}\label{eq:starQAA}
    \gqaanaive&\simeq 2\Omega |\braG \hdrive \ketE| \nonumber\\
    &\simeq 2\Omega \bigg(\frac{1}{\sqrt{\ell/4+1}}\sin\Big(\frac{\pi}{\ell/2+2}\Big)\bigg)^\nbranch. 
\end{align}
Terms that are higher-order in $\Omega/\delta$ do not affect the scaling of $\gqaanaive$ with $\nbranch$, as shown in Appendix~\ref{subsec:star_graph_runtime}. 

\begin{figure*}[th!]
    \centering
    \includegraphics[width=\textwidth]{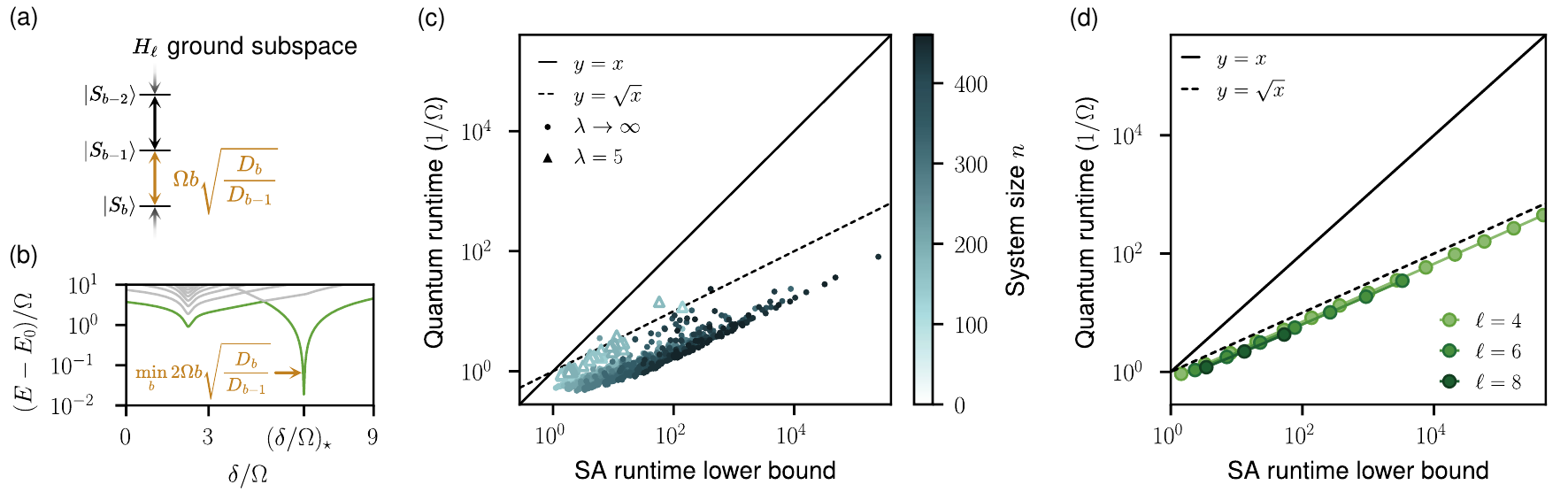}
    \caption{
    Quantum speedup over simulated annealing. 
    (a)~When $\lambda \to \infty$, the dynamics of the modified QAA~[Eq.~\eqref{eq:speedup_hamiltonian}] are restricted to the degenerate ground states of $\hlaplace$, which are the uniform superpositions $\barket{\setsize}$ of each independent set size $\setsize$~[Eq.~\eqref{eq:equal_superposition}].
    The matrix elements of $\hdrive$ (gold) between $\barket{\setsize}$ and $\barket{\setsize-1}$ are coherently enhanced over the analogous rate at which SA transitions from independent sets of size $\setsize-1$ to $\setsize$.
    (b)~The energy spectrum minus the ground state energy $E_0$ of an example $720$-vertex instance, restricted to the ground subspace of $\hlaplace$. 
    The minimum gap $\gqaa$ of the modified QAA is set by the smallest coupling (gold).
    (c)~The modified QAA runtime  $\gqaa^{-1}$ scales as the square root of the SA runtime for the same instances as in Fig.~\ref{fig:speedup_2} when dynamics are restricted to the ground subspace of $\hlaplace$ (circles). 
    The speedup is also obtained for finite $\lambda = 5$ (triangles). 
    (d)~The modified QAA obtains a quadratic speedup over SA for the star graphs with branch length $\ell=4, 6, 8$ when $\lambda = 2.2, 4.1, 6.5$, respectively. 
    }
    \label{fig:speedup_4}
\end{figure*}

Figure~\ref{fig:speedup_3}(e) plots the numerically computed QAA runtime $\gqaa^{-1}$ versus the SA runtime lower bound for $\tsa(1/4)$ for branch lengths \mbox{$\ell = 2, 4, 6,$ and $8$}. 
When $\ell$\,$=$\,$2$, $\ketE$ delocalizes evenly among all domain wall configurations (Eq.~\eqref{eq:wavefunction_e} and Fig.~\ref{fig:speedup_3}(c), bottom), yielding a quadratic quantum speedup because $\gqaa^{-1} = 2\Omega \sqrt{2}^\nbranch \lesssim \sqrt{\tsa(\varepsilon)}$.
As $\ell$ increases, according to Eq.~\eqref{eq:wavefunction_e}, $\ketE$ unfavorably localizes away from $\ketG$, on sets with the domain wall located near the center of each branch~(Fig.~\ref{fig:speedup_3}(d), bottom). 
Expanding Eq.~\eqref{eq:wavefunction_e} for small angles, we find that this results in a \textit{slowdown} for QAA when $\ell\gg 1$, as $\gqaa^{-1} \simeq \tsa(\varepsilon)^{3/2}$.\\ 

\tocless\section{Quantum speedup from delocalization}
\tocless\subsection{Quantum speedup over simulated annealing}
So far, our results show that the optimized QAA achieves a quadratic speedup over SA when its low-energy eigenstates are delocalized, due to the coherent enhancement of the couplings $\bra{S_{\setsize-1}} \hdrive \barket{\setsize}$ in Eq.~\eqref{eq:delocalized_coupling}. 
It is thus natural to ask whether instances with unfavorable localization can be remedied by modifying QAA to force the eigenstates to delocalize. 
We achieve this result by designing a Hamiltonian $\hlaplace$ whose degenerate ground subspace is spanned by the uniform superpositions $\{\barket{\setsize}\}$ \mbox{($\setsize=0,1,\ldots,\alpha$)}, and adding it to the QAA Hamiltonian with a time-independent energy scale $\lambda$,
\begin{align}
    H=\hqaa+\lambda \hlaplace.\label{eq:speedup_hamiltonian}
\end{align} 
In contrast to prior approximate approaches to  favoring delocalization~\cite{dickson_amin, lanting_king}, this approach provably enforces delocalization under certain conditions on the flat energy landscape, which we will state.

To design $\hlaplace$, we draw inspiration from the single-particle quantum kinetic energy operator, the ground state of which is maximally delocalized.
Since the single-particle kinetic energy is the negative of the continuum Laplacian $-\nabla^2$, we let $\hlaplace$ be the discrete Laplacian of the configuration graph in Fig.~\ref{fig:speedup_1}, restricted to each degenerate manifold of $\hcost$, where vertices represent independent sets and edges represent spin exchanges. 
The discrete Laplacian is the negative of the adjacency matrix ($\hse$), plus a diagonal term that counts the degree for that configuration, i.e., the number of possible spin exchanges, 
\begin{align}\label{eq:hlaplace}
    \hlaplace =
    -\hse + \sum_{u\in V}\sum_{(u, v)\in E} n_u (\mathds{1}-n_v)\prod_{\substack{(y,v)\in E\\y\neq u}} (\mathds{1}-n_y),
\end{align}
where \mbox{$G=(V, E)$} is the original problem graph.
Crucially, the diagonal term prevents the ground states of $H_\ell$ from localizing on independent sets with larger degrees on the configuration graph. 
This differs from the perturbative spin-exchange term in the unmodified QAA Hamiltonian $H^{(2)}$ [Eq.~\eqref{eq:second_order_ham}], which energetically favors configurations with more possible spin exchanges. 
We emphasize that $\hlaplace$ can be efficiently constructed using only local  information about the problem graph.
For unit-disk graphs embedded on a square grid, the terms in $\hlaplace$ only involve a constant number of spins, which allows for its implementation in near-term experiments.

To develop some intuition, let us first analyze the modified QAA when the energy scale of $\hlaplace$, $\lambda$, is large. 
If there exists a path between any two configurations in a degenerate manifold under spin exchanges, then each block $H_\setsize$ of \mbox{$H_\ell = H_0\oplus H_1\oplus \ldots \oplus H_\alpha$} has a unique ground state equal to $\barket{\setsize}$ with eigenvalue zero~\cite{chung_1997}. 
Since the QAA dynamics are restricted to this ground subspace when \mbox{$\lambda, U\to\infty$}, the modified QAA Hamiltonian in Eq.~\eqref{eq:speedup_hamiltonian} reduces to a one-dimensional tight-binding Hamiltonian,
\begin{align}
    \htb=-\sum_{\setsize=1}^{\alpha}\delta \setsize\barket{\setsize}\barbra{\setsize}+\Omega \setsize\sqrt{\frac{D_\setsize}{D_{\setsize-1}}}(\barket{\setsize}\barbra{\setsize-1}+\mathrm{h.c.}),\label{eq:tight_binding_hamiltonian}
\end{align}
which has an electric field gradient of strength $\delta$ and site-dependent couplings $\Omega\setsize\sqrt{D_\setsize/D_{\setsize-1}}$ [see Fig.~\ref{fig:speedup_4}(a)].
If the minimum energy gap $\gqaa$ of $\htb$ is set by the smallest coupling, as shown in Fig.~\ref{fig:speedup_4}(b) for an example 720-vertex unit-disk graph, then $\gqaa^{-1}$ is quadratically smaller than the SA runtime lower bound. 
We confirm the trend \mbox{$\gqaa \simeq \min_{\setsize}(\Omega\setsize\sqrt{D_\setsize/D_{\setsize-1}})$} numerically for hundreds of hard instances of the Maximum Independent Set problem on unit-disk graphs in Fig.~\ref{fig:speedup_4}(c). 
To explain these observations, we show in Appendix~\ref{subsec:infinite_lambda} that \mbox{$\gqaa \simeq \Omega\alpha\sqrt{D_\alpha/D_{\alpha-1}}$} on the vast majority of studied instances, for which the smallest coupling is between independent sets of size $\alpha-1$ and $\alpha$ and the remaining couplings are a smooth function of $\setsize$. 
We additionally argue in Appendix~\ref{subsec:finite_lambda} that the same result holds when a small number of configurations within a degenerate manifold are disconnected under spin exchanges, which occurs for a small fraction of instances.

To achieve the quantum speedup in practice, however, $\gqaa^{-1}$ must scale more favorably than the SA runtime when the energy scales of the modified QAA Hamiltonian are measured in units of $\lambda$, when $\lambda$ is the largest energy scale of $H$. 
To investigate the scale of $\lambda/\Omega$ required to obtain the quadratic enhancement of $\gqaa$, in Fig~\ref{fig:speedup_4}(c) we plot $\gqaa^{-1}$ for the top $1\%$ hardest instances with up to $n=80$ vertices, computed using the density matrix renormalization group method (DMRG)~\cite{white_density_1992, Fishman_2022}.
With the modest overhead of $\lambda/\Omega=5$, we observe a clear quadratic scaling advantage over the SA runtime lower bound in Eq.~\eqref{eq:tsa_lower_bound}. Furthermore, the modified QAA with $\lambda/\Omega = 1$ substantially outperforms the unmodified QAA on the same instances (see Fig.~\ref{fig:speedup_finite_lambda}(a) of Appendix~\ref{subsec:finite_lambda}).

We complement our numerical observations with sufficient, though not necessary, conditions on the $\lambda$ which yield a quadratic quantum speedup. 
In Appendix~\ref{subsec:finite_lambda}, we show analytically that a sufficient condition for achieving the quadratic enhancement of $\gqaa$ is $\lambda/\Omega,\lambda/\delta \gtrsim \glaplace{b}^{-1},\glaplace{b-1}^{-1}$, where $\glaplace{b},\glaplace{b-1}$ are the spectral gaps of the delocalizing Hamiltonian $\hlaplace$ restricted to the manifolds $b$ and $b-1$ that share the smallest tight-binding coupling $\min_{\setsize}(\Omega b\sqrt{D_b/D_{b-1}})$.
In Fig.~\ref{fig:speedup_4}(d), we confirm that the modified QAA with $\lambda = \glaplace{\alpha-1}^{-1} = \mathcal{O}(1)$ has a quadratic speedup for the family of star graphs. 
We show in Fig.~\ref{fig:speedup_finite_lambda}(b) of Appendix~\ref{subsec:finite_lambda} that typically $\glaplace{\setsize},\glaplace{\setsize-1}>1/n$ for the unit-disk graphs we study; accordingly, $\lambda\gqaa^{-1}\sim n\min_{\setsize}(\Omega\setsize\sqrt{D_{\setsize-1}/D_\setsize})$. 
Therefore, when $\glaplace{b}^{-1},\glaplace{b-1}^{-1}$ grow at most polynomially in $n$, the modified QAA's runtime is (sub)exponentially faster than the runtime of Grover's search ($\sqrt{2}^{n}$) for the hard unit-disk graphs we study: numerically, the SA runtime goes like $c^{\sqrt{n}}$ for some $c\in(1, 2)$, whereas the modified QAA runtime is $\sqrt{c}^{\sqrt{n}}$ up to polynomial factors in $n$ (see~Appendix~\ref{sec:runtime_system_size}). \\

\tocless\subsection{Quantum speedup over Quantum Monte Carlo}
As the modified QAA does not suffer from a sign problem, path-integral QMC can be used to sample independent sets from its thermal Gibbs distribution \mbox{$\pi_z = \bra{z}e^{-\beta H}\ket{z}/\mathcal{Z}_\beta$}. 
In general, path-integral QMC works by stochastically sampling trajectories from a discretized imaginary-time path integral of the partition function \mbox{$\mathcal{Z}_\beta=\text{Tr}(e^{-\beta H})$}. 
Several prior exponential speedups for QAA over SA have been recovered by sampling from the QMC path integral at low temperatures as the Hamiltonian is varied adiabatically in real time~\cite{Crosson_2016, Isakov_2016}.
It is thus natural to ask whether this procedure, also called \textit{simulated quantum annealing}, can match the modified QAA runtime. 

In Appendix~\ref{subsec:qmc_runtime}, we derive a lower bound for the QMC runtime $\tqmc(\varepsilon)$ of both the modified and unmodified QAA. 
Analogous to the SA runtime $\tsa(\varepsilon)$, $\tqmc(\varepsilon)$ is the number of QMC updates, normalized by $n/M$, where $M$ is the number of imaginary time slices, needed to sample from $\pi$ with total variation distance $\varepsilon<1/2$~\footnote{ 
We include the factor of $M$ in the definition of $\tqmc(\varepsilon)$ to reflect the space-time complexity of a single QMC update. 
Because we allow our QMC update rule to modify all $M$ Trotter slices, this complexity of $\mathcal{O}(M)$. Our results are unchanged if the normalization factor on $\tqmc(\varepsilon)$ is taken to be $n/m$ instead of $n/M$, where $m$ is the number of Trotter slices modified during a single update.}.
We consider any QMC algorithm which alters up to $k$ spins in each imaginary time slice per update, where $k$ is restricted to be constant in $n$. 

Crucial to our argument is the fact that before QMC encounters an independent set $\ket{z}$ with \mbox{$\hcost(z)\leq -\delta \setsize$}, it effectively samples from a restricted Hilbert space of only independent sets with \mbox{$\hcost(z)\geq -\delta (\setsize-1)$}. 
At any point during the adiabatic ramp, we let $H^{\restrb}$ denote the Hamiltonian in this restricted Hilbert space, with corresponding Gibbs populations $\pi_z^{\restrb}$. 
We let $\ket{z_{\max}}$ denote the configuration in this restricted Hilbert space within $k$ spin flips of an independent set of size $\setsize$ with the maximum Gibbs population $\pi_{z_{\max}}^{\restrb}$. 
Further, we let $e_{\max}^{\restrb} = \pi_{z_{\max}}^{\restrb} D_{\setsize-1}$ describe relative enhancement or suppression of its population compared to the uniform superposition state $\barket{\setsize-1}$. 

Analogous to SA, we then apply the Cheeger inequality to derive an upper bound on the QMC Markov chain spectral gap $\gqmc$, which gives a lower bound on $\tqmc(\varepsilon).$ 
This allows us to relate $\gqmc$ to the flow from populations in the Gibbs distribution of $\pi_z^{\restrb}$ to independent sets of size $\geq \setsize.$ 
This flow is proportional to $e_{\max}^{\restrb}D_{\setsize}/D_{\setsize-1}$, which gives us
\begin{align}
    \tqmc(\varepsilon)\geq\frac{\ln\big(\frac{1}{2\varepsilon}\big)}{2nkn^k}\max \frac{D_{\setsize-1}}{e_{\max}^{\restrb} D_\setsize}.\label{eq:tqmc_bound}
\end{align}
Eq.~\eqref{eq:tqmc_bound} shows that when the Gibbs distribution of the restricted Hilbert space is delocalized, i.e., when the $\hlaplace$ energy scale $\lambda$ is sufficiently large, the modified QAA has a quadratic speedup over QMC. 
In this case, $ e_{\max}^{\restrb} \leq 1$, so the QMC runtime goes like $\max_{\setsize} (D_{\setsize - 1}/D_\setsize)$, whereas the modified QAA runtime goes like $\max_{\setsize}\sqrt{D_{\setsize-1}/D_\setsize}$.
To match the modified QAA runtime, the restricted Gibbs distribution must be \textit{exponentially} favorably localized, so that $e_{\max}^{\restrb} = \sqrt{D_{\setsize-1} / D_\setsize}$. 
In this case, however, we expect the QAA runtime to be similarly enhanced under Eq.~\eqref{eq:off_diag_expansion} due to favorable localization.
Thus, QMC does not recover the quadratic speedup due to delocalization, which crucially stems from the quantum coherent enhancement of the coupling $\barbra{\setsize-1}\hdrive \barket{\setsize}$. \\

\begin{figure}[th!]
    \centering
    \includegraphics[width=.5\textwidth]{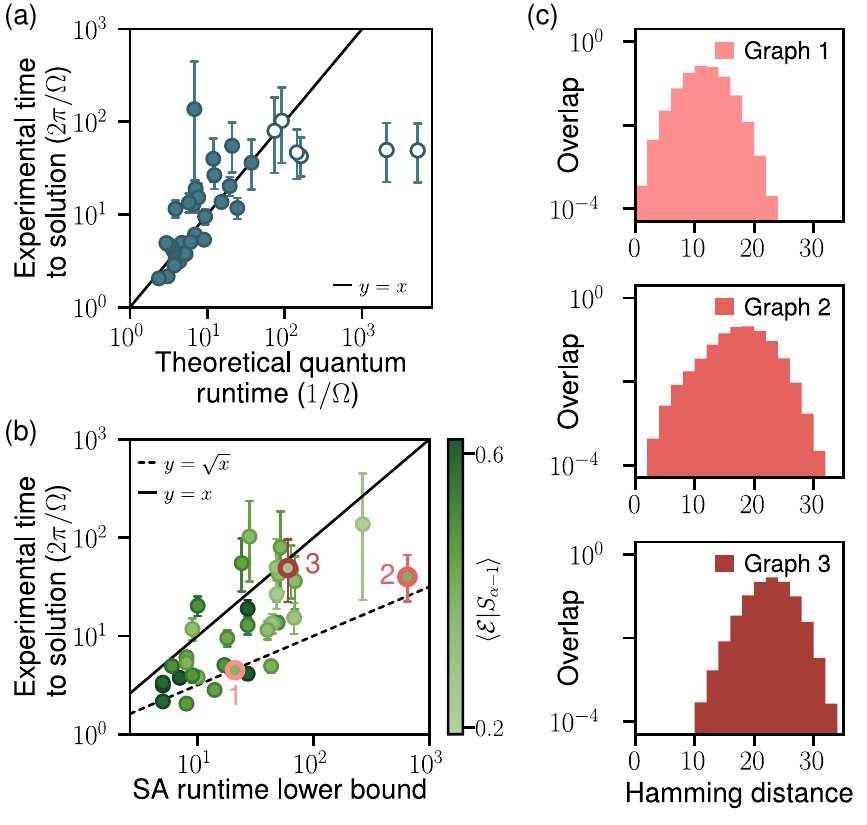}
    \caption{
    Analysis of the experimental performance.
    (a)~The experimental optimized time to solution correlates with the theoretical QAA runtime $\gqaa^{-1}$ on instances where the maximum experimental evolution time $T_\mathrm{max}$ can resolve the minimum gap ($T_\mathrm{max} \leq \gqaa^{-1}$, teal-filled points). 
    Instances for which the evolution time is too short to maintain adiabaticity deviate from the trend ($T_\mathrm{max} \geq \gqaa^{-1}$, white points).
    (b)~The experimental time to solution correlates less strongly with the SA runtime lower bound on instances where $\ketE$ is localized (light green points). 
    On the most delocalized instances (dark green points), the QAA runtime is similar to the square root of the SA runtime.
    (c)~We plot the distribution of Hamming distances between $\ketG$ and $\ketE$ for three localized graphs.  
    The pairwise Hamming distances are larger for the instance where QAA performs poorly relative to SA (bottom), and smaller for the instances where QAA outperforms SA (top, middle).
    }
    \label{fig:speedup_5}
\end{figure}

\tocless\section{Understanding the experimental observations} 
We now apply our framework to interpret recent  experiments on Rydberg atom arrays~\cite{Ebadi_2022} using the aforementioned hardware-efficient encoding of the Maximum Independent Set problem on unit-disk graphs. 
\citet{Ebadi_2022} observed that the experimental optimized QAA outperformed SA on certain hard unit-disk graph instances with a large ratio of $D_{\alpha-1}/D_\alpha$ \mbox{($n=39$\,--\,$80$)}. 
We compute the experimental optimized time to solution as~\cite{albash_lidar} 
\begin{align}
    \text{TTS}_{\text{opt}} = \min_T \frac{T}{\ln[1-p(T)]},
\end{align}
where $p(T)$ is the probability of QAA finding the optimal solution at evolution time $T$. 
In Fig.~\ref{fig:speedup_5}(a), we confirm that $\text{TTS}_\text{opt}$ goes like the theoretical runtime $\gqaa^{-1}$ computed numerically for the Rydberg Hamiltonian [Eqs.~\eqref{eq:hdrive} and~\eqref{eq:hryd}]. 

However, in Fig.~\ref{fig:speedup_5}(b), we find that $\text{TTS}_\text{opt}$ correlates less strongly with the SA runtime lower bound. 
To understand $\gqaa$, and therefore the experimental time to solution, we obtain perturbative estimates for the eigenstates at the avoided level crossing, $\ketG$ and $\ketE$, in the manifold of independent sets of size $\alpha$ and \mbox{$\alpha-1$}, respectively. 
On the more delocalized instances, $\gqaa^{-1}$ is similar to the square root of the SA runtime (Fig~\ref{fig:speedup_5}(b), dark green points), as expected from perturbation theory [Eq.~\eqref{eq:delocalized_coupling}]. 

In contrast, for more localized instances (light green points), we find that $\text{TTS}_\text{opt}$ is less correlated with the SA runtime. 
By Eq.~\eqref{eq:off_diag_expansion}, we expect $\gqaa$ to be small when the Hamming distance between $\ketG$ and $\ketE$ is large and $\loc$ is small, which we verify numerically in Appendix~\ref{subsec:exp_analysis}. 
For illustration, in Fig.~\ref{fig:speedup_5}(c) we examine three localized instances with vastly different SA and QAA runtimes. 
We plot the distribution of the product of populations $\mathcal{G}_z \mathcal{E}_{z'}$ of spin configurations $\ket{z}, \ket{z'}$ in $\ketG, \ketE$ over their Hamming distances.
The instance where SA outperforms QAA is highly localized (bottom, $n=80$), with large Hamming distances compared to the two other instances where QAA outperforms SA (top and middle, $n=65$). 
Due to favorable localization, these instances obtain a significant speedup over SA. 
Thus, the instance-dependent characteristics of $\ketG$ and $\ketE$ can be used to predict the experimental performance. \\

\tocless\section{Outlook\label{sec:outlook}} 
In this work, we have shown that the optimized QAA has a quadratic speedup over a wide class of classical Markov chain algorithms when the low-energy eigenstates are delocalized across a flat energy landscape. 
To promote delocalization on generic problem instances~\cite{altshuler_2010}, we modified QAA by adding a local Hamiltonian $\hlaplace$ with no sign problem, with a time-independent energy scale $\lambda$. 
To observe the corresponding quadratic speedup on near-term devices, the algorithm must be efficiently encoded in hardware~\cite{Stilck_Fran_a_2021}.  
The modified QAA is amenable to direct experimental implementation via hybrid digital-analog Trotterized evolution~\cite{Bluvstein2022}, by generating spin-exchange interactions with excitation into $S$ and $P$ Rydberg states or microwave driving~\cite{Browaeys_2016, Browaeys2020}, and decomposing the diagonal component of $\hlaplace$ into multiqubit controlled phase gates. 
Local detunings can generate the diagonal component of $\hlaplace$ on certain instances with structured configuration graphs, such as when the suboptimal configurations correspond to the motion of a domain wall~\cite{nguyen_2023}. 

Similar to other problems involving Grover-type quadratic speedups~\cite{Roland_2002, krovi_2010}, our approach requires optimizing the QAA evolution to maintain adiabaticity.   
Optimizing QAA evolution in general is an open problem; however, recent work has shown that it is possible to optimize a wide class of QAA algorithms which use the reflection about the uniform superposition state, \mbox{$\mathds{1}-\frac{1}{2^n}\sum_{z, z'}\ket{z}\bra{z'}$}, to drive the evolution instead of $\hdrive$~\cite{jarret_2019}. 
In Appendix~\ref{subsubsec:optimization}, we describe approaches to optimizing the modified QAA when $\lambda\to\infty$, which retain a quadratic speedup. 
Future work could attempt to generalize these results to finite $\lambda$.  
At the same time, one could circumvent the need for optimization by identifying instances with an exponential, rather than quadratic, speedup over SA. 
One approach could be to characterize instances where the low-energy eigenstates are favorably localized at small Hamming distance~\cite{farhi_sa_2002}. 
However, QAA may not generically provide a speedup over QMC on these instances~\cite{farhi_sa_2002, Crosson_2016, Isakov_2016}. 
It remains an open question whether instances exist with an exponential speedup over both SA and QMC, despite optimistic results in the black-box setting~\cite{Hastings_2021, gilyen}. 

It would also be interesting to extend our results beyond flat energy landscapes to problems with the Overlap Gap Property, whose optimal solutions are provably hard to approximate for large classes of both quantum and classical algorithms~\cite{ogp, farhi_2020}. 
In these instances, independent sets of the same size form ``clusters'' separated by large Hamming distances.
As the clusters are disconnected under spin-exchange operations, they independently delocalize, such that the effective Hamiltonian is a tree-like version of the one-dimensional tight-binding Hamiltonian $\htb$ when $\lambda\to\infty$ [Eq.~\eqref{eq:tight_binding_hamiltonian}]. 
Future work could investigate the modified QAA on instances with the Overlap Gap Property using this framework. 
Particularly interesting is the prospect of studying QAA performance in the \textit{diabatic} regime, which can outperform both SA and QMC in finding approximate solutions on certain problem instances~\cite{King2023}.
Utilizing non-adiabatic phenomena via quantum quench algorithms may provide an alternative mechanism for quantum speedup~\cite{Bernien2017, Hastings_2019, Crosson_2021, Schiffer2023}. \\

\tocless\section{Acknowledgements}
We would like to thank Tameem Albash, Lisa Bombieri, Dolev Bluvstein, Sepehr Ebadi, Nicholas Ezzell, Aram Harrow, Marcin Kalinowski, Andrew King, Daniel Lidar, Subir Sachdev, Benjamin Schiffer, Juspreet Singh Sandhu, Lei Wang, and Zhongda Zeng for helpful discussions. 
This work was supported by the US Department of Energy [DE-SC0021013 and DOE Quantum Systems Accelerator Center (contract no. 7568717)], the Defense Advanced Research Projects Agency (grant no. W911NF2010021), the Army Research Office (grant No. W911NF-21-1-0367), the National Science Foundation, the Harvard-MIT Center for Ultracold Atoms, and the European Research Council (grant No.~101041435). 
M.C. acknowledges support from Department of Energy Computational Science Graduate Fellowship under Award Number (DESC0020347).  
S.C. is grateful for support from the NSF under Grant No. DGE-1845298. R.S. is supported by the Princeton Quantum Initiative Fellowship. \\

\appendix 

\section{Runtime lower bounds for classical Markov chain algorithms \label{sec:mcmc_runtime}}

\subsection{Simulated annealing\label{subsec:sa_runtime}}
In this section, we establish a runtime lower bound on all simulated annealing (SA) algorithms using the Metropolis-Hastings update rule. 
Although we focus on the Maximum Independent Set problem in our proof, we will show that our bound applies to generic combinatorial optimization problems. The goal of SA is to sample from an equilibrium probability distribution $\pi$, which we take to be the thermal Gibbs distribution of $\hcost$ at temperature $1/\beta$,
\begin{align}
    \pi_z = \frac{e^{-\beta \hcost(z)}}{\mathcal{Z}_\beta}, \qquad \mathcal{Z}_\beta=\sum_{b}D_{b}e^{\beta\delta b},
\end{align}
where $\pi_z$ is the probability of spin configuration \mbox{$\ket{z}\in\{\ket{0}, \ket{1}\}^n$}, $\mathcal{Z}_\beta$ is the partition function, and $D_\setsize$ is the number of independent sets of size $\setsize$. 
SA stochastically updates a spin configuration $\ket{z}$ to $\ket{z'}$ according to the Markov chain transition  probabilities $P_{z, z'}$. We consider $P_{z, z'}$ given by the Metropolis-Hastings update rule~\cite{Metropolis, Hastings},
\begin{align}
    P_{z, z'} = p_{z, z'}\min\Big( 1, e^{-\beta [\hcost(z')-\hcost(z)]}\Big),\label{eq:mh_update}
\end{align}
where $p_{z, z'} = p_{z', z}$ is the probability of proposing to update from $\ket{z}$ to $\ket{z'}$, and the remaining factor is the probability of accepting the proposed update. 
One can check that for $p_{z, z'} = p_{z', z}$, the update rule satisfies the detailed balance condition,
\begin{align}
    P_{z, z'}\pi_z = P_{z', z} \pi_{z'}.\label{eq:detailed_balance_si}
\end{align}

The \textit{mixing time} of SA is defined as the minimum number of proposed updates per spin to prepare the Gibbs distribution with error (measured in total variation distance, see~\cite{markov_chain_mixing}) less than or equal to $\varepsilon,$  starting from any initial probability distribution $\mu$. 
The total variation distance between two distributions is equal to half the $l_1$ norm of $\pi-\mu$~\cite{markov_chain_mixing}.
We define the SA runtime at inverse temperature $\beta$, $\tsa(\varepsilon, \beta)$, as the mixing time normalized by the Gibbs population of the optimal independent sets of size $\alpha$. 
Explicitly, we let (see~\cite{markov_chain_mixing}, Eqs. 4.2 and 4.30)
\begin{align}
    \tsa(\varepsilon, \beta) = \frac{1}{n\pi_{\alpha}} \min\Big\{t: \max_\mu \sum_{z\in\{0, 1\}^n} |\pi_z - P^t\mu_z|\leq \varepsilon \Big\}, \label{eq:tsa_beta_definition}
\end{align}
where $\mu$ is the initial distribution, $P=(P_{z, z'})$ is the matrix of Markov chain transition probabilities, and
\begin{align}
    \pi_\setsize = \sum_{z: \hcost(z) = -\delta\setsize} \pi_z  
\end{align}
is the Gibbs population of independent sets of size $\setsize$. 
$\tsa(\varepsilon, \beta)$ represents the time to sample an optimal solution from the Gibbs distribution. 
The normalization factor of $1/\pi_\alpha$ is necessary because at high temperatures, the mixing time may be small, but the Gibbs population of the optimal solutions is correspondingly very small. 
At low temperatures, the normalization factor is unnecessary because optimal solutions have high Gibbs population. 
As a result, the mixing time is directly related to the time to find an optimal solution, maximized over all optimal solutions (i.e., the \textit{hitting time})~\cite{peres_sousi}.
We define the SA runtime $\tsa(\varepsilon)$ as the minimum runtime over all temperatures,
\begin{align}
    \tsa(\varepsilon) = \min_\beta \tsa(\varepsilon, \beta).\label{eq:tsa_minimum_over_beta}
\end{align}
Our main result, stated next, is an analytic lower bound on $\tsa(\varepsilon)$.

\begin{theorem} \label{thm:sa_runtime_bound}
    Consider any Metropolis-Hastings SA algorithm that prepares the Gibbs distribution of the Maximum Independent Set cost Hamiltonian $\hcost$. Suppose the SA update rule alters at most $k$ of the $n$ total spins.  
    Define a cutoff independent set size $\setsize^\star$, such that the number of larger independent sets is decreasing, i.e. $D_{\setsize-1}/D_\setsize\geq 1$ for  $\setsize > \setsize^\star$.  
    Then for any error $\varepsilon < 1/2$, the SA runtime $\tsa(\varepsilon)$ can be lower-bounded as
\begin{equation}
    \tsa(\varepsilon) \geq  \frac{\ln\big(\frac{1}{2\varepsilon}\big)}{2nk} \max_{\setsize > \setsize^\star} \frac{D_{\setsize-1}}{D_\setsize}.\label{eq:tsa_lower_bound_si}
\end{equation}
\end{theorem}

Before proceeding, we note that the restriction $\setsize > \setsize^\star$ appearing in Theorem~\ref{thm:sa_runtime_bound} is not necessary when the independence polynomial of the graph is \textit{unimodal}, meaning that $D_0\leq D_1\leq \dots \leq D_{\setsize^\star}\geq \dots\geq D_{\alpha-1}\geq D_{\alpha}$. 
This condition is met for every unit-disk graph we study in Appendix~\ref{sec:runtime_system_size}.\\ 

\noindent{\textit{Proof.}} The SA runtime at temperature $1/\beta$ can be lower-bounded by the inverse of the spectral gap $\gsa = \gsa(\beta)$ between the largest and second largest eigenvalue of the corresponding Markov chain matrix $P$ with transition probabilities $P_{z, z'}$ as~(\cite{markov_chain_mixing}, Eq.~12.14)
\begin{align}
    \tsa(\varepsilon, \beta)\geq \frac{\ln\big(\frac{1}{2\varepsilon}\big)}{n\pi_{\alpha}}\left(\frac{1}{\gsa}-1\right)\label{eq:sa_gap_bound}.
\end{align}
Because $\frac{1}{\gsa}\gg 1$, we will ignore the second term.
This bound applies to any Markov chain transition matrix $P$ which satisfies detailed balance and is \textit{lazy}, meaning that the outwards transition probability $\sum_{z': z'\neq z}P_{z, z'}\leq 1/2$ for any $\ket{z}$. 
Any Markov chain $P$ can be made lazy by taking $(P+\mathds{1})/2$ (i.e., adding weight-1/2 self-loops to each $\ket{z}$). 
This transformation does not substantially affect the mixing time because it reduces the outwards transition probability by at most a factor of $2$, so we will analyze $P$ instead of $(P+\mathds{1})/2$. 
Note that in Eq.~\eqref{eq:sa_gap_bound} we divided the standard definition of mixing time by $n$ because we allow the SA algorithm to ``parallelize" updates over different spins.

We can therefore lower-bound $\tsa(\varepsilon, \beta)$ by upper bounding $\gsa$ and $\pi_\alpha$. To do this, we use the Cheeger inequality~\cite{cheeger_ref_diaconis}, which can be used to establish an upper bound on $\gsa$ for any Markov chain satisfying detailed balance. 
The idea in a Cheeger bound is to bipartition the state space of the Markov chain into two sets, $S$ and $S^c$, such that in the Gibbs distribution $\pi$, very little probability flows from $S$ to $S^c$ during one update of the Markov chain. 
The spectral gap is then upper bounded by this probability flow $Q_{S,S^c}$ normalized by the total Gibbs population $\pi_S$ in $S$. Explicitly, the Cheeger inequality states
\begin{align}
    \gsa\leq \frac{2 Q_{S,S^c}}{\pi_S}, \qquad \pi_S = \sum_{z\in S}\pi_z, \label{eq:cheeger_bound}
\end{align}
for any $S$ with $\pi_S<\frac{1}{2}$, where 
\begin{align}
    Q_{S,S^c}&=\sum_{z\in S, z'\in S^c}\pi_z P_{z, z'}\label{eq:flow_from_S}\\
    &=\sum_{z\in S, z'\in S^c}\pi_{z'} P_{z', z}\nonumber\\
    &= Q_{S^c, S}\nonumber
\end{align}
is the flow from $S$ to $S^c$. Note that $Q_{S,S^c}=Q_{S^c,S}$ follows from the detailed balance condition on $P$ in Eq.~\eqref{eq:detailed_balance_si}. 
When $Q_{S, S^c}$ is small, $\gsa$ is correspondingly small by Eq.~\eqref{eq:cheeger_bound} and the SA runtime is large by Eq.~\eqref{eq:sa_gap_bound}.

We will first consider the low temperature case \mbox{$\pi_S < 1/2$}, and obtain an upper bound on $Q_{S,S^c}/\pi_S$. 
Let \mbox{$k\in \{1, 2, \dots, n\}$} denote the maximum number of spins altered during a proposed update, and $\setsize\in \{\setsize^\star, \setsize^\star+1, \dots, \alpha\}$ represent a particular independent set size satisfying $\setsize > \setsize^\star$. 
We define the set
\begin{align}
    S=\{z: \hcost(z)\geq -\delta (\setsize-1)\}
    \label{eq:cheeger_S}
\end{align}
of independent sets of size $\setsize-1$ or smaller. 
We first replace all the probabilities $\pi_z$ in Eq.~\eqref{eq:flow_from_S} with $e^{\beta\delta (\setsize-1)}/\mathcal{Z}_\beta$.
This gives an upper bound on $Q_{S, S^c}/\pi_S$, because $\hcost=-\delta(\setsize-1)$ is the smallest value of $\hcost$ present in $S$:
\begin{align}
    \frac{Q_{S,S^c}}{\pi_S}&\leq\frac{e^{\beta\delta(\setsize-1)}}{\pi_S\mathcal{Z}_\beta}\sum_{\substack{\hcost(z)\geq-\delta(\setsize-1) \\ \hcost(z')\leq-\delta\setsize}}P_{z,z'}.
\end{align}
Now, plugging in the Metropolis-Hastings update rule from Eq.~\eqref{eq:mh_update}, we have
\begin{align}
    \frac{Q_{S,S^c}}{\pi_S}&\leq\frac{e^{\beta\delta(\setsize-1)}}{\pi_S\mathcal{Z}_\beta} \hspace*{-0.4cm} \sum_{\substack{z': \hcost(z')\leq-\delta\setsize \\
    z: \hcost(z)\geq-\delta(\setsize-1)}} \hspace{-0.4cm} p_{z,z'} \min\bigg( 1, \frac{e^{-\beta [\hcost(z')]}}{e^{-\beta\hcost(z)]}}\bigg)\nonumber\\
    &=\frac{e^{\beta\delta(\setsize-1)}}{\pi_S\mathcal{Z}_\beta}\sum_{\hcost(z')\leq-\delta\setsize}\bigg(\sum_{\hcost(z)\geq-\delta(\setsize-1)}p_{z',z} \bigg).
    \label{eq:flow_boundable}
\end{align}
where in the second line we have used that $p_{z, z'} = p_{z', z}$ under the detailed balance condition [Eq.~\eqref{eq:detailed_balance_si}].
The inner summation over configurations $\ket{z}$ at fixed $\ket{z'}$ is equal to the probability of proposing an update from $\ket{z'}$ to any configuration $\ket{z}$ with $\hcost(z)\geq-\delta(\setsize-1)$.
This probability is at most one because the total transition probability out of $\ket{z'}$ into $S$ is at most one, and is strictly zero if $\hcost(z')< -\delta(\setsize+k-1)$ (because we have assumed that we update at most $k$ spins). This constraint yields
\begin{align}
    \frac{Q_{S,S^c}}{\pi_S}&\leq\frac{e^{\beta\delta(\setsize-1)}}{\pi_S\mathcal{Z}_\beta}\sum_{-\delta\min(\alpha,\setsize+k-1)\leq \hcost(z')\leq-\delta\setsize}1\nonumber\\
    &=\frac{e^{\beta\delta(\setsize-1)}}{\pi_S\mathcal{Z}_\beta}\sum_{\setsize'=\setsize}^{\min(\alpha,\setsize+k-1)}D_{\setsize'}\nonumber\\
    &\leq \frac{kD_\setsize e^{\beta\delta(\setsize-1)}}{\pi_S\mathcal{Z}_\beta} \nonumber\\
    &\leq \frac{kD_\setsize }{D_{\setsize-1}}.
    \label{eq:flow_bounded}
\end{align}
In the third step we used the fact that $D_{\setsize}\geq D_{\setsize'}$ for any $\setsize' > \setsize^\star$, and in the fourth step we have used \mbox{$\pi_S=\sum_{\setsize'=0}^{\setsize-1} D_{b'}e^{\beta\delta b'}/\mathcal{Z}_\beta>D_{\setsize-1}e^{\beta\delta(\setsize-1)}/\mathcal{Z}_\beta$}.
From Eq.~\eqref{eq:cheeger_bound}, the SA spectral gap $\gsa$ is thus bounded as
\begin{align}
    \gsa \leq  \frac{2Q_{S, S^c}}{\pi_S} \leq \frac{2 k D_\setsize}{D_{\setsize-1}}. \label{eq:gsa_bound_high_temp}
\end{align}
Combining this with the lower bound on runtime $\tsa(\varepsilon, \beta)$ [Eq.~\eqref{eq:sa_gap_bound}], and plugging in $\pi_\alpha \leq 1$, we have for any $\beta$ such that $\pi_S < 1/2,$
\begin{align}
    \tsa(\varepsilon, \beta)\geq \frac{\ln\big(\frac{1}{2\varepsilon}\big)}{2nk}\frac{D_{\setsize-1}}{D_\setsize}.
\end{align}

On the other hand, at high temperatures $\pi_S>1/2$, we must swap $S$ with $S^c$ in the Cheeger bound [Eq.~\eqref{eq:cheeger_bound}], 
\begin{align}
\gsa \pi_\alpha &\leq \frac{2Q_{S^c,S}\pi_{\alpha}}{\pi_{S^c}} = \frac{2Q_{S,S^c}\pi_{\alpha}}{\pi_{S^c}} 
\end{align}
where we have used the fact that $Q_{S, S^c} = Q_{S^c, S}.$
By Eq.~\eqref{eq:flow_bounded} we have $Q_{S,S^c}\leq kD_be^{\beta\delta(b-1)}/\mathcal{Z}_\beta$, so we find
\begin{align}
    \gsa\pi_\alpha &\leq \frac{2k D_\setsize e^{\beta\delta(\setsize-1)}\pi_\alpha}{\mathcal{Z}_\beta \pi_{S^c}} \leq \frac{2k D_\setsize e^{\beta\delta(\setsize-1)}}{\mathcal{Z}_\beta},
\end{align}
using $\pi_\alpha\leq \pi_{S^c}$ (because sets of size $\alpha$ are contained in $S^c$). Now, since $\mathcal{Z}_\beta>D_{\setsize-1}e^{\beta\delta(\setsize-1)},$ we are left with
\begin{align}
    \gsa\pi_\alpha &\leq \frac{2k D_\setsize}{D_{\setsize-1}},
\end{align}
which gives the same bound as in the low-temperature case via Eq.~\eqref{eq:sa_gap_bound}. 
Because the same bound holds for all temperatures and for any $\setsize > \setsize_\star$, we can use Eq.~\eqref{eq:tsa_minimum_over_beta} to obtain a lower bound on $\tsa(\varepsilon)$, which gives us Theorem~\ref{thm:sa_runtime_bound}.

Finally, we note that Theorem~\ref{thm:sa_runtime_bound} can be applied to general combinatorial optimization problems with discrete cost Hamiltonian energies. 
Our proof does not change if we replace the energies of $\hcost$, $\{-\delta\setsize\}_{\setsize = 0, 1, \dots, \alpha}$, with energies $\{E_\setsize\}_{\setsize = 0, 1, \dots, \alpha}$ for any generic cost function with $\alpha+1$ discrete energy levels, and let $D_\setsize$ represent the number of spin configurations with energy $E_\setsize.$ 
As a result, Theorem~\ref{thm:sa_runtime_bound} can be applied to generic discrete cost functions beyond Maximum Independent Set.

\subsection{Parallel tempering\label{subsec:pt_runtime}}
We now derive a runtime lower bound for a wide class of parallel tempering algorithms using the Metropolis-Hastings update rule. 
Because our bound uses identical techniques to the runtime lower bound for SA, we recommend the reader read Appendix~\ref{subsec:sa_runtime} before proceeding.
In parallel tempering there are $M$ copies, or \textit{replicas}, of the $n$-spin system of SA, each equilibrating to the Gibbs distribution of $\hcost$ at temperatures $1/\beta_1, \dots, 1/\beta_M$. 
The state space is the product of states over all the replicas $\{z_1 \dots z_M\}$, where $z_i\in \{0, 1\}^n$ represents the spin configuration of the $i$th replica. 
Similar to SA, the state of a single replica can be updated based on proposing an update to at most $k$ spins. 
However, in parallel tempering collective updates involving multiple replicas are also possible. 
We will consider collective Metropolis-Hastings update rules, 
\begin{align}\label{eq:pt_mh_update}
    & P_{z_1\dots z_M, z_1' \dots z_M'} \\
    & \quad = p_{z_1\dots z_M, z_1' \dots z_M'} \min\Big(1, e^{-\sum_{i=1}^M\beta_i [\hcost(z_i')-\hcost(z_i)] }\Big), \nonumber
\end{align}
where $p_{z_1\dots z_M, z_1' \dots z_M'}$ is the probability of proposing an update to the configuration $z_1' \dots z_M'$ given that the current configuration is $z_1\dots z_M$.
Note that this update rule satisfies the detailed balance condition in~Eq.~\eqref{eq:detailed_balance_si}. 
The equilibrium distribution is therefore the Gibbs distribution,
\begin{align}
    \pi_{z_1\dots z_M}=\frac{e^{-\sum_{i=1}^M\beta_i\hcost(z_i)}}{\prod_{i=1}^M\mathcal{Z}_{\beta_i,i}}, \quad  \mathcal{Z}_{\beta_i,i}=\sum_{b=0}^{\alpha}D_be^{\beta_i\delta b}.
\end{align}

We define the parallel tempering runtime $\tpt(\varepsilon)$ as
\begin{align}
    \tpt(\varepsilon) = \min_{\beta_1\dots\beta_M} \tpt(\varepsilon, \beta_1\dots \beta_M),
\end{align}
where $\tpt(\varepsilon, \beta_1\dots \beta_M)$ is the runtime lower bound for replica temperatures $\beta_1\dots\beta_M$ defined similarly to SA~[Eq.~\eqref{eq:tsa_beta_definition}]:
\begin{align}
    &\tpt(\varepsilon, \beta_1\dots \beta_M) \label{eq:tpt_beta_definition}\\ &\qquad \quad = \frac{M}{n\pi_{\alpha}} \min\Big\{t: \max_\mu \hspace*{-0.2cm} \sum_{z\in\{0, 1\}^n} \hspace*{-0.2cm} |\pi_z - P^t\mu_z|\leq \varepsilon \Big\},\nonumber
\end{align}
where $P$ is the parallel tempering Markov chain, $\mu$ is the initial probability distribution, and 
\begin{align}
    \pi_\alpha = \sum_{i=1}^M\sum_{\substack{z_1\dots z_M : \\ \hcost(z_i) = -\delta\setsize }} \pi_{z_1\dots z_M}
\end{align}
is now the probability that the configuration of at least one replica is an independent set of size $\setsize$. 
Note that $\tpt(\varepsilon, \beta_1\dots\beta_M)$ in Eq.~\eqref{eq:tpt_beta_definition} has a factor of $M$ in the numerator. 
This is because we allow the parallel tempering update rule to update the spin configuration on all $M$ replicas; thus, the time complexity to perform an update is $\mathcal{O}(M).$
This also excludes the possibility of a trivial ``speedup'' from making $M$ exponentially large, at the expense of, e.g., $M=\mathcal{O}(2^n)$ space-time complexity.

\subsubsection{Replica exchange, arbitrary single-replica updates, and constant-sized collective updates}

We first consider parallel tempering algorithms that include the following update rules: single-replica updates that can update an arbitrary number of spins $k$ on a single replica, collective-replica updates that modify $k'$ spins on each replica, where $k'$ is restricted to be constant in $n$, and replica exchange updates. 
Replica exchange updates are defined as proposing to exchange the states $z_i$ and $z_j$ of two replicas $i$ and $j$. 
Our runtime lower bound is stated next in Theorem~\ref{thm:pt_runtime_bound}. We will generalize our result to include non-local \textit{isoenergetic cluster updates}~\cite{Houdayer_2001} later in Theorem~\ref{thm:pt_runtime_bound_isoenergetic}.

\begin{theorem}\label{thm:pt_runtime_bound}
    Consider a parallel tempering algorithm with $M$ replicas and any update rule as described above.  
    Define a cutoff independent set size $\setsize^\star$, such that the number of larger independent sets is decreasing, i.e. $D_{\setsize-1}/D_\setsize\geq 1$ for  $\setsize > \setsize^\star$. 
    Then for any error $\varepsilon < 1/2$, the parallel tempering runtime $\tpt(\varepsilon)$ is bounded as
    \begin{align}
        \tpt(\varepsilon) \geq \frac{\ln\big(\frac{1}{2\varepsilon}\big)}{2nk'n^{k'}} \max_{\setsize > \setsize_\star}\frac{D_{\setsize-1}}{D_\setsize}. \label{eq:tpt_runtime_bound}
    \end{align}
\end{theorem}

\noindent \textit{Proof. } Define the set $\mathcal{S}$ as the set of states with all the replicas having independent set size less than $\setsize$, for $\setsize > \setsize_\star$,
\begin{align} 
    \mathcal{S} &= \{ z_1 \dots z_M : \forall i\in \{1, \dots, M\}, \hcost(z_i) \geq -\delta (\setsize-1) \} \nonumber \\
    &= S_1 \times \dots \times  S_M, \label{eq:pt_partition}
\end{align}
where $S_i$ is the partition defined for a single replica as defined in Eq.~\eqref{eq:cheeger_S}. 
As with the SA runtime lower bound in  Appendix~\ref{subsec:sa_runtime}, our goal is to bound the flow of probability $Q_{\mathcal{S}, \mathcal{S}^c}$ from $\mathcal{S}$ to $\mathcal{S}^c$ in the Gibbs distribution, 
\begin{align}
    Q_{\mathcal{S},\mathcal{S}^c}&=\sum_{\substack{z_1\dots z_M\in \mathcal{S}\\ z_1'\dots z_M'\in \mathcal{S}^c}}\pi_{z_1\dots z_M} P_{z_1\dots z_M, z_1'\dots z_M'},
\end{align}
to obtain a Cheeger bound on the spectral gap of the parallel tempering Markov chain $\gpt = \gpt(\beta_1\dots\beta_M)$,  
\begin{align}
    \gpt \leq \frac{2Q_{\mathcal{S},\mathcal{S}^c}}{\pi_\mathcal{S}}&=\frac{2\sum_{\substack{z_1\dots z_M\in \mathcal{S}\\ z_1'\dots z_M'\in \mathcal{S}^c}}\pi_{z_1\dots z_M} P_{z_1\dots z_M, z_1'\dots z_M'}}{ \prod_{i=1}^M \pi_{S_i}} \label{eq:flow_pt} 
\end{align}
where $\pi_{S_i}$ is the Gibbs population of $S_i$ on a replica $i$, as in Eq.~\eqref{eq:cheeger_bound}, and $\pi_{\mathcal{S}}$ is the Gibbs population of $\mathcal{S}$. 
Eq.~\eqref{eq:flow_pt} gives a lower bound on the runtime via Eq.~\eqref{eq:sa_gap_bound}. 
From Eq.~\eqref{eq:flow_pt} we can immediately see that replica exchange updates do not contribute to $Q_{\mathcal{S}, \mathcal{S}^c}$ because swapping the states of two replicas in $\mathcal{S}$ does not transfer probability from $\mathcal{S}$ to $\mathcal{S}^c$. 
In addition, arbitrary updates to a single replica are subject to the same bound as SA [Eq.~\eqref{eq:tsa_lower_bound_si}]. 
Therefore, it only remains to bound collective updates that update at most $k'$ spins on each replica, where $k'$ is constant in $n$. 
The runtime lower bound is then given by the minimum of the runtime lower bounds on collective updates and single-replica updates. 
We will find that the runtime lower bound for collective updates is smaller than for single-replica updates; hence, Theorem~\ref{thm:pt_runtime_bound} reflects the collective update bound.

As before, we will obtain an upper bound on $Q_{\mathcal{S},\mathcal{S}^c}$. 
Notice that only transitions from configurations in $\mathcal{S}$ within $k'$ spin flips of some $z_1\dots z_M\in\mathcal{S}^c$ can contribute to $Q_{\mathcal{S},\mathcal{S}^c}$. 
We denote these configurations as $\partial \mathcal{S}$. 
Configurations in $\partial \mathcal{S}$ must have least one replica $j$ within $k'$ spin flips of $S^c_j$, whereas all other replicas $i$ may be in any configuration in $S_i$. 
We let $\partial S_j$ denote configurations $z_j \in S_j$ within $k'$ spin flips of $S^c_j$, and $\pi_{\partial S_j} = \sum_{z_j\in \partial S_j}\pi_{z_j}$. We then can show
\begin{align}
    Q_{\mathcal{S},\mathcal{S}^c}
    &\leq\sum_{\substack{z_1\dots z_M\in \partial\mathcal{S}}}\pi_{z_1\dots z_M}\nonumber \\
    &\leq \sum_{j=1}^M \pi_{\partial S_j} \prod_{\substack{i=1\\ i\neq j}}^M \pi_{S_i} = \sum_{j=1}^M \pi_{\partial S_j} \frac{\pi_\mathcal{S}}{\pi_{S_j}}\nonumber \\
    &\leq (k')^2{n\choose k'}\sum_{j=1}^M\frac{D_\setsize e^{\delta \beta_j (\setsize-1)}}{\mathcal{Z}_{\beta_j, j}}\frac{\pi_\mathcal{S}}{\pi_{S_j}} \nonumber\\
    &\leq k'n^{k'}\sum_{j=1}^M\frac{D_\setsize e^{\delta \beta_j (\setsize-1)}}{\mathcal{Z}_{\beta_j, j}}\frac{\pi_\mathcal{S}}{\pi_{S_j}},
    \label{eq:pt_flow_bound}
\end{align}
where in the third line we have used the fact that the Gibbs population of any configuration in $\partial S_j$ is \mbox{$\leq e^{\delta\beta_j(b-1)}/\mathcal{Z}_{\beta_j, j}$}, and the fact that there are \mbox{$\leq (k')^2{n \choose k'} D_b$} such configurations.
Then we may write
\begin{align}
    \frac{Q_{\mathcal{S}, \mathcal{S}^c}}{\pi_\mathcal{S}} &\leq k'n^{k'}\sum_{j=1}^M\frac{D_\setsize e^{\delta \beta_j (\setsize-1)}}{\mathcal{Z}_{\beta_j, j}\pi_{S_j}}\nonumber\\
    &\leq k'n^{k'}\sum_{j=1}^M\frac{D_\setsize }{D_{\setsize-1}}= k' n^{k'}M\frac{D_\setsize }{D_{\setsize-1}}, \label{eq:pt_cheeger_bound}
\end{align}
where in the second line we have used that $\pi_{S_j}\geq D_{\setsize-1}e^{\beta_j\delta (\setsize-1)}/\mathcal{Z}_{\beta_j, j}$. 
We can combine Eqs.~\eqref{eq:pt_cheeger_bound} and~\eqref{eq:cheeger_bound} and use the fact that $\pi_\alpha\leq 1$ to get 
\begin{align}
    \gpt \leq 2k'n^{k'}M\min_{\setsize > \setsize_\star}\frac{D_\setsize}{D_{\setsize-1}}
\end{align}
for the spectral gap of the parallel tempering Markov chain $\gpt.$
This, combined with Eq.~\eqref{eq:sa_gap_bound}, yields a runtime lower bound for parallel tempering given by Eq.~\eqref{eq:tpt_runtime_bound}. 
We emphasize that this bound is only restrictive when $\max_{\setsize > \setsize_\star} (D_\setsize / D_{\setsize-1})$ is much larger than $n^{k'}$.
Because $\max_{\setsize > \setsize_\star} (D_\setsize / D_{\setsize-1})$ grows exponentially in $\sqrt{n}$ in the worse case for the Maximum Independent Set problem on unit-disk graphs (see Appendix~\ref{sec:runtime_system_size}), this bound is only useful when $k'$ does not grow with $n$.

The above result holds when $\pi_\mathcal{S} <1/2$. 
Just as in the case of SA, we can derive the same bound on the runtime when $\pi_\mathcal{S} > 1/2$.  
Using the fact that $Q_{\mathcal{S}, \mathcal{S}^c} = Q_{\mathcal{S}^c, \mathcal{S}}$ [see Eq.~\eqref{eq:flow_from_S}], we  use Eq.~\eqref{eq:pt_flow_bound} to receive:
\begin{align}
    \gpt \pi_{\alpha}&\leq \frac{2Q_{\mathcal{S}, \mathcal{S}^c} \pi_\alpha}{\pi_{\mathcal{S}^c}}\leq 2Q_{\mathcal{S}, \mathcal{S}^c} \nonumber \\
    &\leq 2 k' n^{k'}M\min_{\setsize > \setsize_\star}\frac{D_\setsize }{D_{\setsize-1}}.
\end{align}
As a result, the same bound Eq.~\eqref{eq:tpt_runtime_bound} holds for the case where $\pi_{\mathcal{S}}> 1/2.$

Finally, we note that  Theorem~\ref{thm:pt_runtime_bound} can be applied to general combinatorial optimization problems with discrete cost Hamiltonian energies. 
Our proof, as in the case of SA, does not change if we replace the energies of $\hcost$, $\{-\delta\setsize\}_{\setsize = 0, 1, \dots, \alpha}$, with   energies $\{E_\setsize\}_{\setsize = 0, 1, \dots, \alpha}$ for any generic cost function with $\alpha+1$ discrete energy levels, and let $D_\setsize$ represent the number of spin configurations with energy $E_\setsize.$ 

\subsubsection{Isoenergetic cluster updates}
Here we obtain a runtime lower bound for all parallel tempering algorithms that use the same update rules as in the previous Appendix~\ref{subsec:pt_runtime}, in addition to \textit{isoenergetic cluster updates}, which are non-local updates designed specifically for optimizing two-dimensional spin glasses~\cite{Houdayer_2001}. 
Isoenergetic cluster updates collectively update a pair of replicas $i, j$ by identifying clusters of spins (vertices) connected by edges for which $z_i$ and $z_j$ differ. 
The update rule then proposes to exchange the configurations of spins within a randomly chosen connected cluster between $z_i$ and $z_j$. 
One can check that this update rule conserves the total energy of the two replicas: $\hcost(z_i) + \hcost(z_j) = \hcost(z_i') + \hcost(z_j')$, where $z_i'$ and $z_j'$ are the spin configurations after an isoenergetic cluster update.
Note that isoenergetic cluster updates are equivalent to replica exchange updates when there is only a single connected cluster of differing spins. 

\begin{figure}[h!]
    \centering
    \includegraphics[]{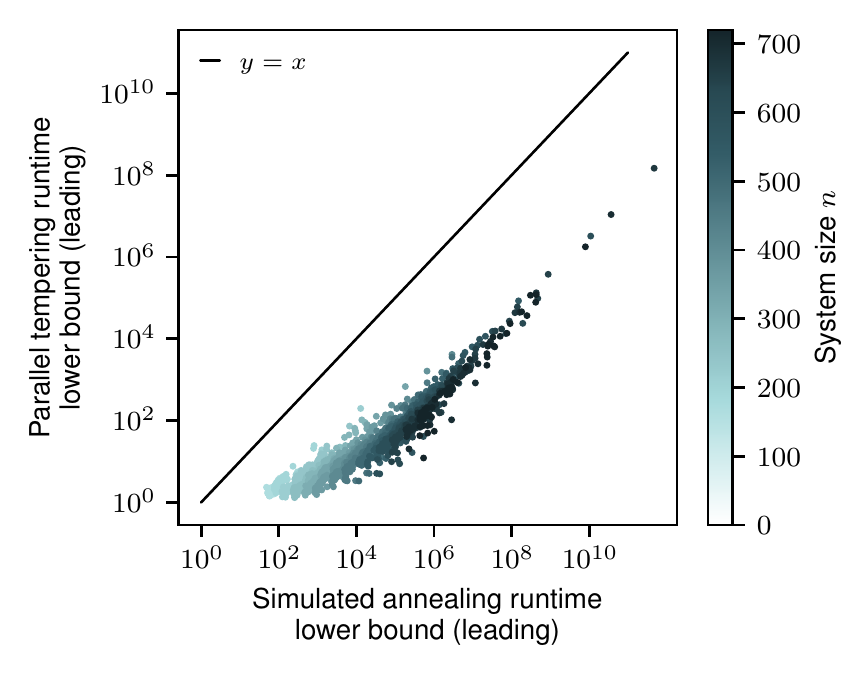}
    \caption{Simulated annealing and parallel tempering runtime lower bounds. 
    We plot $\max_{\setsize > \setsize^\star} \big[D_\setsize / D_{\setsize-1} +   \sum_{b_1'=b}^{\alpha}\sum_{b_2'=0}^{2(b-1)-b_1}\sum_{k= b_1'-b+1}^{b-1-b_2'}(D_{b_1'}D_{b_2'})(D_{b_1'-k}D_{b_2'+k})\big]^{-1}$ versus $\max_{\setsize > \setsize^\star}(D_{\setsize-1}/D_\setsize)$ for the top 5\% hardest instances of each system size studied in Appendix~\ref{sec:runtime_system_size}. 
    These quantities are equal to the SA and  parallel tempering runtime lower bounds in Theorems~\ref{thm:sa_runtime_bound} and~\ref{thm:pt_runtime_bound_isoenergetic}, respectively, up to subleading polynomial factors in $n$.}
    \label{fig:speedup_si_pt}
\end{figure}

The bound that we will derive in Theorem~\ref{thm:pt_runtime_bound_isoenergetic} is similar to the parallel tempering runtime lower bound previously derived in Theorem~\ref{thm:pt_runtime_bound} when $\min_{\setsize > \setsize^\star} (D_{\setsize} / D_{\setsize-1})$ is small compared to the other ratios $ D_{\setsize} / D_{\setsize-1}$, i.e. when there is a single smallest coupling that limits the runtime. 
In Fig.~\ref{fig:speedup_si_pt} we numerically find that the scaling of our bound, stated next in Theorem~\ref{thm:pt_runtime_bound_isoenergetic}, is similar to the SA runtime lower bound in Theorem~\ref{thm:sa_runtime_bound} for the top 5\% hardest instances of each system size studied in Appendix~\ref{sec:runtime_system_size}. 
In particular, Fig.~\ref{fig:speedup_si_pt} plots $\max_{\setsize > \setsize^\star} \big[D_\setsize / D_{\setsize-1} +   \sum_{b_1'=b}^{\alpha}\sum_{b_2'=0}^{2(b-1)-b_1}\sum_{k= b_1'-b+1}^{b-1-b_2'}(D_{b_1'}D_{b_2'})(D_{b_1'-k}D_{b_2'+k})\big]^{-1}$  versus the quantity $\max_{\setsize > \setsize^\star}(D_{\setsize-1}/D_\setsize)$. 
These quantities are equal to the parallel tempering and SA runtime lower bounds in Theorems~\ref{thm:sa_runtime_bound} and~\ref{thm:pt_runtime_bound_isoenergetic}, respectively, up to subleading polynomial factors in $n$.

\begin{theorem}
    \label{thm:pt_runtime_bound_isoenergetic}
    Consider a parallel tempering algorithm with $M$ replicas using isoenergetic cluster updates as described above, in combination with the updates described in Theorem~\ref{thm:pt_runtime_bound}. 
    Then for any error $\varepsilon < 1/2$, the parallel tempering runtime $\tpt(\varepsilon)$ is bounded as
    \begin{align}\label{eq:pt_runtime_bound_isoenergetic}
        \tpt(\varepsilon)\geq &  \frac{\ln\left(\frac{1}{2\varepsilon}\right)}{2n}\max_{b>b^{\star}}\bigg[ k'n^{k'}\frac{D_b}{D_{b-1}}\nonumber \\
        &\quad +\sum_{b_1'=b}^{\alpha}\sum_{b_2'=0}^{2(b-1)-b_1}\sum_{k= b_1'-b+1}^{b-1-b_2'}\frac{D_{b_1'}D_{b_2'}}{D_{b_1'-k}D_{b_2'+k}}\bigg]^{-1}.
    \end{align}
\end{theorem}

\noindent\textit{Proof.}
As before, we bound the flow of probability $Q_{\mathcal{S}, \mathcal{S}^c}$ from $\mathcal{S}$ to $\mathcal{S}^c$ in the Gibbs distribution, 
\begin{align}
    Q_{\mathcal{S},\mathcal{S}^c}&=\sum_{\substack{z_1\dots z_M\in \mathcal{S}\\ z_1'\dots z_M'\in \mathcal{S}^c}}\pi_{z_1\dots z_M} P_{z_1\dots z_M, z_1'\dots z_M'},
\end{align}
We define $\mathcal{S}$ identically to Eq.~\eqref{eq:pt_partition}. 
As a result, the runtime lower bound we will derive in Theorem~\ref{thm:pt_runtime_bound_isoenergetic} automatically applies to the same update rules from Theorem~\ref{thm:pt_runtime_bound}, and it only remains to upper bound $Q_{\mathcal{S},\mathcal{S}^c}$ for isoenergetic cluster updates. 
The total $Q_{\mathcal{S},\mathcal{S}^c}$ will then be bounded by the sum of the bounds on $Q_{\mathcal{S},\mathcal{S}^c}$ derived here for isoenergetic cluster updates and on the bound in Theorem~\ref{thm:pt_runtime_bound}. 
The inverse of this sum of bounds will yield the bound in Theorem~\ref{thm:pt_runtime_bound_isoenergetic}.

An isoenergetic cluster update first proposed to update the configurations of a pair of replicas, which are chosen according some probability distribution. 
Without loss of generality, we will call these replicas $1$ and $2$, and denote the probability they are proposed as $p_{12}$. 
Once a pair of replicas is proposed, the quantity $Q_{\mathcal{S}, \mathcal{S}^c} / \pi_{\mathcal{S}}$ is independent of the remaining replicas. Thus, we may consider the flow $Q^{(12)}_{\mathcal{S}, \mathcal{S}^c}$ on only replicas $1$ and $2$.
We may bound the flow as
\begin{align}
    Q^{(12)}_{\mathcal{S},\mathcal{S}^c}&=p_{12}\sum_{\substack{z_1 z_2\in \mathcal{S}\\ z_1'z_2'\in \mathcal{S}^c}}\pi_{z_1'z_2'} P_{z_1'z_2', z_1z_2} \nonumber\\
    &=p_{12}\sum_{\substack{z_1 z_2\in \mathcal{S}\\ z_1'z_2'\in \mathcal{S}^c}}\pi_{z_1'z_2'} p_{z_1'z_2', z_1z_2}\min\left(1,\frac{\pi_{z_1z_2}}{\pi_{z_1'z_2'}}\right) \nonumber \\
    &\leq p_{12}\sum_{\substack{z_1z_2\in \mathcal{S}\\ z_1'z_2'\in \mathcal{S}^c}}\pi_{z_1z_2} p_{z_1'z_2', z_1z_2},
\end{align}
where $P_{z_1'z_2', z_1z_2}$ is the probability of updating to $z_1'z_2'$ given that the current configuration of the two replicas we have chosen to update is $z_1 z_2.$
Since $z_1'z_2'\in\mathcal{S}^c$, at least one of $z_1'$ or $z_2'$ must be in $S_1^c$ or $S_2^c$.
We assume without loss of generality that it is $z_1'$, so that $\hcost(z_1')\leq -\delta b$.
Then, if $z_1',z_2'$ can isoenergetically update to $z_1,z_2\in\mathcal{S}$, we must have $\hcost(z_2')\geq -2\delta(b-1)-\hcost(z_1')$, because the combined energy of $z_1',z_2'$ must be at least $-2\delta(b-1)$.
Furthermore, the number of spins that can be exchanged between the two replicas is lower-bounded by the restriction that $z_1\in S_1$ and upper-bounded by the restriction that $z_2\in S_2$. 
As a result, the sum can be parameterized as
\begin{align}
    &Q^{(12)}_{\mathcal{S},\mathcal{S}^c}\leq p_{12}\sum_{\substack{z_1 z_2\in \mathcal{S}\\ z_1'z_2'\in \mathcal{S}^c}}\pi_{z_1z_2} p_{z_1'z_2', z_1z_2}\label{eq:pt_iso_beta_bound} \\
    &\leq p_{12}\sum_{b_1'=b}^{\alpha}\hspace{-0.1cm}\sum_{b_2'=0}^{2(b-1)-b_1}\hspace{-0.1cm}\sum_{k= b_1'-b+1}^{b-1-b_2'}\hspace{-0.2cm}\sum_{\substack{\hcost(z_1')=-\delta b_1'\\ \hcost(z_2')=-\delta b_2' \\ \hcost(z_1)=-\delta(b_1'-k) \\ \hcost(z_2)=-\delta(b_2'+k)}}\hspace{-0.6cm}\pi_{z_1z_2} p_{z_1'z_2', z_1z_2} \nonumber\\
    &\leq p_{12} \sum_{b_1'=b}^{\alpha}\hspace{-0.1cm}\sum_{b_2'=0}^{2(b-1)-b_1}\hspace{-0.1cm}\sum_{k= b_1'-b+1}^{b-1-b_2'}\hspace{-0.3cm}D_{b_1'}D_{b_2'}\frac{e^{\beta_1\delta(b_1'-k)+\beta_2\delta(b_2'+k)}}{\mathcal{Z}_{\beta_1, 1}\mathcal{Z}_{\beta_2, 2}}.\nonumber
\end{align}
In the third line, we used the facts that $\pi_{z_1z_2}=e^{\beta_1\delta(b_1'-k)+\beta_2\delta(b_2'+k)}/(\mathcal{Z}_{\beta_1, 1}\mathcal{Z}_{\beta_2, 2})$ and \mbox{$\sum_{z_1z_2}p_{z_1'z_2',z_1z_2}\leq 1$}, then replaced $\sum_{z_1'z_2'}$ with $D_{b_1'}D_{b_2'}$. 
To remove the factors of $\beta_{1}$ and $\beta_2$, we may also use the fact that $\mathcal{Z}_{\beta_1, 1}$ contains a $D_{b_1'-k}e^{\beta_1\delta (b_1'-k)}$ term and $\mathcal{Z}_{\beta_2, 2}$ contains a  $D_{b_2'+k}e^{\beta_2\delta(b_2'+k)}$ term, to obtain
\begin{align}
    Q^{(12)}_{\mathcal{S},\mathcal{S}^c}&\leq p_{12}\sum_{b_1'=b}^{\alpha}\sum_{b_2'=0}^{2(b-1)-b_1}\sum_{k= b_1'-b+1}^{b-1-b_2'}\frac{D_{b_1'}D_{b_2'}}{D_{b_1'-k}D_{b_2'+k}}.
\end{align}
Now summing over all replicas (not just $1,2$) that could be proposed for replica updates and using $\sum_{ij}p_{ij}\leq 1$, we arrive at
\begin{align}
    Q_{\mathcal{S},\mathcal{S}^c}&\leq \sum_{b_1'=b}^{\alpha}\sum_{b_2'=0}^{2(b-1)-b_1}\sum_{k= b_1'-b+1}^{b-1-b_2'}\frac{D_{b_1'}D_{b_2'}}{D_{b_1'-k}D_{b_2'+k}}.
\end{align}
For reasons analogous to those given in Appendix~\ref{subsec:sa_runtime}, this is sufficient to establish the bound in Theorem~\ref{thm:pt_runtime_bound_isoenergetic} when $\pi_\mathcal{S}>1/2$.
When $\pi_\mathcal{S}<1/2$, we instead revert to Eq.~\eqref{eq:pt_iso_beta_bound} and compute the bound as
\begin{align}
    &\frac{Q^{(12)}_{\mathcal{S},\mathcal{S}^c}}{\pi_{\mathcal{S}}}\leq p_{12}\pi_{\mathcal{S}}^{-1} \nonumber\\
    &\times \sum_{b_1'=b}^{\alpha}\sum_{b_2'=0}^{2(b-1)-b_1}\sum_{k= b_1'-b+1}^{b-1-b_2'}D_{b_1'}D_{b_2'}\frac{e^{\beta_1\delta(b_1'-k)+\beta_2\delta(b_2'+k)}}{\mathcal{Z}_1\mathcal{Z}_2}
    \nonumber \\
    &\leq p_{12}(\sum_{b_1=0}^{b}D_{b_1}e^{\beta_1\delta b_1})^{-1}(\sum_{b_2=0}^{b_1}D_{b_2}e^{\beta_2\delta b_2})^{-1} \nonumber\\
    &\times\sum_{b_1'=b}^{\alpha}\sum_{b_2'=0}^{2(b-1)-b_1}\sum_{k= b_1'-b+1}^{b-1-b_2'}D_{b_1'}D_{b_2'}e^{\beta_1\delta(b_1'-k)+\beta_2\delta(b_2'+k)}.
\end{align}
At this point, we again use the fact that for every $(b_1',b_2',k)$ term in the numerator, the denominator contains a term $D_{b_1'-k}e^{\beta_1\delta(b_1'-k)}D_{b_2'+k}e^{\beta_2\delta(b_2'+k)}$, allowing us to arrive at
\begin{align}
    \frac{Q^{(12)}_{\mathcal{S},\mathcal{S}^c}}{\pi_{\mathcal{S}}}&\leq p_{12}\sum_{b_1'=b}^{\alpha}\sum_{b_2'=0}^{2(b-1)-b_1}\sum_{k= b_1'-b+1}^{b-1-b_2'}\frac{D_{b_1'}D_{b_2'}}{D_{b_1'-k}D_{b_2'+k}},
\end{align}
from which we can
establish the bound in Theorem~\ref{thm:pt_runtime_bound_isoenergetic} after summing over all choices of $1,2$.

\subsection{Quantum Monte Carlo\label{subsec:qmc_runtime}}

We now establish a runtime lower bound for a wide class of QMC algorithms. 
Our bound uses identical techniques to the analytic runtime lower bounds of SA (Appendix~\ref{subsec:sa_runtime}) and parallel tempering (Appendix~\ref{subsec:pt_runtime}), which we recommend the reader read first for context. 
We consider path-integral QMC algorithms which are designed to sample from the populations of the Gibbs distribution of the modified QAA Hamiltonian, 
\begin{align}
    \rho_{zz} = \frac{\bra{z}e^{-\beta (\hqaa + \lambda \hlaplace)}\ket{z}}{\mathcal{Z}_\beta}, \ \mathcal{Z}_\beta = \text{Tr}\big(e^{-\beta (\hqaa + \lambda \hlaplace)}\big).\label{eq:qmc_population}
\end{align}
We can write the partition function  $\mathcal{Z}_\beta$ in the $z$-basis by Trotterizing $H=\hqaa+\lambda\hlaplace$ and inserting copies of the identity matrix. 
Although we do not assume a particular form of Trotterization of $\mathcal{Z}_\beta$, we may take, for example,
\begin{align}
    \mathcal{Z}_\beta &= \sum_{z_1}\bra{z_1}e^{-\beta (\hqaa + \lambda \hlaplace)}\ket{z_1} \label{eq:qmc_partition_function_trotterization}\\
    &\simeq \sum_{z_1} \bra{z_1} \big(e^{-\beta H_\text{o.d.}/M} e^{-\beta H_\text{d}/M}\big)^{M}\ket{z_1} \nonumber\\
    &=\sum_{z_1\dots z_M}\hspace{-0.2cm}\bra{z_1} e^{-\beta H_\text{o.d.}/M} \ket{z_2}  \bra{z_2}e^{-\beta H_\text{d}/M} \ket{z_2} \nonumber \\
    & \times \bra{z_2} \dots\ket{z_M} \bra{z_M} e^{-\beta H_\text{o.d.} / M} \ket{z_1} \bra{z_1} e^{-\beta H_\text{d}/M} \ket{z_1},\nonumber
\end{align}
where $H_\text{o.d.}$ contains only off-diagonal terms of $H$ in the computational basis, and $H_\text{d}$ contains only diagonal terms in the computational basis. 
When the number of Trotter steps $M$ is  sufficiently large, the marginal probability of configuration $\ket{z_1}$ approximates its population in the Gibbs distribution,
\begin{align}\label{eq:qmc_to_gibbs}
    \pi_{z_1} &= \sum_{z_2\dots z_M} \pi_{z_1\dots z_M}\\
    &= \rho_{z_1 z_1} \text{\ as\ } M\to\infty\nonumber,
\end{align}
where
\begin{align}\label{eq:qmc_gibbs_population}
    & \pi_{z_1\dots z_M} = \frac{1}{\mathcal{Z}_{\beta}}\bra{z_1}e^{-\beta H_\text{o.d.}/M} \ket{z_2}\bra{z_2}e^{-\beta H_\text{d}/M} \ket{z_2} \nonumber \\
    & \quad \times \bra{z_2}  \dots \ket{z_M}\bra{z_M} e^{-\beta H_\text{o.d.} / M} \ket{z_1}\bra{z_1}e^{-\beta H_\text{d}/M}\ket{z_1}
\end{align}
under the particular Trotterization in Eq.~\eqref{eq:qmc_partition_function_trotterization}. 
Since the number of Trotter steps needed to obtain a good approximation of $\mathcal{Z}_\beta$ is typically polynomial in $\beta$ and the norm of $H$, we consider finite but large $U\gg |\delta|, \beta$. 

Path-integral QMC can be used to sample configurations from the distribution $\pi_{z_1\dots z_M}$. 
The Metropolis-Hastings update rule updates configuration $z_1\dots z_M$ to $z_1'\dots z_M'$ with probability 
\begin{align}\label{eq:qmc_mh_update}
    P_{z_1\dots z_M, z_1' \dots z_M'} = p_{z_1\dots z_M, z_1' \dots z_M'} \min\Big(1, \frac{\pi_{z_1'\dots z_M'}}{\pi_{z_1\dots z_M}}\Big),
\end{align}
where $p_{z_1\dots z_M, z_1' \dots z_M'}$ is the probability of proposing an update to $z_1'\dots z_M'$ given that the current configuration is $z_1\dots z_M$.

We define the QMC runtime analogously to parallel tempering, as
\begin{align}
    \tqmc(\varepsilon) = \min_{\beta} \tqmc(\varepsilon),
\end{align}
where $\tqmc(\varepsilon, \beta)$ is the runtime lower bound for QMC at temperature $1/\beta$,
\begin{align}
    \tqmc(\varepsilon, \beta) = \frac{M}{n\pi_{\alpha}} \min\Big\{t: \max_\mu \hspace{-0.1cm}\sum_{z\in\{0, 1\}^n}\hspace{-0.1cm} |\pi_z - P^t\mu_z|\leq \varepsilon \Big\}.\label{eq:tqmc_beta_definition}
\end{align}
where $P$ is the QMC Markov chain, $\mu$ is the initial probability distribution, and 
\begin{align}
    \pi_\alpha = \sum_{i=1}^M\sum_{\substack{z_1\dots z_M : \\ \hcost(z_i) \leq -\delta\setsize }} \pi_{z_1\dots z_M}
\end{align}
is now the probability that the configuration of at least one replica is an independent set of size $\setsize$.
As with the definition of parallel tempering runtime in Eq.~\eqref{eq:tpt_beta_definition}, we include a factor of $M$ in the numerator of Eq.~\eqref{eq:tqmc_beta_definition}.
This decision is justified because we allow the update rule to alter all $M$ Trotter slices, which takes $\mathcal{O}(M)$ time complexity. 
It also excludes the trivial ``speedup'' that one might obtain by using exponentially many time slices to enumerate an exponential number of low-energy configurations, at the expense of exponential space complexity. 
The inclusion of this factor makes our runtime lower bound, stated next in Theorem~\ref{thm:qmc_runtime_bound}, independent of the parameter $M$. 
We will remark in our proof of Theorem~\ref{thm:qmc_runtime_bound} that if the number of Trotter slices modified in a single update is $m < M$, then $m$ can be substituted for $M$ in our definition of $\tqmc(\varepsilon, \beta)$ in Eq.~\eqref{eq:tqmc_beta_definition}.

\begin{theorem}\label{thm:qmc_runtime_bound}
    Consider any path-integral QMC algorithm which uses a Metropolis-Hastings update rule to modify at most $k$ spins on each of $M$ imaginary time slices, where $k$ is a constant in $n$. 
    For a given $\setsize$, let $H^{\restrb}$ denote the modified QAA Hamiltonian $H = \hqaa+\lambda\hlaplace$ restricted to the space of configurations $z$ with $\hcost(z)>-\delta \setsize$, and let $\pi^{\restrb}$ be the QMC equilibrium distribution associated with $H^{\restrb}$ at inverse temperature $\beta$. 
    Let $\ket{z_{\max}}$ denote the configuration within $k$ spin flips of an independent set $\ket{z}$ with $\hcost(z) = -\delta\setsize$ with the maximum Gibbs population $\pi_{z_{\max}}^{\restrb}$, and let $e_{\max}^{\restrb}=\pi_{z_{\max}}^{\restrb}   D_{\setsize-1}$ describe relative enhancement or suppression of its population compared to the uniform superposition state $\barket{\setsize-1}$. 
    Then the QMC runtime $\tqmc(\varepsilon)$ for any error $\varepsilon <1/2$  is bounded as 
    \begin{align}
        \tqmc(\varepsilon)\geq \frac{\ln\left(\frac{1}{2\varepsilon}\right)}{2nkn^k }\max_{\setsize>\setsize_\star}\frac{D_{\setsize-1}}{e^{\restrb}_{\max}D_\setsize}
    \label{eq:qmc_runtime_bound_lambda_finite}.
    \end{align} 
\end{theorem}

We first comment on the implications of Theorem~\ref{thm:qmc_runtime_bound} before proceeding to its proof. 
Denote the Gibbs distribution of $H^{\restrb}$ as \mbox{$\rho^{\restrb}=e^{-\beta H^{\restrb}}/\tr(e^{-\beta H^{\restrb}})$}. For the purpose of discussion, assume that $M$ is large enough such that $\pi_z^{\restrb}$ is a good approximation for $\rho_{zz}^{\restrb}.$
Now, when $\lambda$ is large enough to ensure that $\rho^{\restrb}$ is delocalized in the manifold of independent sets of size $\setsize-1,$ we have $e_{\max}^{\restrb}\leq 1$. 
Thus, Eq.~\eqref{eq:qmc_runtime_bound_lambda_finite} recovers the parallel tempering runtime bound in Theorem~\ref{thm:pt_runtime_bound}. 
Additionally, the QMC runtime is quadratically larger than the QAA runtime.
Conversely, if $\rho^{\restrb}$ is favorably localized among sets of size $\leq\setsize-1$ within Hamming distance $k$ of sets of size $\setsize$, Eq.~\eqref{eq:qmc_runtime_bound_lambda_finite} yields a weak bound.
In particular, if $e_{\max}^{\restrb} \gtrsim \sqrt{D_{\setsize-1}/D_{\setsize}}$, then Theorem~\ref{thm:qmc_runtime_bound} suggests that QMC recovers the modified QAA's quadratic speedup.
In such a scenario, however, it is likely that QAA itself also favorably localizes on configurations which are close in Hamming distance to solutions of size $\geq\setsize$.
In such a situation, adding a large $\lambda$ to the QAA likely does \textit{not} enhance its performance, because it already benefits from (exponentially) favorable localization in the absence of $\lambda$.
In other words, the only scenario where QMC can recover the QAA's quadratic speedup is one in which the quadratic speedup is irrelevant due to favorable localization, which can be exploited by both QAA and QMC.
We note also that there is no reason \textit{a priori} to expect such favorable localization to occur (and indeed, Fig.~\ref{fig:speedup_finite_lambda}(a) of Appendix~\ref{subsec:finite_lambda} suggests that it typically does not), although we cannot strictly exclude it from formal arguments.\\

\noindent \textit{Proof. }
As before, we will use the Cheeger inequality to prove an upper bound on the spectral gap of the QMC Markov chain $\gqmc$. This gives us a lower bound on the QMC runtime via Eq.~\eqref{eq:sa_gap_bound}.
We will adopt identical notation and similar techniques to the parallel tempering proof in Appendix~\ref{subsec:pt_runtime}. As in Eq.~\eqref{eq:pt_partition}, let $\mathcal{S}$ be the set of configurations with $\hcost(z_i)>-\delta\setsize$ for all $i$. 
Let $\partial S_i$ represent the configurations $z_i\in S_i$ for which QMC can transition into $S_i^c$ in a single update of at most $k$ spins.

We will first consider the regime where $\pi_{\mathcal{S}}< 1/2$.
We can compute 
\begin{align}               
    Q_{\mathcal{S},\mathcal{S}^c}&=\sum_{\substack{z_1\dots z_M\in\mathcal{S} \\ z_1'\dots z_M'\in\mathcal{S}^c}}\pi_{z_1\dots z_M}P_{z_1\dots z_M,z_1'\dots z_M'} \nonumber\\
    &\leq \sum_{i=1}^M\sum_{\substack{z_i\in\partial S_i\\  z_j\in S_j, j\neq i }}\pi_{z_1\dots z_M} \nonumber\\
    &\leq M\sum_{\substack{z_1\in\partial S_1\\  z_j\in S_j, j\neq 1 }}\pi_{z_1\dots z_M}. \label{eq:qmc_cheeger0}
\end{align}
In the second line we used the fact that if $z_1\dots z_M\in\mathcal{S}$ can transition into $\mathcal{S}^c$, then $z_i\in \partial S_i$ for at least one replica $i$. 
The third line uses the standard cyclic permutation property of QMC Gibbs populations. 
We note that strictly speaking, one can choose to Trotterize the path integral in QMC in such a way that the cyclic permutation property is modified. 
For instance, if instead of the $H_{\text{o.d.}}/H_{\text{d}}$ decomposition above, we apply $\hcost$, $\hlaplace$ and $\hdrive$ in separate imaginary time slices, the QMC Gibbs weights will only be invariant under ``even'' cyclic shifts $z_i\to z_{i+2a\mod M}$ for $a\in\mathbb{Z}$. 
This does not affect the result because in such cases, the transition between $\mathcal{S}$ and $\mathcal{S}^c$ must still happen in one ``block'' of the cycle (e.g. one $\hcost,\hlaplace,\hdrive$ block in this example), and all the configurations within a single such block must be within a constant Hamming distance from each other. 
Finally, we remark that if at most $m<M$ Trotter slices are modified during a QMC update, then the factor of $M$ in Eq.~\eqref{eq:qmc_cheeger0} can be replaced with $m$. This can be seen by writing the $Q_{\mathcal{S}, \mathcal{S}^c}$ as a sum over proposed updates to $m$ replicas, then only summing over configurations with one of those $m$ replicas $i$ in $\partial S_i$. 

Therefore, we have
\begin{align}\label{eq:qmc_cheeger1}
    \frac{Q_{\mathcal{S},\mathcal{S}^c}}{\pi_{\mathcal{S}}}\leq M\frac{\sum_{\substack{z_1\in\partial S_1\\  z_j\in S_j, j\neq 1 }}\pi_{z_1\dots z_M}}{\sum_{z_1\dots z_M\in\mathcal{S}}\pi_{z_1\dots z_M}}.
\end{align}
Now notice that the summations in the numerator and denominator of Eq.~\eqref{eq:qmc_cheeger1} are only over configurations in $\mathcal{S}$. 
Thus, they can be related to the Gibbs state of $\hqaa+\lambda\hlaplace$ in a restricted Hilbert space that includes no configurations in $\mathcal{S}^c$. 
We will denote quantities in this restricted Hilbert space with a superscript $\restrb$, so that $H^{\restrb}=\hqaa^{\restrb}+\lambda\hlaplace^{\restrb}$.
Note now that because $H^{\restrb}$ is identical to $H$ on this restricted space, the populations that one would compute with QMC in this restricted space are related to their values in the full Hilbert space by an overall normalization factor:
\begin{align}
    \pi_{z_1\dots z_M}^{\restrb}&=\frac{\mathcal{Z}_\beta}{\mathcal{Z}_\beta^{\restrb}}\pi_{z_1\dots z_M}.
\end{align}
As a result, we may write
\begin{align}
    \frac{Q_{\mathcal{S},\mathcal{S}^c}}{\pi_{\mathcal{S}}}&\leq M\frac{\sum_{\substack{z_1\in\partial S_1\\  z_j\in S_j, j\neq 1 }}\pi_{z_1\dots z_M}}{\sum_{z_1\dots z_M\in\mathcal{S}}\pi_{z_1\dots z_M}}\nonumber\\
    &=M\frac{\sum_{\substack{z_1\in\partial S_1\\  z_j\in S_j, j\neq 1 }}\pi^{\restrb}_{z_1\dots z_M}}{\sum_{z_1\dots z_M\in\mathcal{S}}\pi^{\restrb}_{z_1\dots z_M}}\nonumber\\
    &=M\frac{\sum_{z_1\in\partial S_1}\pi_{z_1}^{\restrb}}{\sum_{z_1\in S_1}\pi_{z_1}^{\restrb}}\nonumber\\
    &\leq Mkn^k\frac{e_{\max}^{\restrb}D_\setsize}{D_{\setsize-1}}.\label{eq:qmc_cheeger2}
\end{align}
In the final line, we have used that there are $(k)^2{n\choose k}D_\setsize\leq k'n^k D_\setsize$ configurations in $\partial S_1$, and the definition \mbox{$e_{\max}^{\restrb}=\pi_{z_{\max}}^{\restrb} D_{\setsize-1}$}.
From Eq.~\eqref{eq:qmc_cheeger2}, we can thus immediately obtain the bound in Eq.~\eqref{eq:qmc_runtime_bound_lambda_finite}.

\begin{figure*}[th!]
    \centering
    \includegraphics[width=4.7in]{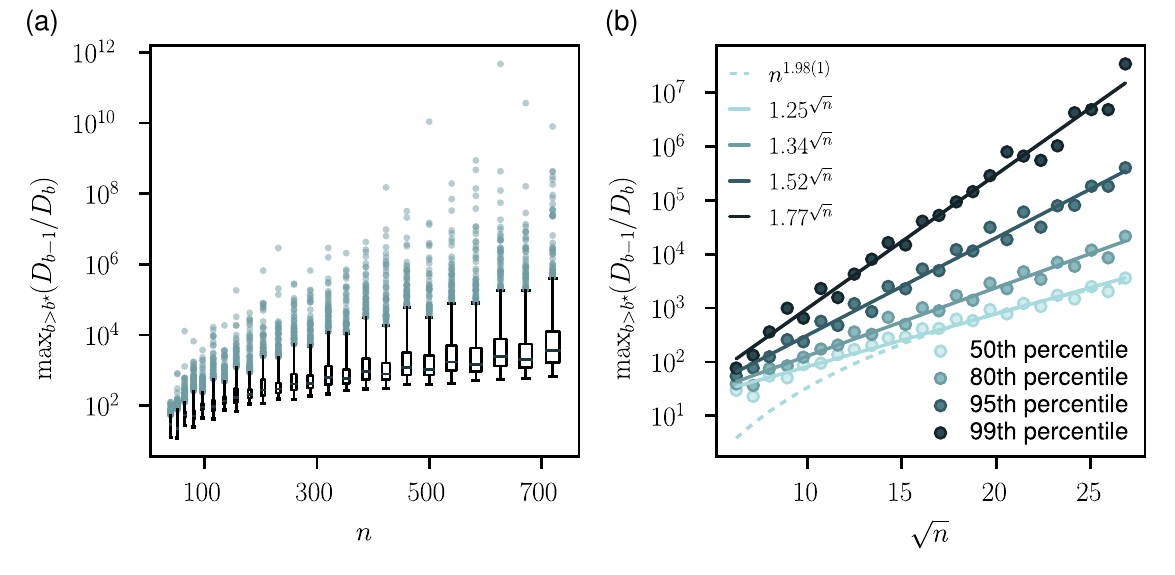}
    \caption{Classical Markov chain runtime versus system size. 
    (a) A box-and-whiskers plot of the classical runtime lower bound versus the system size $n$. 
    The box endpoints are the $25$th and $75$th percentiles, and the whiskers are the $0$th and $95$th percentiles. 
    (b) The runtime lower bounds at the 80th, 95th and 99th percentiles scale exponentially in $\sqrt{n}$. 
    The 50th percentile runtime lower bound is also consistent with exponential scaling in $\sqrt{n}$.}
    \label{fig:speedup_si_runtime_scaling_vs_n}
\end{figure*}

The discussion so far has assumed $\pi_\mathcal{S}<1/2$.
When $\pi_\mathcal{S}>1/2$, we must instead compute $Q_{\mathcal{S},\mathcal{S}^c}/\pi_{\mathcal{S}^c}$ for the Cheeger bound in Eq.~\eqref{eq:cheeger_bound}. 
We multiply this quantity by $\pi_\alpha$ to obtain the quantity that appears in the QMC runtime definition in Eq.~\eqref{eq:tqmc_beta_definition},
\begin{align}
    \frac{Q_{\mathcal{S},\mathcal{S}^c}\pi_\alpha}{\pi_{\mathcal{S}^c}}\leq Q_{\mathcal{S},\mathcal{S}^c}\leq M\sum_{\substack{z_1\in\partial S_1\\  z_j\in S_j, j\neq 1 }}\pi_{z_1\dots z_M}.
    \label{eq:qmc_cheeger4}
\end{align}
By the above arguments, we may then write
\begin{align}
    Q_{\mathcal{S},\mathcal{S}^c}&\leq M\frac{\mathcal{Z}_{\beta}^{\restrb}}{\mathcal{Z}_{\beta}}\sum_{z_1\in\partial S_1}\pi_{z_1}^{\restrb}.
\end{align}
We now note that $\mathcal{Z}_{\beta}^{\restrb}\leq\mathcal{Z}_{\beta}$, because the Gibbs weights contained in $\mathcal{Z}_{\beta}^{\restrb}$ are a subset of the weights contained in $\mathcal{Z}_{\beta}$. 
Note that we use the fact that the Hamiltonian does not have a sign problem, which ensures the positivity of the Gibbs weights. 
Thus, we have 
\begin{align}
    \frac{Q_{\mathcal{S},\mathcal{S}^c}\pi_\alpha}{\pi_{\mathcal{S}^c}}\leq Mkn^k\frac{e_{\max}^{\restrb}D_\setsize}{D_{\setsize-1}},
\end{align}
using the same reasoning as in Eq.~\eqref{eq:qmc_cheeger2}.
As our bounds hold at any point during the adiabatic ramp and at any temperature $1/\beta$, we have thus shown Theorem~\ref{thm:qmc_runtime_bound}.

\section{Runtime scaling with system size \label{sec:runtime_system_size}}
Here, we numerically study the runtime lower bounds for the classical Markov Chain Monte Carlo algorithms studied in Appendix~\ref{sec:mcmc_runtime} as a function of the number of vertices $n$, and compare the bounds against leading exact classical algorithms. 
The runtime lower bounds for these algorithms are equal to the quantity $\max_{\setsize >\setsize^\star}(D_{\setsize-1}/D_{\setsize})$ up to polynomial factors in $1/n$, where $D_\setsize$ is the number of independent sets of size $\setsize$, and $\setsize^\star$ is the cutoff independent set size as defined in Appendix~\ref{subsec:sa_runtime}. 
This quantity is large when there are many independent sets of some size $\setsize-1$ compared to independent sets of size $\setsize.$  
We are interested in determining how this quantity scales with $n$.

We randomly generate unit-disk graph instances with up to $720$ vertices embedded on a two-dimensional square lattice with random $80\%$ filling (see Fig.~\ref{fig:speedup_1}(a), main text). 
We study 1000 instances at each system size and compute $\max_{\setsize > \setsize^\star} (D_{\setsize-1}/D_\setsize)$ using the tensor-network algorithm for computing solution-space properties of combinatorial optimization problems detailed in Ref.~\cite{liu_tensor_network_2022}. 
We find that the independence polynomial of every single instance is \textit{unimodal}, i.e., $D_0\leq D_1\leq \dots \leq D_{\setsize^\star}\geq \dots \geq D_{\alpha-1} \geq D_\alpha,$ which may be of independent interest~\cite{Levit2005TheIP}. 
This means that for the unit-disk graphs we study, in practice it is not strictly necessary to have a cutoff independent set size $\setsize^\star$ in the runtime lower bound $\max_{\setsize >\setsize^\star}(D_{\setsize-1}/D_{\setsize})$: any $\setsize$ with $D_{\setsize-1}/D_\setsize \geq 1$ can be used in the maximization. 
The vast majority (99.87\%) of instances we study have $\max_{\setsize}(D_{\setsize-1}/D_\setsize) = D_{\alpha - 1}/D_\alpha$, and the remainder have $\max_{\setsize}(D_{\setsize-1}/D_\setsize) = D_{\alpha - 2}/D_{\alpha-1}$.

Figure~\ref{fig:speedup_si_runtime_scaling_vs_n}(a) shows a box-and-whiskers plot of the full distribution of runtime lower bounds as a function of $n$. 
The variance of runtimes spans several orders of magnitude and increases with $n$, and the largest runtime over all the studied graphs is nearly $10^{12}$. 
In Fig.~\ref{fig:speedup_si_runtime_scaling_vs_n}(b), we plot various percentiles of $\max_{\setsize >\setsize^\star}(D_{\setsize-1}/D_{\setsize})$ versus $\sqrt{n}$. 
We find that the runtime is exponential in $\sqrt{n}$ for instances in the 80th percentile and above. The 50th percentile runtime also appears to scale exponentially in $\sqrt{n}$ rather than polynomially.  
Therefore, the classical runtime lower bounds are (sub)exponentially faster than black-box search, which has an expected runtime of $\mathcal{O}(2^n/D_\alpha)$, which is exponential in $n$ instead of $\sqrt{n}$.

We can compare the scaling of the runtime lower bound with system size to leading exact classical algorithms, which are guaranteed to return the largest independent set.
The best exact classical algorithms for solving the unit-disk Maximum Independent Set problem find the solution in time $\mathcal{O}(c^{\sqrt{n}})$, for some constant $c\in (1, 2)$.
This scaling can be achieved using dynamic programming~\cite{fomin2013exact} or tensor-network methods~\cite{liu_tensor_network_2022}. 
Numerical evidence for the system sizes studied (see Fig.~\ref{fig:speedup_2} in the main text) suggests that the actual SA runtime is linearly related to the SA runtime lower bound, suggesting that the typical SA runtime also scales as  $\mathcal{O}(c^{\sqrt{n}})$. 
If this result holds as $n\rightarrow \infty$, then the scaling of both classical Markov chain algorithms and the modified QAA are  typically polynomially related to the best classical algorithms. 
In particular, if the SA runtime scaling is $\mathcal{O}(c^{\sqrt{n}})$, then the runtime of our modified QAA  scales roughly as $\mathcal{O}(\sqrt{c}^{\sqrt{n}})$.

\section{Resolvent method for the minimum gap}\label{sec:resolvent}
\subsection{Derivation of the minimum gap formula\label{subsec:resolvent_derivation}}
Here we will derive an exact method to pertubatively compute the minimum gap $\gqaa$ of \mbox{$\hqaa = \hcost - \hdrive$} when the avoided level crossing location $\loc\ll 1$. 
In the main text we used degenerate perturbation theory to compute, to leading order in $\Omega/\delta$, the orthogonal states
$\ketG, \ketE$ which approximate the ground and first excited eigenstates at \mbox{$\Omega/\delta \lesssim \loc \ll 1$} (see Eq.~\eqref{eq:second_order_ham}, main text). 
Here we will exactly compute $\gqaa$ in terms of the matrix elements of an effective Hamiltonian $\heff(z)$ acting on the subspace spanned by $\ketG, \ketE$, defined by the projector \mbox{$P = \ketG\braG + \ketE\braE$}.
Our main results are in Eq.~\eqref{eq:heff_linear_si}, which gives $\gqaa$ exactly in terms of the matrix elements of $\heff(z)$, and Eq.~\eqref{eq:gqaa_gqaanaive_correction}, which simplifies the result under a motivated approximation.

$\heff(z)$ can be derived by rewriting the eigenvalue equation \mbox{$\hqaa \ket{\psi} = z\ket{\psi}$ as \mbox{$\hqaa (P+Q)\ket{\psi} = z\ket{\psi}$},} where \mbox{$Q = \mathds{1}-P$}, then multiplying by $P$ and $Q$ to obtain a system of equations for the eigenvector $\ket{\psi}$:
\begin{equation}
    \begin{bmatrix}
        Q\hqaa Q & Q\hqaa P\\
        P\hqaa Q & P\hqaa P
    \end{bmatrix}
    \begin{bmatrix}
        Q\ket{\psi}\\
        P \ket{\psi}
    \end{bmatrix} = 
    z\begin{bmatrix}
        Q\ket{\psi}\\
        P \ket{\psi}
    \end{bmatrix}.
\end{equation}
These equations can then be written in terms of $P\ket{\psi}$ as
\begin{align}
    \underbrace{\bigg[P \hqaa P \hspace{-0.05cm} + \hspace{-0.05cm} P\hqaa  \frac{Q}{z-Q\hqaa Q} \hqaa P \bigg]}_{\heff(z)}\hspace{-0.1cm}\ket{\psi}\hspace{-0.05cm}= \hspace{-0.05cm} z P \ket{\psi}\hspace{-0.05cm}. \label{eq:heff_eigenvalue_si}
\end{align}
The left hand side of the equation defines $\heff(z)$, the effective Hamiltonian in the subspace spanned by $\ketG, \ketE$. 
The second term in $\heff(z)$ can be interpreted as a perturbative addition to original Hamiltonian, $P\hqaa P$, due to higher-order couplings in $\Omega/\delta$ that come from the $Q$ subspace, which is energetically separated from the $P$ subspace. 
Expanding the denominator using the matrix Taylor expansion $(A+B)^{-1} = A^{-1}\sum_{l=0}^{\infty}(-BA^{-1})^l$, we receive
\begin{align}
    \heff(z) &= P\hcost P - \sum_{l=0}^{\infty} P  \bigg(-\hdrive\frac{Q}{z - \hcost}\bigg)^l \hdrive P,\label{eq:heff_expanded_si}
\end{align}
where we have used  $P\hcost Q = 0$ because $\ketG, \ketE$ are eigenstates of $\hcost$. 
This form of $\heff(z)$ has an intuitive interpretation: each order $l$ applies a factor of $\hdrive$, but is suppressed by a factor of $\mathcal{O}(\Omega/\delta)$.

\begin{figure*}[th!]
    \includegraphics[width=\textwidth]{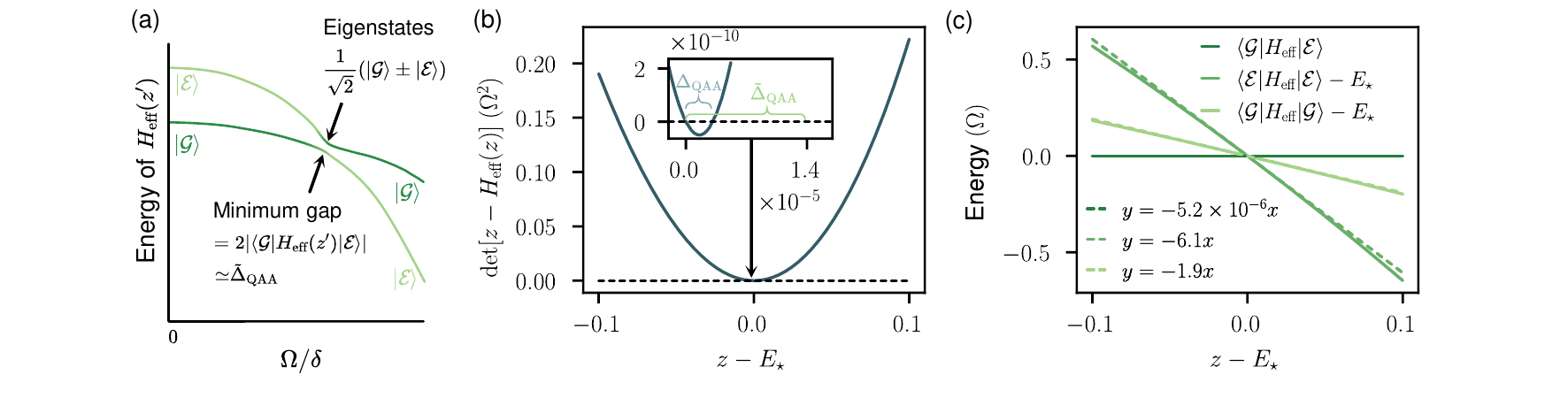}    \caption{\label{fig:speedup_si_resolvent} Computing the minimum gap using the resolvent formalism. 
    (a) When the avoided level crossing location $\loc \ll 1,$ the avoided level crossing can be understood in terms of Landau-Zener physics between $\ketG, \ketE$ under $\heff(z')$.
    At $\Omega/\delta = \loc$, $\ketE$ and $\ketG$ have the same on-diagonal energy under $\heff(z)$, and the minimum gap of $\heff(z')$ is given by their off-diagonal coupling $2|\braG \heff(z') \ketE|.$
    (b) $\gqaa$ equals the difference of the first two zeroes of $\det[z-\heff(z)]$, which occur at $z=\ecrit$ and $ \ecrit + \gqaa$ (light blue, inset). 
    The estimated gap $\gqaanaive = 2|\braG \heff(\ecrit)\ketE|$ overestimates the minimum gap $\gqaa$ by a factor of $4.53$ for this instance (light green, inset). 
    (c)  When $z-\ecrit$ is small, matrix elements of $\heff$ (solid lines) are well-approximated by a linear function of $z$ (dashed lines). 
    For the star graph with $b=40, \ell=2$, the $\braG\heff(z) \ketE$ matrix element changes as a function of $z-\ecrit$ with a slope of $\mge = -5.2\times 10^{-6}$. 
    The matrix elements $\braG\heff(z) \ketG$ and $\braE\heff(z) \ketE$ change at much higher rates of $\mgg = -1.9$ and $\mee = -6.1$, respectively.  }
\end{figure*}

Prior works have estimated $\gqaa$ from the off-diagonal matrix element of $\heff(\ecrit)$ evaluated at $\loc$ as~\cite{altshuler_2010,choi_first_order}
\begin{align} 
    \gqaanaive = 2 |\braG \heff(\ecrit) \ketE|,\label{eq:gqaanaive_matrix_element_si}
\end{align}
which we analyzed in the main text [see Eq.~\eqref{eq:off_diag_expansion}]. 
This equation has an intuitive interpretation under the assumption of Landau-Zener physics on $\heff(z)$, which we illustrate in Fig.~\ref{fig:speedup_si_resolvent}(a). 
At $\Omega/\delta=0$, $\ketG, \ketE$ are eigenstates of $\hqaa$ with eigenenergies given by the on-diagonal entries of $\heff(z)$ ($\braG \hcost \ketG$ and $\braE \hcost \ketE$, respectively). 
At the avoided level crossing $\Omega/\delta = \loc$, we expect the  on-diagonal eigenenergies of $\ketG, \ketE$ in $\heff(z)$ to cross at a value close to $\ecrit$ for some value of \mbox{$z\simeq \ecrit$}, which we denote by $z'$. The gap of $\heff(z')$ at $\loc$ is then given by the off-diagonal coupling \mbox{$2 |\braG \heff(z') \ketE| \simeq \gqaanaive$}. $\gqaanaive$ indeed captures the correct qualitative physics, but is quantitatively inaccurate. Here we show that $\gqaa$ can be computed exactly from the matrix elements of $\heff(z)$ in Eqs.~\eqref{eq:heff_linear_si} and~\eqref{eq:gqaa_gqaanaive_correction}. 

$\gqaanaive$ does not equal $\gqaa$ in general because of the $z$-dependence of $\heff(z)$, which prevents it from being interpreted as a true Hamiltonian. 
The only guaranteed relationship between $\heff$ and the spectrum of $\hqaa$ is that each eigenvalue $z$ of $\hqaa$ is also an eigenvalue of $\heff(z)$ [see Eq.~\eqref{eq:heff_eigenvalue_si}], i.e., 
\begin{align}
    \det[z-\heff(z)]=0 \label{eq:heff_determinant_si}
\end{align}
whenever $z$ is an eigenvalue of $\hqaa$. 
$\gqaa$ can therefore be obtained exactly from taking the difference between the first two values of $z$ that solve Eq.~\eqref{eq:heff_determinant_si}, which are the two lowest energy eigenvalues at $z=\ecrit, \ecrit + \gqaa$. 
We show an example of numerically using this method to exactly reconstruct $\gqaa$ in Fig.~\ref{fig:speedup_si_resolvent}(b) for a star graph with $b=40$ branches of length $\ell=2$. 
In contrast, we find that $\gqaanaive$, computed numerically, overestimates $\gqaa$ for the same instance by a factor of $4.53$ (Fig~\ref{fig:speedup_si_resolvent}(b), inset).
This discrepancy is due  to the $z$-dependence of $\heff$, which we show in Fig.~\ref{fig:speedup_si_resolvent}(c) for the same instance. 

To account for this $z$-dependence, we will consider $z$ in the neighborhood of $\ecrit$, and compute the leading order, linear dependence of $\heff$ on $z$. 
We adopt the following ansatz by expanding $\heff(z)$ around a reference point $z=z_0$:
\begin{widetext}
    \begin{align}\label{eq:heff_linear_si}
        \heff(z)&=\begin{bmatrix} \braE \heff(z_0) \ketE+\mee(z-z_0) & \eg{z_0} +\mge(z-z_0) \\ \eg{z_0} +\mge(z-z_0) & \braG \heff(z_0) \ketG+\mgg(z-z_0)
        \end{bmatrix},
    \end{align}
\end{widetext}
where $\mge = m_{eg}$ because $\heff$ is real. 
$\gqaa$ can then be obtained from solving Eq.~\eqref{eq:heff_determinant_si} using the ansatz for $\heff$ in Eq.~\eqref{eq:heff_linear_si}, which gives
\begin{widetext}
    \begin{align}
        \gqaa=&2\Bigg[\eg{z_0} ^2f_{ee}f_{gg}+\frac{1}{4}\left[\left(f_{ee}+f_{gg}\right)\eone+\left(f_{gg} - f_{ee}\right)(\ezero-z_0)\right]^2 \nonumber\\
        &+\eg{z_0} \mge\left[\left(f_{ee} + f_{gg}\right)(\ezero-z_0)+\left(f_{gg}-f_{ee}\right)\eone\right]+\mge^2[(\ezero-z_0)^2-\eone^2]\Bigg]^{1/2}/\left[f_{gg}f_{ee}-\mge^2\right],\label{eq:corrected_gap_si}
    \end{align}
\end{widetext}
where we have defined the mean and difference of the on-diagonal energies, 
\begin{align}
    \ezero(z_0) &= \frac{1}{2}(\braE \heff(z_0) \ketE + \braG \heff(z_0) \ketG) \nonumber\\
    \eone(z_0) &= \frac{1}{2}(\braE \heff(z_0) \ketE - \braG \heff(z_0) \ketG),
\end{align}
and let
\begin{align}
    f_{ee} &= 1-\mee\nonumber\\
    f_{gg} &= 1-\mgg.
\end{align}
Eq.~\eqref{eq:corrected_gap_si} therefore gives $\gqaa$ in terms of the matrix elements of $\heff$ and their first-order derivatives in $z$. 
In the absence of $z$-dependence (\mbox{$\mee=\mgg=\mge=0$}), one can check that this expression reduces the result one would obtain from directly diagonalizing $\heff(z_0)$. 
Therefore, as expected, when $\heff$ is independent of $z$, it can be treated as a true Hamiltonian acting on $\ketG, \ketE$ and diagonalized to find $\gqaa$. 

Although Eq.~\eqref{eq:corrected_gap_si} is exact, we can vastly simplify it using intuition from Landau-Zener theory. 
Suppose, to good approximation, there exists a $z'$ such that the diagonal entries of $\heff(z')$ intersect at $z'$ for $\Omega/\delta=\loc$: \mbox{$\braE \heff(z') \ketE = \braG \heff(z') \ketG=z'$}. 
Because $\heff$ is independent of the point of expansion $z_0$ in the regime where the linear approximation is valid, we may choose $z_0 = z'.$
Using our assumption that the diagonal entries of $\heff(z')$ intersect at $\loc$, we then have $\eone(z_0) = 0$ and $\ezero(z_0) = z_0.$ 
Under this choice of $z_0$, $\gqaa$ simplifies to
\begin{align}
    \label{eq:gqaa_no_z0} \gqaa=&\frac{2\sqrt{\eg{z'} ^2f_{ee}f_{gg}}}{f_{gg}f_{ee}-\mge^2}.
\end{align}
We may further simplify this expression using the fact that we expect $|\mge|\ll |f_{gg}|,|f_{ee}|$. 
To see this, we compute $d\heff /dz$ for $z\in \mathbb{R}$ as
\begin{align}
    &\frac{d\heff}{dz} = -\Omega^2P \hdrive \left(\frac{Q}{z-Q \hqaa Q}\right)^2\hdrive P \label{eq:dHeff_by_dz} \\
    &\quad =-\Omega^2\left[\frac{Q}{z-Q\hqaa Q}\hdrive P\right]^{\dagger}\hspace{-0.1cm}\left[\frac{Q}{z-Q\hqaa Q}\hdrive P\right],\nonumber
\end{align}
which is similar to the second term of  $\heff(z)$ in Eq.~\eqref{eq:heff_eigenvalue_si}. 
By expanding Eq.~\eqref{eq:dHeff_by_dz} in powers of $\Omega/\delta$, as in Eq.~\eqref{eq:heff_expanded_si}, one can see that the on-diagonal entries can in general be  large because they connect either $\ketG$ or $\ketE$ to itself via even multiples of $\hdrive$. 
On the other hand, the off-diagonal entries should be smaller by $\mathcal{O}(\gqaa)$ because they connect $\ketG$ to $\ketE$ via odd multiples of $\hdrive$, similar to the off-diagonal entries of $\heff(z).$ 
Therefore, we expect that $|\mge|/|f_{ee}|,|\mge|/|f_{gg}| = \mathcal{O}( \gqaa)$. We verify numerically that $|\mge| = \mathcal{O}(\gqaa)$ and $f_{gg}, f_{ee} = \mathcal{O}(1)$ in Fig.~\ref{fig:speedup_si_resolvent}(c) for an example star graph. 
Therefore, to good approximation we have 
\begin{align}
    \gqaa =&\frac{2|\eg{\ecrit}|}{\sqrt{{f_{gg}f_{ee}}}}  = \frac{2\gqaanaive}{\sqrt{{f_{gg}f_{ee}}}}.\label{eq:gqaa_gqaanaive_correction}
\end{align}
Note that by the form of Eq.~\eqref{eq:dHeff_by_dz}, $d\heff / dz$ can be written as $-\Omega^2$ times a positive semidefinite operator, so all the derivatives of $\heff$ are negative.  
Therefore, $f_{ee}, f_{gg} \geq 1$, so $\gqaanaive$ is an overestimate of the gap, consistent with our numerical results on the star graph in Fig.~\ref{fig:speedup_si_resolvent}(b).
We verify numerically in Appendix~\ref{subsec:star_graph_level_crossing_params}, Fig.~\ref{fig:speedup_si_star}(d) that Eq.~\eqref{eq:gqaa_gqaanaive_correction} recovers the  $\gqaa$ for the star graph to high accuracy.

\subsection{Validity of the resolvent method\label{subsec:resolvent_validity}}
For $\gqaanaive$ to be a good qualitative predictor of $\gqaa$ via Eq.~\eqref{eq:gqaa_gqaanaive_correction}, the factors $f_{gg}, f_{ee}$ cannot be large compared to the minimum gap as to change its leading-order scaling behavior with $n$. 
The $z$-dependence of $\heff(z)$ comes from the factor of $(z-Q\hqaa Q)^{-1}$ in Eq.~\eqref{eq:heff_eigenvalue_si}, which creates a pole at every eigenvalue of $Q\hqaa Q$. 
Although this creates significant $z$-dependence in $\heff(\ecrit)$ if $Q\hqaa Q$ has an eigenvalue close to the ground state energy $\ecrit\equiv E_0$, the $z$-dependence will be modest if the ground state energy of $Q\hqaa Q$ is significantly larger than $E_0$. 
We expect this to occur when $Q$ is a sufficiently good projector out of the ground and first-excited states of $\hqaa$, and the second-excited state energy of $\hqaa$ is much larger than $\gqaa$. 
To formalize this intuition, in the following Theorem~\ref{thm:resolvent_condition} we relate the energy difference between $E_0$ and the ground state energy of $Q\hqaa Q$, denoted $\delta E$, to the overlap of $\ketG$ and $\ketE$ with the ground and first excited eigenstates of $\hqaa$.
\begin{theorem}\label{thm:resolvent_condition}
    Denote the eigenstates of $\hqaa$ at $\loc$ as \mbox{$\ket{\psi_0}, \ket{\psi_1}\dots, \ket{\psi_{2^n}}$} with corresponding eigenvalues \mbox{$E_0\leq E_1\leq \dots \leq E_{2^n}$}, and assume that $\gqaa=E_1-E_0\ll E_2-E_1$.
    Denote the ground state energy of $Q\hqaa Q$ by $E_0+\delta E$.
    Then, if $|\braket{\psi_0|\mathcal{G}}\braket{\psi_1|\mathcal{E}}-\braket{\psi_0|\mathcal{E}}\braket{\psi_1|\mathcal{G}}|^2\gg \gqaa$, $\delta E$ is bounded as 
    \begin{align}
        \delta E\geq \frac{1}{4}(E_2-E_1)|\braket{\psi_0|\mathcal{G}}\braket{\psi_1|\mathcal{E}}-\braket{\psi_0|\mathcal{E}}\braket{\psi_1|\mathcal{G}}|^2.
    \end{align}
\end{theorem}

Hence, if $|\braket{\psi_0|\mathcal{G}}\braket{\psi_1|\mathcal{E}}-\braket{\psi_0|\mathcal{E}}\braket{\psi_1|\mathcal{G}}|$ and $E_2-E_1$ are at worst polynomially small in $n$, we will have $\delta E$ at worst polynomially small in $n$. 
This will make the correction factors $f_{ee},f_{gg}$ at most polynomially large in $n$, and thus subleading when $\gqaa$ is exponentially small in $n$.
The quantity $\braket{\psi_0|\mathcal{G}}\braket{\psi_1|\mathcal{E}}-\braket{\psi_0|\mathcal{E}}\braket{\psi_1|\mathcal{G}}$ can be interpreted as the area of the parallelogram defined by $\ketG,\ketE$ in the $\ket{\psi_0},\ket{\psi_1}$ subspace. 
If we define $\mathcal{P}=\ket{\psi_0}\bra{\psi_0}+\ket{\psi_1}\bra{\psi_1}$, the condition that $(\braket{\psi_0|\mathcal{G}}\braket{\psi_1|\mathcal{E}}-\braket{\psi_0|\mathcal{E}}\braket{\psi_1|\mathcal{G}})$ is large is thus both a statement about the size of the overlaps $\braG\mathcal{P}\ketG,\braE\mathcal{P}\ketE$ and also a statement about the linear independence of $\mathcal{P}\ketG,\mathcal{P}\ketE$. 
Intuitively, $\ketG,\ketE$ must have good overlap with the span of $\ket{\psi_0}$ and $\ket{\psi_1}$, and must furthermore capture sufficiently different directions within this space.
We expect $\ketG, \ketE$ to satisfy this condition when $\loc \ll 1$ because $\ketG$ approximates $\ket{\psi_0}$ and $\ketE$ approximates $\ket{\psi_1}$ in perturbation theory, and $\ketG, \ketE$ are orthogonal. \\

\noindent\textit{Proof.} We first find the ground state energy of $Q\hqaa Q$ by computing $\min_{{\phi}}\bra{\phi}\hqaa\ket{\phi}$ subject to $\braket{\phi|\phi}=1$ and $P\ket{\phi}=0$. 
This can be formulated as the minimization of $\bra{\phi}\hqaa\ket{\phi}-\zeta_0\braket{\phi|\mathcal{G}}-\zeta_1\braket{\phi|\mathcal{E}}-\mu(\braket{\phi|\phi}-1)$ where $\zeta_{0,1}, \mu$ are Lagrange multipliers. 
Writing $\ket{\phi}$ in the eigenbasis $\ket{\psi_i}$ of $\hqaa$ and setting the derivatives with respect to $\braket{\phi|\psi_i}$ of this expression to zero  yields the condition
\begin{align}\label{eq:wf_constr_min}
    \braket{\psi_i|\phi}&=\frac{1}{2}\frac{1}{E_i-\mu}(\zeta_0\braket{\psi_i|\mathcal{G}}+\zeta_1\braket{\psi_i|\mathcal{E}}).
\end{align}
Plugging Eq.~\eqref{eq:wf_constr_min} into the constraints \mbox{$\braket{\mathcal{G}|\phi}=\braket{\mathcal{E}|\phi}=0$} yields two equations involving $\zeta_{0,1},\mu$, and $\braket{\psi_i|\mathcal{G}}, \braket{\psi_i|\mathcal{E}}$. 
Solving one equation for $\zeta_0$ and substituting into the other yields, after simplification,
\begin{align}
    0&=\sum_{i,j}\frac{|\braket{\psi_i|\mathcal{G}}\braket{\psi_j|\mathcal{E}}-\braket{\psi_i|\mathcal{E}}\braket{\psi_j|\mathcal{G}}|^2}{(E_i-\mu)(E_j-\mu)} \nonumber \\
    &=\sum_{ij}\frac{|\braket{\psi_i\psi_j|\Phi}|^2}{(E_i-\mu)(E_j-\mu)},\label{eq:wf_overlap_constraint}
\end{align}
where we have defined the wavefunction $\ket{\Phi}=\ketG\otimes\ketE-\ketE\otimes\ketG$, which exists in a doubled Hilbert space. 
Now plugging Eq.~\eqref{eq:wf_constr_min} into $\bra{\phi}\hqaa\ket{\phi}/\braket{\phi|\phi}$ and simplifying using Eq.~\eqref{eq:wf_overlap_constraint} yields the conclusion that the minimum energy of $Q\hqaa Q$ is $\mu$. 
So we must use Eq.~\eqref{eq:wf_overlap_constraint}, which gives a constraint on $\mu$, to constrain $\delta E=\mu-E_0$.

Before doing so we briefly note that although we must have $\mu\geq E_0$ by definition, we must exclude the possibility that $E_0\leq \mu\leq E_1$ by contradiction: were this to happen, we could rewrite Eq.~\eqref{eq:wf_overlap_constraint} as
\begin{align}
    &2\frac{|\braket{\psi_0\psi_1|\Phi}|^2}{(\mu-E_0)(E_1-\mu)}+2\frac{1}{\mu-E_0}\sum_{i\neq 0,1}\frac{|\braket{\psi_0\psi_i|\Phi}|^2}{E_i-\mu}\nonumber \\ 
    &\ =2\frac{1}{E_1-\mu}\sum_{i\neq 0,1}\frac{|\braket{\psi_1\psi_i|\Phi}|^2}{E_i-\mu}+\hspace{-0.2cm}\sum_{i,j\neq 0,1}\frac{|\braket{\psi_i\psi_j|\Phi}|^2}{(E_i-\mu)(E_j-\mu)}.
\end{align}
Here all terms are positive, but the first term on the left hand side is $\mathcal{O}(|\braket{\psi_0\psi_1|\Phi}|^2\gqaa^{-2})$ (because $\mu - E_0, E_1 \leq E_1 - E_0  = \gqaa$), whereas the first term on the right hand side is only $\mathcal{O}(\gqaa^{-1})$.
The equation is thus impossible to satisfy under the assumption \mbox{$|\braket{\psi_0|\mathcal{G}}\braket{\psi_1|\mathcal{E}}-\braket{\psi_1|\mathcal{G}}\braket{\psi_0|\mathcal{E}}|^2\gg\gqaa$}.

Therefore, to constrain $\delta E$, we can assume that $E_1\leq \mu\leq E_2$, and use Eq.~\eqref{eq:wf_overlap_constraint} to write
\begin{widetext}
\begin{align}
    \frac{|\braket{\psi_0\psi_1|\Phi}|^2+|\braket{\psi_1\psi_0|\Phi}|^2}{(\mu-E_0)(\mu-E_1)}&\leq \frac{1}{\mu-E_0}\sum_{i\neq 0,1}\frac{|\braket{\psi_0\psi_i|\Phi}|^2+|\braket{\psi_i\psi_0|\Phi}|^2}{E_i-\mu}+\frac{1}{\mu-E_1}\sum_{i\neq 0,1}\frac{|\braket{\psi_1\psi_i|\Phi}|^2+|\braket{\psi_i\psi_1|\Phi}|^2}{E_i-\mu} \nonumber \\
    &\leq \frac{1}{(\mu-E_0)(E_2-\mu)}(2|\braket{\psi_0|\mathcal{G}}|^2+2|\braket{\psi_0|\mathcal{E}}|^2-|\braket{\psi_0\psi_1|\Phi}|^2-|\braket{\psi_1\psi_0|\Phi}|^2) \nonumber \\
    &+\frac{1}{(\mu-E_1)(E_2-\mu)}(2|\braket{\psi_1|\mathcal{G}}|^2+2|\braket{\psi_1|\mathcal{E}}|^2-|\braket{\psi_0\psi_1|\Phi}|^2-|\braket{\psi_1\psi_0|\Phi}|^2)
\end{align}
\end{widetext}
Between the first and second lines we used $E_2-\mu\leq E_i-\mu$ for $i\geq 2$, and
\begin{align}
    \sum_{i\neq 0,1}|\braket{\psi_0\psi_i|\Phi}|^2=|\braket{\psi_0|\mathcal{G}}|^2+|\braket{\psi_0|\mathcal{E}}|^2-|\braket{\psi_0\psi_1|\Phi}|^2.
\end{align}
Now multiplying out all the denominators and using $E_0=\ecrit,E_1=\ecrit+\gqaa$, we obtain
\begin{widetext}
\begin{align}\label{eq:projected_energy_bound}
    \mu-\ecrit &\geq (E_2-\ecrit)\frac{|\braket{\psi_0\psi_1|\Phi}|^2+|\braket{\psi_1\psi_0|\Phi}|^2}{2|\braket{\psi_0|\mathcal{G}}|^2+2|\braket{\psi_0|\mathcal{E}}|^2+2|\braket{\psi_1|\mathcal{G}}|^2+2|\braket{\psi_1|\mathcal{E}}|^2-(|\braket{\psi_0\psi_1|\Phi}|^2+|\braket{\psi_1\psi_0|\Phi}|^2)}\nonumber \\
    &+\gqaa\frac{2|\braket{\psi_0|\mathcal{G}}|^2+2|\braket{\psi_0|\mathcal{E}}|^2-(|\braket{\psi_0\psi_1|\Phi}|^2+|\braket{\psi_1\psi_0|\Phi}|^2)}{2|\braket{\psi_0|\mathcal{G}}|^2+2|\braket{\psi_0|\mathcal{E}}|^2+2|\braket{\psi_1|\mathcal{G}}|^2+2|\braket{\psi_1|\mathcal{E}}|^2-(|\braket{\psi_0\psi_1|\Phi}|^2+|\braket{\psi_1\psi_0|\Phi}|^2)}.
\end{align}
\end{widetext}
The final term can be dropped because it is small, by the assumption $|\braket{\psi_0|\mathcal{G}}\braket{\psi_1|\mathcal{E}}-\braket{\psi_1|\mathcal{G}}\braket{\psi_0|\mathcal{E}}|^2\gg\gqaa$.
The bound in Theorem~\ref{thm:resolvent_condition} then follows from maximizing the denominator in the first term.

\subsection{Conditions for a perturbative avoided level crossing\label{subsec:late_crossing}}
By the arguments in Appendix~\ref{subsec:resolvent_validity}, the  formula in Eq.~\eqref{eq:corrected_gap_si} for the minimum gap of \mbox{$\hqaa = \hcost - \hdrive$} converges when the location of the avoided level crossing $\loc \ll 1$. 
Here we establish a condition for when this occurs, given in Eq.~\eqref{eq:perturbative_crossing_si}, and motivate why we expect this condition to hold for problem instances with flat energy landscapes. 
We refer the reader to Ref.~\cite{werner2023bounding} for a detailed framework to predict $\loc$ for general combinatorial optimization problems.

Recall that the perturbed eigenstates (energy shifts) are the eigenvectors (eigenvalues) of the perturbed Hamiltonian
\begin{align} \label{eq:second_order_ham_si}
    H^{(2)}
    =-\frac{\Omega^2}{\delta}\bigg(&\hse+\sum_{u\in V}n_u -H_{fv}\bigg),
\end{align}
where 
\begin{align}
    H_{fv} = \sum_{u\in V}(\mathds{1}-n_u)\prod_{(u,v)\in E}(\mathds{1}-n_v)\label{eq:hfv},
\end{align}
counts the number of free vertices for each independent set $\ket{z}$ (vertices which can be added to $\ket{z}$ without violating the independent set constraint). 
$\ketG$ is the ground state of $H^{(2)}$ in the $\hcost = -\delta\alpha$ manifold because this is the instantaneous ground state of the system as $\Omega/\delta\to 0$ (see Fig.~\ref{fig:speedup_3}(b), main text). 
Its perturbed energy under $H^{(2)}$ is
\begin{align}\label{eq:g_energy}
    \bra{\mathcal{G}}\hcost + H^{(2)}\ketG = -\delta\alpha - \frac{\Omega^2}{\delta} (\alpha +\bra{\mathcal{G}} \hse \ketG),
\end{align}
where we have used that $\bra{\mathcal{G}}H_{fv}\ketG = 0$ because no vertices can be added to $\ketG$.
The last term counts the expected number of spin exchanges possible between neighboring vertices in $\ketG$. 

$\ketE$ can be found by determining the $\hcost$ manifold whose ground state energy intersects $\ketG$ first at finite $\Omega/\delta$ (see Fig.~\ref{fig:speedup_3}(b) and Ref.~\cite{choi_first_order} for a discussion). Suppose $\ketE$ is the ground state of Eq.~\eqref{eq:second_order_ham_si} in the $\hcost = -\delta\setsize$ manifold, for some unknown $\setsize$. Then the perturbed eigenenergy of $\ketE$ is 
\begin{align}\label{eq:e_energy}
    &\braE\hcost + H^{(2)}\ketE \nonumber \\
    &\qquad = -\delta\setsize - \frac{\Omega^2}{\delta} (\setsize +\braE \hse \ketE - \braE H_{fv} \ketE).
\end{align}
Note we have assumed that the ground state of each manifold of $H^{(2)}$ is nondegenerate, so that $\ketG$ and $\ketE$ can be uniquely identified. 
On instances where the degeneracy is not broken, or the energy splitting between the ground and first excited state of a manifold is 
too small to accurately identify $\ketG$ or $\ketE$, one can compute $\ketG$ and $\ketE$ by going to higher order in perturbation theory.

$\loc$ can then be estimated by computing the value of $\Omega/\delta$ where Eqs.~\eqref{eq:g_energy} and~\eqref{eq:e_energy} intersect, which is 
\begin{align}\label{eq:gap_closing_loc_general}
    \loc = \sqrt{\frac{\alpha - \setsize}{\braE\hspace{-0.05cm}\hse\hspace{-0.05cm} \ketE - \braG\hspace{-0.05cm}\hse \hspace{-0.05cm}\ketG - \braE \hspace{-0.05cm}H_{fv}\hspace{-0.05cm} \ketE-\alpha+\setsize}}
\end{align}
to second order in $\Omega/\delta$.
If $\loc \ll 1$, then the numerator of Eq.~\eqref{eq:gap_closing_loc_general} must be much smaller than the denominator:
\begin{align}
    \bra{\mathcal{E}}\hse \ketE - \bra{\mathcal{G}}\hse \ketG  &\gg 2(\alpha-\setsize) + \bra{\mathcal{E}}H_{fv}\ketE.
\end{align}
We can use the bound $\alpha - \setsize \geq \bra{\mathcal{E}}H_{fv} \ketE$ to get the following condition for when $\loc \ll 1$:
\begin{align} \label{eq:perturbative_crossing_si}
     \bra{\mathcal{E}}\hse \ketE - \bra{\mathcal{G}}\hse \ketG  \gg 3(\alpha-\setsize).
\end{align}

Therefore, $\loc\ll 1$ when $\ketE$ has a large number of expected possible spin exchanges compared to $\ketG$, and $\ketE$ is comprised of near-optimal independent sets. 
This is exactly the case on problem instances with flat energy landscapes at near-optimal independent set size $\setsize \simeq \alpha$. 
Because there are many independent sets of size $\setsize$ with freedom to spin-exchange (see e.g. the configuration graph in Fig.~\ref{fig:speedup_1}, main text), we might expect $\bra{\mathcal{E}}\hse \ketE$ to be large. 
For example, if each vertex in independent sets of size $\setsize$ has $k$ possible spin exchanges, then $\bra{\mathcal{E}}\hse \ketE = k\setsize$ is extensively large with $n$. 
We  similarly expect $\bra{\mathcal{G}}\hse\ketG = k'\alpha$, if vertices in independent sets of size $\alpha$ have $k'$ possible spin exchanges (in the case where there is a unique largest independent set, $k' = 0$). 
Since there are far fewer independent sets of size $\alpha$, and larger independent sets may have less freedom to spin-exchange under the independent set constraint, we might expect that $k>k'$ and therefore that $\bra{\mathcal{E}}\hse \ketE \gg \bra{\mathcal{G}}\hse \ketG$ for large systems. 
Therefore, on problem instances with flat energy landscapes, we expect the avoided level crossing location to occur near the end of the ramp, $\loc\ll 1$.

We verify that this interpretation is correct for the family of star graphs in Appendix~\ref{sec:star_graph}. 
We consider the case of fixed branch length $\ell$, and look at the avoided level crossing location as the number of branches (and therefore $n$) grows. 
The largest independent set is unique, so $\bra{\mathcal{G}}\hse \ketG = 0.$
$\ketE$ is in the $\hcost = -\delta (\alpha-1)$ manifold, so the right hand side of Eq.~\eqref{eq:perturbative_crossing_si} is equal to $3$. 
Typical independent sets in $\ketE$ can participate in $\mathcal{O}(n)$ spin exchanges. 
Therefore, by Eq.~\eqref{eq:gap_closing_loc_general} the avoided level crossing location $\loc$ goes like $\mathcal{O}(1/\sqrt{\nbranch}) = \mathcal{O}(1/\sqrt{n})$ as $n\to\infty$.

\begin{figure*}[th!]
    \centering
    \includegraphics[width=\textwidth]{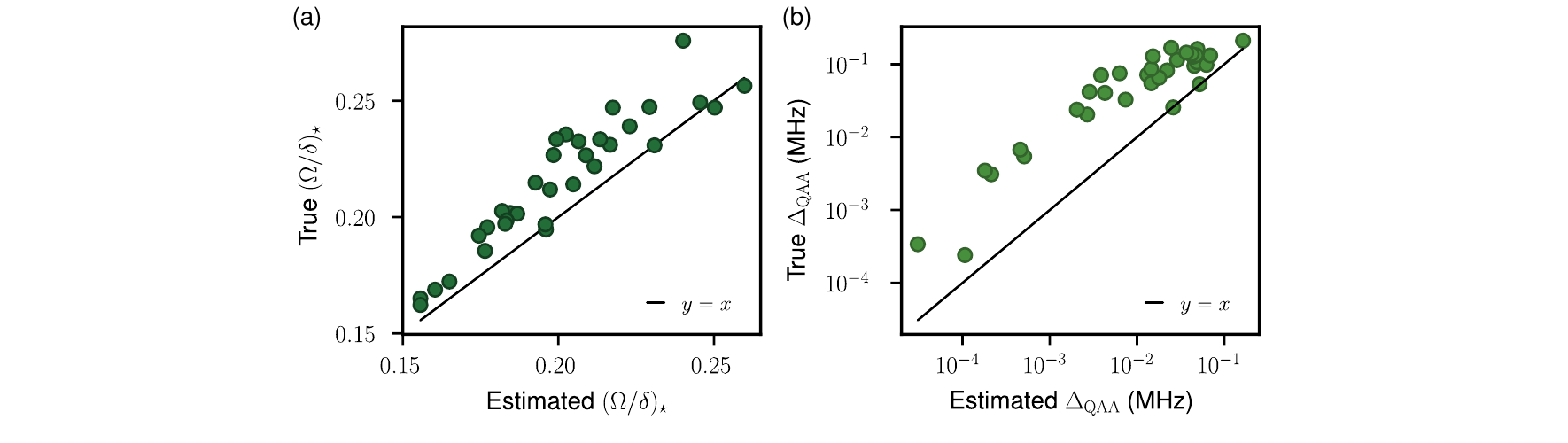}
    \caption{Perturbation theory on the Rydberg Hamiltonian. 
    (a) The true location of the avoided level crossing $\loc$ is close to the predicted value from perturbation theory, particularly for small $\loc$. 
    Here we use Hamiltonian energy scales identical to those used in the experimental implementation. 
    (b) The true minimum gap $\gqaa$ can be estimated by only considering low-order terms in the resolvent formalism. 
    }
    \label{fig:speedup_si_rydberg}
\end{figure*}

\subsection{Experimental Rydberg Hamiltonian resolvent gaps\label{subsec:exp_analysis}}
Here we analyze the performance of the Rydberg atom array experiment~\cite{Ebadi_2022} using the resolvent gap formalism described in Appendix~\ref{subsec:resolvent_derivation}. 
Because the Rydberg Hamiltonian $\hryd$ (Eq.~\eqref{eq:hryd}, main text) has long-range interactions not present in the  Maximum Independent Set cost function $\hcost$, we must modify our perturbative formalism developed to predict the minimum gap $\gqaa$. 
Here we describe our method to perturbatively compute $\gqaa$ for the Rydberg Hamiltonian. 
We then verify that the resolvent gap formalism qualitatively captures the experimental performance.

In the main text, we estimated the parameters of the avoided level crossing $\ketG, \ketE, \loc,$ and $ \ecrit$  (see Fig.~\ref{fig:speedup_3}(b), main text) by solving for the perturbative  Hamiltonian $H^{(2)}$ approximating the system Hamiltonian $\hqaa$ at small $\Omega/\delta$ (Eq.~\eqref{eq:second_order_ham}, main text). 
To find $H^{(2)}$ we performed second-order perturbation theory in the degenerate manifolds of $\hcost$, each of which contained independent sets of a fixed size. 
These manifolds become non-degenerate when exchanging $\hcost$ for the Rydberg Hamiltonian,
\begin{align}
    \hryd &= -\delta\sum_{u\in V} n_u + \sum_{u, v} V_{uv} n_u n_v,\label{eq:hryd_si}
\end{align}
due to the long-range interactions  $V_{uv}$\,$\sim$\,$ 1/|r_u-r_v|^6$. At sufficiently large distances $|r_u-r_v|$, $V_{u, v}$ is small and has negligible effect. 
However, to safely perform perturbation theory in $\loc$, we must carefully handle the Rydberg interaction energy at short distances.

In the experimental implementation, the avoided level crossing occurs at a detuning of $\delta_\star\simeq 7$--$13$\,MHz. 
The energy scale for $\hdrive$ is $|\Omega| = 2$\,MHz (note that our definition of $\Omega$ differs from the standard definition of Rabi frequency by a factor of 2). 
Although the resulting value of $\loc\ll 1$,  the necessary condition to perform perturbation theory is that the energy difference under $\hryd$  between independent sets connected via $\hdrive$ is large compared to $\Omega.$ 
Therefore, in addition to $\delta$, we must consider the interaction energy $V_{uv}$, which is $107$\,MHz for nearest-neighbors on the square lattice and $13.6$\,MHz for next-nearest neighbors (see Fig.~\ref{fig:speedup_1}, main text, and  Supplementary Information of Ref.~\cite{Ebadi_2022}). 
Suppose we take an independent set and add a vertex via $\hdrive$, creating an independent set violation between nearest-neighbors. 
This interaction can be treated perturbatively because the energy difference between an independent set with and without a single nearest-neighbor violation under $\hryd$ is $\geq 94$\,MHz $\gg|\Omega|$. 
However, suppose we instead add a vertex that creates a single independent set violation with a next-nearest-neighbor. 
The new energy under $\hryd$ increases by at least $13.6$\,MHz due to $V_{uv}$, and decreases by $13$\,MHz due to $\delta$, meaning that this transition can be near-resonant under $\hdrive$. 
Therefore, we must treat single independent set violations between next-nearest neighbors non-perturbatively. 
We find that for most instances, removing a vertex from an independent set via a spin flip can be treated perturbatively, and discuss rare exceptions later.

We will use the standard Schrieffer-Wolff transformation to compute $H^{(2)}$ for the Rydberg Hamiltonian. 
By the above arguments, $|\Omega|$ is perturbatively small compared to the energy difference between near-degenerate manifolds of states that include: 
\begin{enumerate}
    \item Valid independent sets of the same size, and
    \item Independent sets with any number of next-nearest neighbor independent set violations (where each vertex  has at most a single next-nearest neighbor in the Rydberg state).
\end{enumerate}
Of course, these configurations are not truly degenerate under $\hryd$ due to long range interactions, but their splitting is comparable to $|\Omega|$, and typically small compared to the energy splitting between adjacent manifolds, which is approximately  13.6\,MHz (up to interactions that are longer-range than next-nearest neighbors). 
Within a near-degenerate manifold, $H^{(2)}$ is given by
\begin{align}
    H^{(2)} = \hryd - \hdrive - \frac{1}{2}[S, \hdrive],
\end{align}
where the $\hdrive$ term implicitly acts only within a near-degenerate manifold (i.e., it only (de)excites single next-nearest-neighbor independent set violations). 
The third term $-\frac{1}{2}[S, \hdrive]$ describes perturbative interactions between neighboring manifolds due to $\hdrive$, where $S$ satisfies
\begin{align}
    -\hdrive + [S, \hryd] = 0.\label{eq:sw_s_condition_si}
\end{align}

Solving Eq.~\eqref{eq:sw_s_condition_si} for $S$ gives 
\begin{align}
    \bra{z}S\ket{z''} = -\frac{\bra{z}\hdrive \ket{z''}}{\hryd(z) - \hryd(z'')}.
\end{align}
Here $z$ and $z''$ are in adjacent manifolds connected by $\hdrive$. 
Therefore we see explicitly that the perturbative condition is $\hryd(z) - \hryd(z'') \gg |\Omega|.$

Inserting $S$ into Eq.~\eqref{eq:sw_s_condition_si}, we find that $H^{(2)}$ is given by
\begin{align}\label{eq:second_order_ham_ryd_si}
    H^{(2)}_{z, z'} =& \bra{z}\hryd\ket{z'} -  \bra{z} \hdrive \ket{z'}  \nonumber\\
    & +\sum_{z'': \bra{z}\hdrive \ket{z''}\bra{z''}\hdrive\ket{z'}\neq 0} 
    \frac{\Omega^2}{2}\Big(\frac{1}{\hryd(z) -\hryd(z'')}\nonumber \\
    & + \frac{1}{\hryd(z') -\hryd(z'')}\Big),
\end{align}
for $z, z'$ in the same near-degenerate manifold (here we have removed couplings involving two vertex additions or removals because they connect different manifolds, and are therefore off-resonant). 
Eq.~\eqref{eq:second_order_ham_ryd_si} is identical to the perturbative Hamiltonian for $\hqaa$ [Eq.~\eqref{eq:second_order_ham}], but with the denominator of the second-order terms replaced with the energy difference under $\hryd$ instead of $\hcost$. 
We note that for a small number of instances, there exists one or more independent sets $\ket{z}$ such that removing a single vertex creates an independent set $\ket{z''}$ for which \mbox{$\hryd(z) -\hryd(z'') \leq |\Omega|$}, because the Rydberg interaction energy from the removed vertex is comparable to $-\delta$. 
We observe that this occurs only when the removed vertex cannot spin-exchange, so only the corresponding contribution to the second-order diagonal energy shift in $H^{(2)}$ is non-perturbative. 
In these rare cases we modify this matrix entry to be the hybridized energy of $\ket{z}$ and $\ket{z''}$. 

Given our expression for $H^{(2)}$, we can now compute the parameters involved in the avoided level crossing.
For each graph instance, we enumerate the independent sets of size $\alpha$ and $\alpha-1$ using a tensor network algorithm~\cite{liu_tensor_network_2022}, which is easily achieved on a laptop for the system sizes we study ($n=39$\,--\,80). 
From the independent sets we construct $H^{(2)}$ and find its lowest energy eigenstate and eigenenergy for a given value of $\Omega/\delta$, which corresponds to the leading order approximation for $\ketG$ or $\ketE$ under $S$.  
From this, we can predict $\loc$ by finding the value of $\Omega/\delta$ where the perturbed energies of $\ketG, \ketE$ intersect. 
The energy where $\ketG, \ketE$ intersect provides an estimate of $\ecrit$.
Figure~\ref{fig:speedup_si_rydberg}(a) shows that the estimated $\loc$ from perturbation theory agrees with the true $\loc$ computed via DMRG, particularly as $\loc$ becomes small. 

Using our perturbatively estimated $\ketG, \ketE$, and $\loc$, we can now estimate the minimum gap $\gqaa$. 
Ideally we would evaluate Eq.~\eqref{eq:off_diag_expansion} in the main text, replacing $\hcost$ with $\hryd$, but this is intractable at the largest system sizes we study ($n=65, 80$). 
Inspired by the form of  Eq.~\eqref{eq:off_diag_expansion}, we instead compute $\gqaaest$, an estimate for $\gqaa$ given by 
\begin{align}
    \gqaaest = 2\sum_{z, z'}\loc^{d(z, z')}\langle z\ketG\langle z'\ketE,
\end{align}
where $d(z, z')$ is the pairwise Hamming distance between $z$ and $z'$.
$\gqaaest$ corresponds to only considering the lowest-order coupling between $\ket{z}\in \ketG$ and  $\ket{z'} \in \ketE$ under $\hdrive$ in Eq.~\eqref{eq:off_diag_expansion}, which approximately occurs at order $\loc^{d(z, z')}$. In Fig.~\ref{fig:speedup_si_rydberg}(b), we show that $\gqaaest$ and $\gqaa$ are similar. This verifies that even low-order estimations can qualitatively predict $\gqaa$. 
We note that $\gqaaest$ can be computed with relatively low space complexity on the order of $\mathcal{O}(D_\alpha + D_{\alpha-1})$, where recall $D_\setsize$ is the number of independent sets of size $\setsize.$

\section{The star graph\label{sec:star_graph}}
Here we analyze the QAA runtime to find the largest independent set of a family of star graphs. 
A star graph has $\nbranch$ branches of even length $\ell$ connected by a central vertex. 
We are interested in the runtime as a function of $\nbranch$ at fixed $\ell$. 

\subsection{Level-crossing parameters\label{subsec:star_graph_level_crossing_params}}
We start by deriving the parameters involved in the avoided level crossing when $\loc\rightarrow 0.$ 
In this limit, we can perturbatively predict $\loc$, the ground state energy at the avoided crossing  $\ecrit$, and the states involved in the avoided  crossing $\ketG, \ketE$. 
We can determine these parameters from the eigenstates and eigenenergies of the second-order perturbed Hamiltonian (Eq.~\eqref{eq:second_order_ham}, main text),
\begin{align}
    H^{(2)}\hspace{-0.05cm}
    =&-\frac{\Omega^2}{\delta}\bigg(\hse+\sum_{u\in V}\bigg[n_u -(\mathds{1}-n_u)\hspace{-0.2cm}\prod_{(u,v)\in E}\hspace{-0.2cm}(\mathds{1}-n_v)\bigg]\bigg).\nonumber
\end{align}

$\ketG$ is the ground state of $H^{(2)}$ in the manifold of independent sets with $\hcost = -\delta\alpha$. 
This corresponds to the unique largest independent set of the star graph with $\alpha = \frac{\ell \nbranch}{2} + 1$ vertices, including the central vertex and alternating vertices on each branch (see Fig.~\ref{fig:speedup_3}(c), main text). 
The eigenenergy of $\ketG$ in $H^{(2)}$  corresponds to the second-order energy shift of $\ketG$. 
It has nonzero contributions only from the second term of $H^{(2)}$, which evaluates to $-\frac{\Omega^2}{\delta}\sum_{u\in V}n_u =-\frac{\Omega^2}{\delta}\alpha$. 
The other terms are zero because no spin-exchange operations are possible and no vertices can be added to the largest independent set. 
Therefore at second-order the energy of $\ketG$ is given by 
\begin{align}\label{eq:perturbed_energy_g}
    \braket{\mathcal{G}|\hcost + H^{(2)}|\mathcal{G}} 
    = -\delta \alpha  - \frac{\Omega^2}{\delta}\alpha.
\end{align}

\begin{figure*}[th!]
    \centering
    \includegraphics[width=\textwidth]{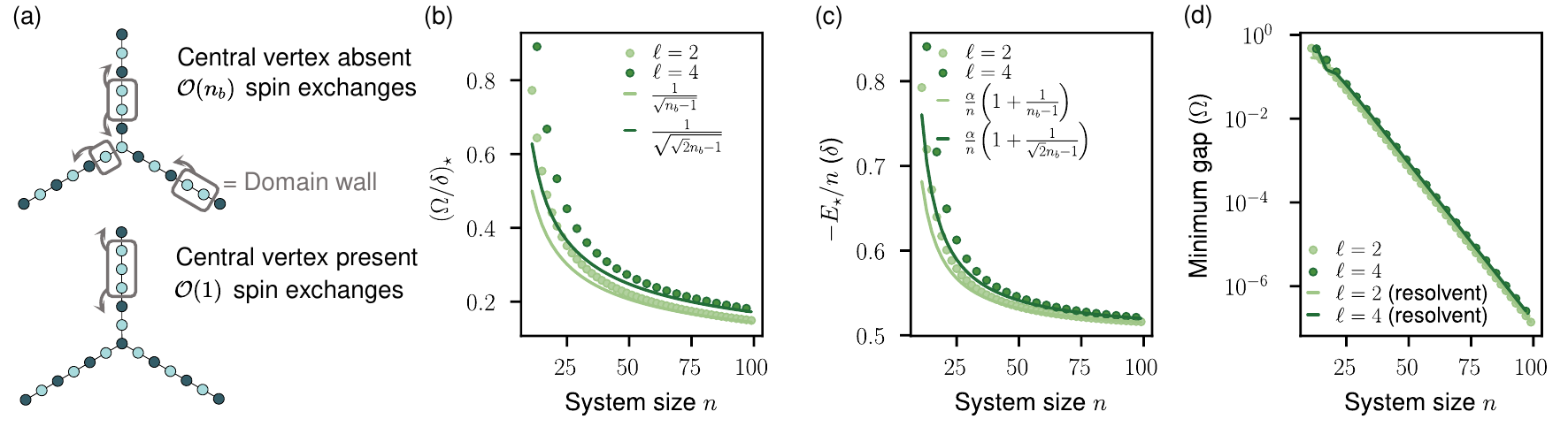}
    \caption{Perturbative avoided level crossing in the star graph. 
    (a) There are two types of suboptimal independent sets of size $\alpha-1$ in the star graph (dark blue vertices are present in the independent set, light blue vertices are absent). 
    Sets with the central vertex absent have a single domain wall on each of $\nbranch$ branches that can hop to neighboring sites via spin exchanges, yielding $\mathcal{O}(\nbranch)$ possible spin exchanges per independent set. 
    When the central vertex is present, only one branch has a domain wall, so there are $\mathcal{O}(1)$ possible spin exchanges. 
    These latter independent sets have negligible amplitude in $\ketE$, which favors independent sets with more possible spin exchanges.
    The predicted (solid lines) and numerically computed (data points) values of $\loc$ (b) and $-\ecrit / n$ (c) match as $n$ increases at fixed branch length $\ell$. 
    As $n\to\infty$, $\loc \rightarrow 0$. 
    (d) The minimum gap computed via exact diagonalization matches the gap computed by numerically evaluating the resolvent method formula Eq.~\eqref{eq:gqaa_gqaanaive_correction} in Appendix~\ref{subsec:resolvent_derivation}. 
    At fixed $\ell$, the minimum gap decreases exponentially as a function of $\nbranch$ (and therefore $n$).
    }
    \label{fig:speedup_si_star}
\end{figure*}

Next we determine $\ketE$ and its corresponding energy shift by finding the ground state of $H^{(2)}$ in the $\hcost = -\delta (\alpha-1)$ manifold. 
In this manifold there are $(\ell/2+1)^\nbranch$ independent sets of size $\alpha-1$ with the central vertex absent, and each branch in one of the $\ell/2+1$ largest independent sets of a one-dimensional length-$\ell$ chain with open boundary conditions (see Fig.~\ref{fig:speedup_si_star}(a), top). 
This degeneracy corresponds to the motion of a single domain wall (two adjacent vertices absent from the independent set) in the antiferromagnetic ordering on each branch. 
There are also a small number of independent sets of size $\alpha-1$ with the central vertex present (see Fig.~\ref{fig:speedup_si_star}(a), bottom). 
In these sets, all but one of the branches has perfect anti-ferromagnetic ordering ($\ell/2$ vertices in the set per branch), and the remaining branch has  $\ell/2-1$ vertices. 
One can count that the number of such independent sets is $3\nbranch(\ell/2 -1)$, meaning they form a vanishingly small fraction of independent sets of size $\alpha-1$ as $\nbranch\rightarrow \infty.$

As $\nbranch$ grows we find that $\ketE$ primarily has support on the independent sets with the central vertex absent. 
We first observe that the first term in $H^{(2)}$, $-\frac{\Omega^2}{\delta} \hse$, determines the ground state of $H^{(2)}$ to good approximation.
To see this, first note that the second term in $H^{(2)}$ acts uniformly on all independent sets of size $\alpha-1$, so it does not affect the eigenvectors. 
The third term gives a small diagonal shift onto an independent set for every vertex that can be added to that set. 
This term is zero for all but $\ell \nbranch/2 + 1$ independent sets that connect to the largest independent set via a single spin flip, where it gives a shift of $\frac{\Omega^2}{\delta}$. 
When $\nbranch$ is large, this energy shift is negligible compared to the ground state energy of the remaining term $-\frac{\Omega^2}{\delta} \hse$, which maximizes the expected number of spin exchanges. 
In particular, the ground state energy of this term is dominated by independent sets with the central vertex absent, which have anywhere between $\nbranch$ and $2\nbranch$ possible spin exchanges, depending on if the domain wall is on the boundary (one possible spin exchange) or in the bulk (two spin exchanges) of that branch. 
In comparison, independent sets with the central vertex present have only one or two total possible spin exchanges. 

Therefore $\ketE$ is well-approximated as the ground state of $-\frac{\Omega^2}{\delta} \hse$ restricted to the independent sets with the central vertex absent. 
On each branch this acts as a one-dimensional hopping Hamiltonian with open boundary conditions for the single domain wall. 
$\ketE$ is therefore the product of the ground state over all $\nbranch$ branches 
\begin{align}\label{eq:wavefunction_e_si}
    \braket{x_1 x_2 \dots x_{\nbranch}|\mathcal{E}} = \prod_{i=1}^\nbranch \frac{1}{\sqrt{\ell/4+1}}\sin\Big(\frac{\pi x_i}{\ell/2+2}\Big),
\end{align}
where $\ket{x_i}, x_i\in \{1, 2, \dots, \ell/2+1\}$ is the state with the domain wall on the $i$th branch between sites $2x_i-2$ and $2x_i-1$. 
We confirm numerically that the overlap of Eq.~\eqref{eq:wavefunction_e_si} with the true ground state of $H^{(2)}$ quickly approaches one as $\nbranch$ grows for $\ell\in \{2, 4, 6, 8\}$.

The corresponding perturbed energy at second-order is then
\begin{align}\label{eq:perturbed_energy_e}
    &\braket{\mathcal{E}|\hcost + H^{(2)}|\mathcal{E}} \nonumber \\
    &\qquad \simeq -\delta (\alpha-1) - \frac{\Omega^2}{\delta}(\alpha-1) - \frac{\Omega^2}{\delta}\braket{\mathcal{E}|\hse|\mathcal{E}}.
\end{align}
By our earlier reasoning, $\braket{\mathcal{E}|\hse|\mathcal{E}} = c_\ell\nbranch$ where \mbox{$c_\ell\in [1, 2]$} is a computable number depending on $\ell.$
For $\ell=2$, each configuration in $\ketE$ can spin-exchange $\nbranch$ times (once on each branch), so $c_\ell=1$. 
As $\ell$ increases, the $\ketE$ localizes on configurations that can spin-exchange $2\nbranch$ times (with the domain walls in the bulk of each branch), so $c_\ell\rightarrow 2$. 

Having computed $\ketG, \ketE$ and their energies as a function of $\Omega/\delta$, we can now estimate $\loc$ and $\ecrit$ to second order in $\Omega/\delta$. 
$\loc$ is the value of $\Omega/\delta$ where the two perturbed energies, Eqs.~\eqref{eq:perturbed_energy_g} and~\eqref{eq:perturbed_energy_e}, intersect, given by
\begin{align}
    \loc = \sqrt{\frac{1}{c_\ell\nbranch - 1}}.
\end{align}
This quantity goes to zero as $\nbranch\rightarrow\infty$, verifying that our perturbation theory converges as $n\to\infty$ at fixed $\ell$. 
Fig.~\ref{fig:speedup_si_star}(b) shows the predicted and numerically computed (via exact diagonalization) value of $\loc$ for $\ell=2$ and $4$, which have $c_\ell = 1$ and $\sqrt{2}$, respectively. 
We reach system sizes of $100$ and $97$ for $\ell = 2$ and $4$, respectively, by symmetrizing the Hamiltonian over the branches of the star graph. 

The corresponding ground state energy is computed by evaluating Eq.~\eqref{eq:perturbed_energy_g} at $\loc$. 
This gives, in units of $\delta$,
\begin{align}
    -\frac{\ecrit}{n} = \frac{\alpha}{n}\left(1+{\frac{1}{c_\ell\nbranch-1}}\right).
\end{align}
Figure~\ref{fig:speedup_si_star}(c) shows the predicted and actual values of $-\ecrit/n$ for the same instances at $\ell=2, 4$. 
As expected, the predicted values converge to the true values as $\nbranch$ increases.

\subsection{Quantum runtime\label{subsec:star_graph_runtime}}
We now compute the minimum gap $\gqaa(\ell, \nbranch)$ and analyze its scaling as a function of $\nbranch$ at fixed $\ell$.
We will show that $\gqaa(\ell, \nbranch)$ scales as
\begin{align}\label{eq:stargraph_asymptotic_gap}
    \gqaa(\ell, \nbranch) = \mathcal{O}\left(\Omega\left[\frac{1}{\sqrt{\ell/4+1}}\sin\left(\frac{\pi}{\ell/2+2}\right)\right]^\nbranch\right),
\end{align}
up to polynomial factors in $\nbranch$, which are subleading compared to Eq.~\eqref{eq:stargraph_asymptotic_gap}, which is exponentially small in $\nbranch$.
This matches the scaling predicted from leading-order perturbation theory  in $\loc$ in Eq.~\eqref{eq:starQAA} from the main text.

Following the resolvent formalism discussed in Appendix~\ref{subsec:resolvent_derivation}, we will evaluate the estimated minimum gap  $\gqaanaive$. 
Recall from Eqs.~\eqref{eq:heff_expanded_si} and~\eqref{eq:gqaanaive_matrix_element_si} that $\gqaanaive$ is given by the off-diagonal matrix element of an effective Hamiltonian $\heff(z)$ acting on the subspace spanned by $\ketG$ and $\ketE$,
\begin{align}\label{eq:stargraph_resolvent}
    \gqaanaive &= 2|\braG \heff(z) \ketE|\\
    &= 2\left|\braG\hcost-\hdrive+\hdrive\frac{Q}{z-Q\hqaa Q}\hdrive\ketE \right|,\nonumber
\end{align}
where $\hqaa=\hcost-\hdrive$ is evaluated at $\loc$, and $z\simeq\ecrit$ is a parameter with dimensions of energy. 
By the resolvent formalism equation for the minimum gap in Eq.~\eqref{eq:gqaa_gqaanaive_correction}, $\gqaanaive(\ell, \nbranch)$ gives $\gqaa(\ell, \nbranch)$ up to a computable proportionality factor that depends on $d\heff(z)/dz$ and which is close to one. 
We numerically verify the correctness of Eq.~\eqref{eq:gqaa_gqaanaive_correction} in Fig.~\ref{fig:speedup_si_star}(d) by computing $\gqaa(\ell, \nbranch)$ for $\ell=2, 4$ via both exact diagonalization and by numerically evaluating Eq.~\eqref{eq:gqaa_gqaanaive_correction}. 
To compute Eq.~\eqref{eq:gqaa_gqaanaive_correction}, we first compute $\heff(z)$, which gives us $\gqaanaive$ by Eq.~\eqref{eq:stargraph_resolvent}.
We compute the proportionality factor by evaluating $d\heff(z)/dz$ using the finite difference method.
When this correction factor is applied to $\gqaanaive(\ell, \nbranch)$, the result matches $\gqaa(\ell,\nbranch)$ computed via exact diagonalization to high accuracy, as expected. 
We observe numerically that \mbox{$\gqaanaive \simeq 4.53\gqaa$ for $\ell=2$}, and \mbox{$\gqaanaive \simeq 7.85 \gqaa$ for $\ell = 4$}, approximately independently of $\nbranch$.
Therefore, as argued in Appendix~\ref{subsec:resolvent_derivation}, $\gqaanaive$ captures the relevant scaling of $\gqaa$ in $\nbranch$.

To simplify the computation of Eq.~\eqref{eq:stargraph_asymptotic_gap}, we use a slightly different choice of $\ketE$ from the previous Appendix~\ref{subsec:star_graph_level_crossing_params}. 
This is allowed as long as $\ketG, \ketE$ are reasonable approximations to the eigenstates involved in the avoided level crossing (see Appendix~\ref{subsec:resolvent_validity}). 
We let $\hqaa^{(i)}$ equal $\hqaa$ restricted to the $i$th  branch of the star graph. 
We let $\ketEi{i}$ be the ground state of $\hqaa^{(i)}$, and choose $\ketE=\otimes_{i=1}^\nbranch\ketEi{i}.$
Note that $\ketE$ is equal to the ground state of $\hqaa$ from second-order degenerate perturbation theory [Eq.~\eqref{eq:wavefunction_e_si}], to leading order in $\loc$. 

We now evaluate Eq.~\eqref{eq:stargraph_resolvent}. 
The first term yields $\braG \hcost \ketE = 0$. 
The second term is of the same order as Eq.~\eqref{eq:stargraph_asymptotic_gap} because  our $\ketE$ is equal to the prediction from second-order degenerate perturbation theory to leading order in $\nbranch$ (see Eq.~\eqref{eq:starQAA}, main text). 
Therefore it remains to compute the third term, $\braG\hdrive\frac{Q}{z-Q\hqaa Q}\hdrive\ketE$. 
We begin by simplifying the outermost factors of $\hdrive$. 
First, we define the state 
\begin{align}
    \kettildeEi{i}=(\mathds{1}-\ketEi{i}\bra{\mathcal{E}_i})(\mathds{1}-\ketG\bra{\mathcal{G}})\hdrive^{(i)}\ketEi{i},
\end{align} 
where $\hdrive^{(i)}$ is $\hdrive$ restricted to a single branch $i$. Then,
\begin{align}
    Q\hdrive \ketE&=\sum_{i=1}^\nbranch\kettildeEi{i}\otimes_{j\neq i}\ketEi{j},
\end{align}
where we have used the fact that $H_q\ketE$ is a sum of $\nbranch$ terms, in each of which $\nbranch-1$ branches are in $\ketEi{i}$.

Meanwhile, when $\hdrive$ acts on $\ketG$ on the left hand side of the third term of Eq.~\eqref{eq:stargraph_resolvent}, one term in $\hdrive$ removes the central vertex from $\ketG$, yielding the state $\otimes_{i=1}^\nbranch\ket{x_i=1}$, where $\ket{x_i=1}$ denotes that the domain wall on the $i$th branch is on the first site (see  Appendix~\ref{subsec:star_graph_level_crossing_params}). 
$\hdrive\ketG$ also contains terms in which vertices are removed from the branches of $\ketG$, while the central vertex is left excited. 
These terms cannot have better scaling with $\nbranch$ than the term with the central vertex removed from $\ketG$, which we confirm numerically. 
They are higher order because to connect to $\ketG$ via these terms, one must first go through $\otimes_{i=1}^\nbranch\ket{x_i=1}$ to add the central vertex. 
Therefore, we have 
\begin{align}
    &\braG\hdrive\frac{Q}{z-Q\hqaa Q}\hdrive\ketE \nonumber \\
    &\quad =   \nbranch\left(\otimes_{i=1}^\nbranch\bra{x_i=1}\right)Q\left[\frac{\mathds{1}}{z-Q\hqaa Q}\right]\hspace{-0.1cm}\kettildeEi{1}\otimes_{i=2}^\nbranch\ketEi{i}.
\end{align}
Here we have specified without loss of generality that the factor of $\kettildeEi{i}$ occurs on $i=1$, which yields the factor of $\nbranch$.

We will now make the approximation that $Q$ factorizes between branches,
\begin{align}
    Q\hqaa Q\approx \sum_{i=1}^\nbranch Q\hqaa^{(i)}Q,
\end{align}
where $\hqaa^{(i)}$ is $\hqaa$ restricted to a single branch $i$. 
This is an approximation because it neglects terms in $\hqaa$ which act on the central vertex of the graph.
This leaves us with (up to polynomial factors in $\nbranch$)
\begin{align}
    &\braG\hdrive\frac{Q}{z-Q\hqaa Q}\hdrive\ketE\sim\nonumber\\
    &\quad \left(\otimes_{i=1}^\nbranch\bra{x_i=1}\right)Q\frac{\mathds{1}}{z-\sum_{i=1}^\nbranch Q\hqaa^{(i)}Q}\kettildeEi{1}\otimes_{i=2}^\nbranch\ketEi{i}.
\end{align}
Note now that $[Q\hqaa^{(i)}Q,Q\hqaa^{(j)}Q]=0$, so that we may use the identity
\begin{align}
    &\frac{\mathds{1}}{z-\sum_{i=1}^\nbranch Q\hqaa^{(i)}Q} \\ 
    &\quad =\frac{1}{(2\pi i)^{\nbranch-1}}\int dz_1\dots dz_\nbranch  \bigg[\delta\bigg(z-\sum_{i=1}^\nbranch z_i\bigg) \nonumber \\
    &\qquad \qquad \qquad \qquad \qquad \qquad\quad  \times \prod_{i=1}^\nbranch\frac{\mathds{1}}{z_i-Q\hqaa^{(i)}Q}\bigg],\nonumber
\end{align}
where the $z_i$ integrals are taken on a contour encircling the real axis.
Therefore, we have
\begin{align}
    &\braG\hdrive\frac{Q}{z-Q\hqaa Q}\hdrive\ketE\nonumber \\
    &\quad \sim\frac{1}{(2\pi i)^{\nbranch-1}}\int dz_1\dots dz_\nbranch \bigg[\delta\bigg(z-\sum_{i=1}^\nbranch z_i\bigg)\nonumber\\
    &\quad \times \left(\otimes_{i=1}^\nbranch\bra{x_i=1}\right) Q \prod_{i=1}^\nbranch\frac{\mathds{1}}{z_i-Q\hqaa^{(i)}Q}\kettildeEi{1}\otimes_{i=2}^\nbranch\ketEi{i}\bigg].
\end{align}
Note now that $Q$ acts trivially on $\kettildeEi{1}\otimes_{i=2}^\nbranch\ketEi{i}$, because $\kettildeEi{1}$ has no overlap with $\ketEi{1}$. 
Furthermore, $\hqaa^{(\nbranch)}$ only changes the state on branch $\nbranch$, so that we can write
\begin{align}
    &\frac{\mathds{1}}{z_\nbranch-Q\hqaa^{(\nbranch)}Q}\kettildeEi{1}\otimes_{i=2}^\nbranch\ketEi{i}\nonumber \\
    &\qquad \qquad =\kettildeEi{1}(\otimes_{i=2}^{\nbranch-1}\ketEi{i})\frac{\mathds{1}}{z_\nbranch-
    \hqaa^{(\nbranch)}}\ketEi{\nbranch}.
\end{align}
We can repeat this process $\nbranch-2$ more times to obtain
\begin{align}
    &\braG\hdrive\frac{Q}{z-Q\hqaa Q}\hdrive\ketE\nonumber\\
    &\sim\frac{1}{(2\pi i)^{\nbranch-1}}\hspace{-0.2cm}\int dz_1\dots dz_\nbranch \bigg[\delta\bigg(z-\sum_{i=1}^\nbranch z_i\bigg)\left(\otimes_{i=1}^\nbranch\bra{x_i=1}\right)\nonumber \\
    &\qquad \times Q\frac{\mathds{1}}{z_1-Q\hqaa^{(1)}Q}\bigg(\kettildeEi{1}\otimes_{i=2}^\nbranch\frac{\mathds{1}}{z_i-\hqaa^{(i)}}\ketEi{i}\bigg)\bigg].
\end{align}
We then make the replacement \mbox{$\frac{\mathds{1}}{z_i-\hqaa^{(i)}}\ketEi{i}\to2\pi i\delta(z_i-\epsilon)\ketEi{i}$}, where $\epsilon$ is the energy of $\ketEi{i}$ under $\hqaa$. 
This is valid by our choice of integration contour and because $\ketEi{i}$ is an eigenstate of $\hqaa^{(i)}$.
Performing the $z_i$ integrals yields
\begin{align}
    &\braG\hdrive\frac{Q}{z-Q\hqaa Q}\hdrive\ketE \sim\left(\otimes_{i=1}^\nbranch\bra{x_i=1}\right) \\
    &\qquad \qquad \times Q\frac{\mathds{1}}{z-(\nbranch-1)\epsilon-Q\hqaa^{(1)}Q}\left(\kettildeEi{1}\otimes_{i=2}^\nbranch\ketEi{i}\right).\nonumber
\end{align}

At this point, formally, the factors of $Q=\mathds{1}-\ketG\braG-\prod_{i=1}^\nbranch\ketEi{i}\bra{\mathcal{E}_i}$ act on all factors in the wavefunction. 
However, since all but one of the $\nbranch$ factors in the tensor product on the right are $\ketEi{i}$, and since $\hqaa^{(1)}$ only changes the state on the branch $i$ which is not in $\ketEi{i}$, we may safely replace $Q$ with $Q_1=\mathds{1}-\ketEi{1}\bra{\mathcal{E}_1}$, and obtain
\begin{align}
    &\braG\hdrive\frac{Q}{z-Q\hqaa Q}\hdrive\ketE\sim\left(\otimes_{i=1}^\nbranch\bra{x_i=1}\right)\nonumber \\
    &\qquad \times Q\frac{\mathds{1}}{z-(\nbranch-1)\epsilon-Q_1\hqaa^{(1)}Q_1}\left(\kettildeEi{1}\otimes_{i=2}^\nbranch\ketEi{i}\right).
\end{align}
At this point, the final factor of $Q$ may be dropped, because $\frac{\mathds{1}}{z-(\nbranch-1)\epsilon-Q_1\hqaa^{(1)}Q_1}\kettildeEi{1}$ has no overlap with $\ketEi{1}$. 
The expression becomes
\begin{align}
    &\braG\hdrive\frac{Q}{z-Q\hqaa Q}\hdrive\ketE \sim(\braket{x_1=1|\mathcal{E}_1})^{\nbranch-1} \nonumber \\
    &\qquad \times\bra{x_1=1} \frac{\mathds{1}}{z-(\nbranch-1)\epsilon-Q_1\hqaa^{(1)}Q_1}\kettildeEi{1}.
\end{align}
The factor of $\bra{x_1=1}\frac{\mathds{1}}{z-(\nbranch-1)\epsilon-Q_1\hqaa^{(1)}Q_1}\kettildeEi{1}$ should scale at most polynomially with $\nbranch$ and is thus subleading, by the arguments presented in Appendix~\ref{subsec:resolvent_derivation}. 
The term $\braket{x_1=1|\mathcal{E}_1})^{\nbranch-1}$ scales as Eq.~\eqref{eq:stargraph_asymptotic_gap}. 
Therefore, we conclude that  $\gqaanaive$ (and therefore $\gqaa$) has the same asymptotic scaling with $\nbranch$ as Eq.~\eqref{eq:stargraph_asymptotic_gap}. 

\begin{figure*}[th!]
    \centering   \includegraphics[width=\textwidth]{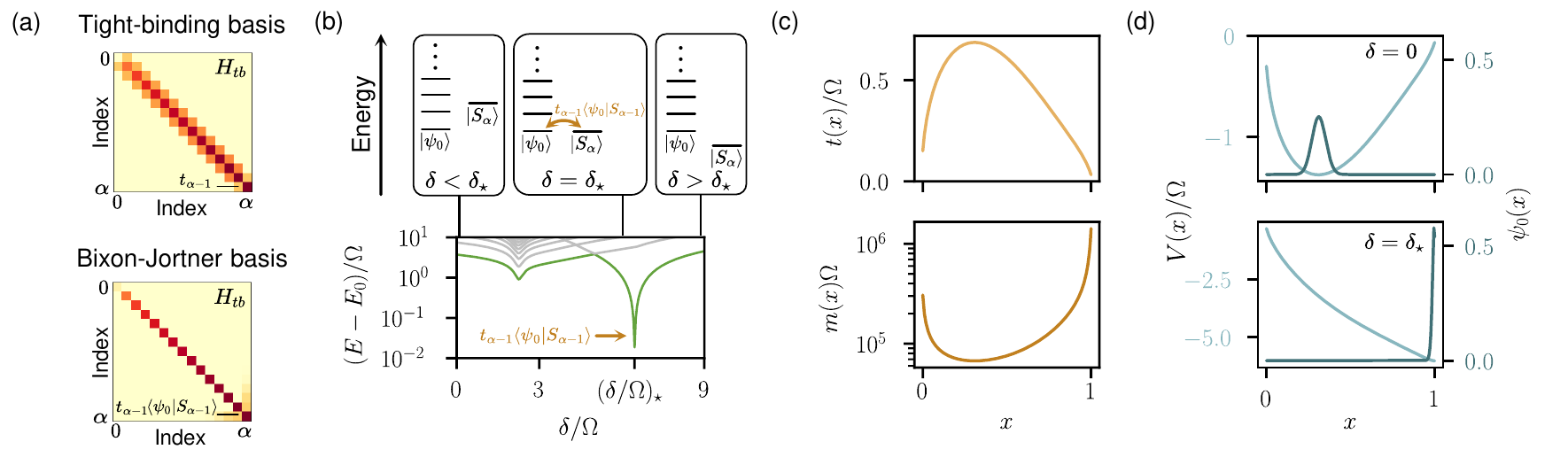}
    \caption{Minimum gap of the modified QAA at infinite $\lambda$. 
    (a) The original one-dimensional tight-binding Hamiltonian $\htb$ has a weak coupling $\bottleneck$ between the last and second-to-last sites (top). 
    $\htb$ can be partially diagonalized to generate an effective Bixon-Jortner model that weakly couples all $\hbulk$ eigenstates to the last site of the tight-binding model (bottom). 
    (b) The lowest energy eigenvalues of $\htb$ as a function of $\delta / \Omega$ for a representative instance with $n=720$ vertices (bottom) are paired with schematic eigenenergies of $\hbulk$ and the last site $\barket{\alpha}$ at three different detunings (top). 
    For $\delta < \delta_\star$, the spectral gap of $\htb$ is equal to the spectral gap of $\hbulk$. At the resonance condition $\delta = \delta_\star$ between the last site and the $\hbulk$ ground state $\ket{\psi_0}$, the weak coupling $\bottleneck$ sets the gap. 
    For $\delta > \delta_\star$, Wannier-Stark localization sets in and the gap is proportional to $\delta$.  
    (c) We show a representative example of the couplings $t(x)$ and position dependent mass $m(x) = \frac{\alpha^2}{t(x)}$ from a $720$-vertex hard unit-disk graph instance. 
    (d) For the same instance, we show the effective potential $V(x)$, neglecting the second-derivative term $\partial^2_x t(x)$, and the $\hbulk$ ground state wavefunction $\psi_0(x)$ for $\delta=0$ and $\delta=\delta_\star$. 
    At $\delta=0$, the ground state is delocalized in the middle of the bulk (top). 
    At $\delta=\delta_\star$, the wavefunction is localized near the weak coupling (bottom). }
    \label{fig:tb_gap_analysis}
\end{figure*}

\section{Runtime of the modified QAA}
In this section, we will analyze the optimized runtime $\gqaa^{-1}$ of the modified QAA  [Eq.~\eqref{eq:speedup_hamiltonian}, main text]. 
We will show that $\gqaa^{-1}$ scales as the square root of the classical Markov chain runtime lower bounds from Appendix~\ref{sec:mcmc_runtime} under motivated assumptions about the energy landscape. 
We numerically verify this when the $\hlaplace$ energy scale $\lambda \to\infty$ for system sizes of up to $460$ vertices in Fig.~\ref{fig:speedup_4}(c) of the main text. 
Here, we provide an analytic arguments supporting these numerical observations. 
We first analyze the case where $\lambda\to\infty$ next in Appendix~\ref{subsec:infinite_lambda}. 
Perturbative corrections to our arguments for the case when $\lambda$ is finite are discussed in  Appendix~\ref{subsec:finite_lambda}.

\subsection{Infinite $\lambda$ case \label{subsec:infinite_lambda}}
In the $\lambda\to\infty$ limit, the adiabatic dynamics are projected onto the ground subspace of $\hlaplace$. 
The ground subspace of $\hlaplace$ is spanned by the uniform superpositions of each independent set size $\{\barket{\setsize}\}_{\setsize = 0, \dots, \alpha}$ [Eq.~\eqref{eq:equal_superposition}, main text] when there exists a path between any two independent sets of the same size via spin exchanges. 
Here we assume this condition is met, and discuss exceptions in Appendix~\ref{subsec:finite_lambda}.
Each uniform superposition $\barket{\setsize}$ experiences an energy shift of $-\delta\setsize$ from $\hcost$ and couples to neighboring independent set sizes under $\hdrive$ with coupling strength $t_\setsize = \barbra{\setsize}\hdrive \barket{\setsize-1} =  \Omega \setsize \sqrt{D_\setsize/D_{\setsize-1}}$. 
Therefore, the effective dynamics are given by the one-dimensional tight-binding Hamiltonian $\htb$ [Eq.~\eqref{eq:tight_binding_hamiltonian}, main text],
\begin{align}
    \htb=-\sum_{\setsize=1}^{\alpha}\left[\delta \setsize\barket{\setsize}\barbra{\setsize}+t_\setsize(\barket{\setsize}\barbra{\setsize-1}+\mathrm{h.c.})\right].\label{eq:tight_binding_hamiltonian_si}
\end{align}

Our goal is to show that the minimum gap $\gqaa$ of $\htb$ goes like the smallest coupling $\min_\setsize t_\setsize$.
For simplicity, we will focus on instances where the smallest coupling is between the largest independent sets of size $\alpha$ and suboptimal independent sets of size $\alpha-1$, i.e.  $\min_\setsize t_\setsize = t_{\alpha-1}$. 
This was overwhelmingly the most common case, representing 99.87\% of the hundreds of instances studied in  Appendix~\ref{sec:runtime_system_size}. 
If $\gqaa\propto t_{\alpha-1}$, then the modified optimized QAA runtime $\gqaa^{-1}\propto t_{\alpha-1}^{-1}$  is quadratically smaller than the classical Markov chain runtime $\propto t_{\alpha-1}^{-2}$, up to polynomial factors in $n$. 
These polynomial factors are insignificant because numerically, $t_{\alpha-1}^{-2}$ is exponentially large in $\sqrt{n}$ for the Maximum Independent Set problem on unit-disk graphs (see Appendix~\ref{sec:runtime_system_size}). 

We first leverage the assumption that $\bottleneck$ is the smallest parameter in $\htb$. 
We bipartition the system into two parts: the last site, corresponding to $\barket{\alpha}$, and the remaining sites which form the ``bulk'' of the chain. 
These two parts are connected by the weakest coupling $\bottleneck$:
\begin{align}\label{eq:split_htb}
    \htb=\hbulk-\delta \alpha \barket{\alpha}\barbra{\alpha} -\bottleneck (\barket{\alpha}\barbra{\alpha-1}+\text{h.c.}),
\end{align}
where 
\begin{align}
    \hbulk =-\sum_{\setsize=0}^{\alpha-1}\left[\delta \setsize \barket{\setsize}\barbra{\setsize}+t_\setsize(\barket{\setsize}\barbra{\setsize-1}+\mathrm{h.c.})\right].\label{eq:hbulk}
\end{align} 
We then diagonalize $\hbulk$ and re-express $\htb$ in terms of its eigenenergies $E_l$ and eigenvectors $\ket{\psi_l}$ (where \mbox{$l = 0, \dots, \alpha-1$} is ordered from lowest to highest energy):
\begin{align}\label{eq:bixon_jortner}
    &\htb= - \delta \alpha \barket{\alpha} \barbra{\alpha} \\
    &\quad + \sum_{l=0}^{\alpha-1} \left[E_l \ket{\psi_l}\bra{\psi_l} -\bottleneck\braket{\psi_l | S_{\alpha-1}} (\ket{\psi_l} \barbra{\alpha}+\text{h.c.})\right].\nonumber
\end{align}
Eq.~\eqref{eq:bixon_jortner} is a Bixon-Jortner model~\cite{Heller}, a standard model in quantum optics where uncoupled levels interact with each other only by coupling to a common mode. 
Here, the uncoupled eigenstates $\ket{\psi_l}$ of $\hbulk$ are each coupled to the last site of the tight-binding chain (the common mode) with strength $\bottleneck\braket{\psi_l |S_{\alpha-1}}$, as in Fig. \ref{fig:tb_gap_analysis}(a). 
The coupling to the common mode is generated by projecting the last site  onto the energy eigenstates of $\hbulk$: $\bra{\psi_l}\hdrive\barket{\alpha} = \bottleneck \bra{\psi_l} S_{\alpha-1}\rangle$.

We now consider what happens to the spectral gap of $\htb$ as we vary the detuning $\delta$ at fixed $\Omega = 1$, which we visualize in Fig.~\ref{fig:tb_gap_analysis}(b).
We let $\delta_\star$ denote the detuning corresponding to when the ground state of $\hbulk$, $\ket{\psi_0}$, and the last site $\barket{\alpha}$ are resonant in energy (i.e., $E_0 = -\delta_\star\alpha$). 
From the canonical solution of the Bixon-Jortner problem~\cite{Heller}, it follows that once $E_0, E_1, \cdots E_{\alpha-1} > -\delta_\star \alpha$, the spectral gap increases due to level repulsion as $\delta$ is increased. 
In the language of the original tight-binding Hamiltonian, when $\delta > \delta_\star$, the electric field $\delta$ dominates so that the instance-specific details of the couplings $t_\setsize$ become irrelevant, and Wannier-Stark localization occurs in the bulk (see Fig.~\ref{fig:tb_gap_analysis}(b), top right). 
Therefore, the spectral gap is set by $\delta$ and the smallest coupling $\bottleneck$ does not play a role in determining the gap for $\delta > \delta_\star$. 
When $\delta < \delta_\star$, the spectral gap of $\htb$ is set by the spectral gap of $\hbulk$, which we denote as $\gbulk$ (see Fig.~\ref{fig:tb_gap_analysis}(b), top left). 

The gap is sensitive to the smallest coupling $\bottleneck$ at $\delta_\star.$ 
Then, the ground state energy of $\hbulk,$ $\ket{\psi_0}$, is resonant with the energy of the last site $\barket{\alpha}$, i.e., $E_{0}(\delta_\star) = - \delta_\star \alpha$. 
Here we show that the minimum gap $\gqaa$ is given by the gap at the resonance, 
\begin{align} \label{eq:gqaa_infinite_lambda}
    \gqaa & = \bottleneck \braket{\psi_{0}|S_{\alpha-1}} + \mathcal{O}(\bottleneck^{2}),
\end{align}
when the following condition holds:
\begin{enumerate}
    \item The spectral gap $\gbulk$ of $\hbulk $ is at least polynomially small in $n$ for all values of $\delta$.
\end{enumerate}
This first condition guarantees two things: first, that two-level Landau-Zener physics occurs at $\delta = \delta_\star$, as the bulk ground state $\ket{\psi_0}$ and the last site $\barket{\alpha}$ are energetically well-separated from higher excited eigenstates of $\htb$. 
This ensures that at $\delta = \delta_\star$, the gap is given by the Bixon-Jortner coupling $\bottleneck \braket{\psi_{0}|S_{\alpha-1}}.$
Second, it guarantees that the avoided level crossing at $\delta = \delta_\star$ is the \textit{minimum} gap, as the gap for $\delta < \delta_\star$, $\gbulk$, is larger than Eq.~\eqref{eq:gqaa_infinite_lambda}. 

$\gqaa$ is thus quadratically smaller than the classical Markov Chain runtime lower bounds up to polynomial factors in $n$ when a second condition holds:
\begin{enumerate}\setcounter{enumi}{1}
    \item $\braket{\psi_{0}|S_{\alpha-1}}$ is, at least, polynomially small in $1/n$. 
\end{enumerate}
This condition guarantees that the magnitude of the Bixon-Jortner matrix coupling $\gqaa = \bottleneck \braket{\psi_0 | S_{\alpha-1}}$ at $\delta_\star$ is set by $\bottleneck$ and not by localization of $\ket{\psi_{0}}$ at sites other than $\barket{\alpha-1}$. 
Therefore, it is sufficient to show that $\ket{\psi_{0}}$ at  $\delta_\star$ has at least polynomial in $n$ overlap near the $(\alpha-1)$st site in the chain. 
If both of these conditions hold, $\gqaa^{-1}$ is quadratically enhanced over the inverse of the classical Markov chain runtime up to polynomial factors in $n$.

\begin{figure*}[th!]
    \centering    \includegraphics[width=\textwidth]{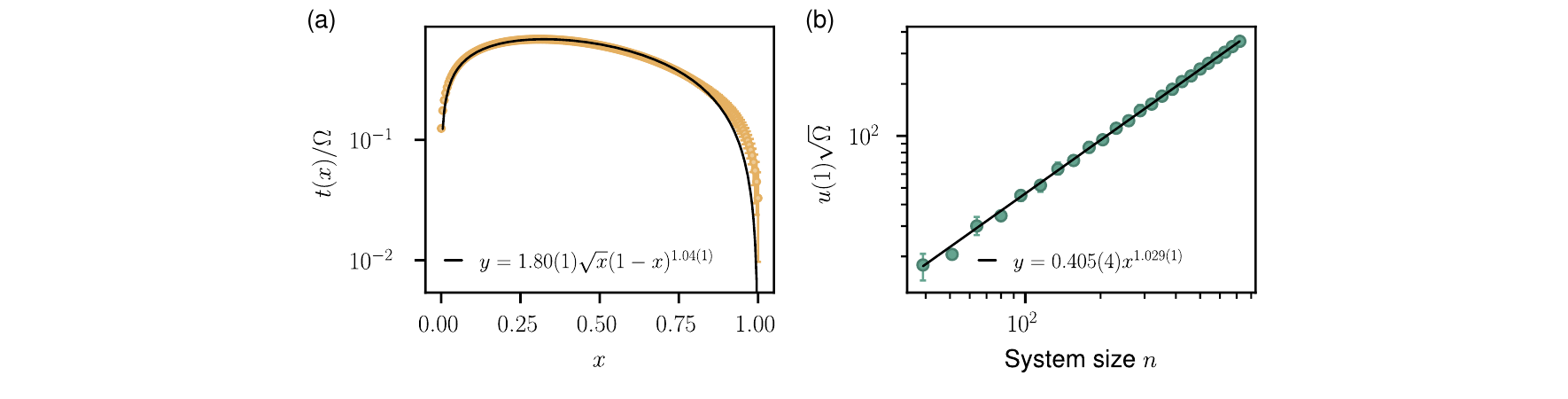}
    \caption{Parameters of the one-dimensional tight-binding Hamiltonian. (a) We show the mean couplings $t(x)$ generated from $229$ unique $720$-vertex unit-disk graph instances with largest independent set size $\alpha = 217$. Error bars give the maximum and minimum coupling over all instances.
    $t(x)$ is well-described by a smooth function with universal behavior for $x \lesssim 0.3$. 
    There are significant instance-by-instance variations in the couplings as $x\to1.$ We find that the mean couplings $t(x)$ are well-fit to the functional form $y=1.80(1)\sqrt{x}(1-x)^{1.04(1)}$ (black line). (b) The mean value of $u(1) = \int_0^1 t(y)^{-1/2}dy$ grows polynomially with the system size $n$. As a result, the gap of $\hbulk$ vanishes at most polynomially in $1/n.$  Error bars give the maximum and minimum value of $u(1)$ over 1000 instances for each system size.}
    \label{fig:tb_numerics}
\end{figure*}

We show next that both of these conditions are met under motivated assumptions about the couplings $t_\setsize$. 
We numerically analyze hundreds of hard Maximum Independent Set instances on large graphs (from $460$ to $720$ vertices). 
Our numerical investigations of the couplings $t_\setsize$ reveal that while the specifics of the couplings vary from instance to instance, \textit{in the bulk} they can, empirically, be well-described by a smooth function of the site index $\setsize$ along the tight-binding chain. 
Note that this condition often breaks down at the interface between the $\alpha$ and $\alpha-1$ because $\bottleneck$ is exponentially small in $\sqrt{n}$, but we have crucially split that term from the bulk. 
In passing we note that the normalized couplings, $t(x) \equiv t_\setsize/\alpha$ appears to converge to a near-universal curve across hundreds of instances as a function of $x = \setsize/\alpha$ for small to intermediate $0<x<0.5$, and as $\frac{1}{\sqrt{\alpha}}$ for small $x$ (one can easily check this for the $x =1/\alpha$ case). 
The normalized couplings peak at a constant value $\simeq 0.69$ before displaying instance-to-instance variation as they become small for $x \to 1$, as displayed in Fig. \ref{fig:tb_numerics}(a).

Motivated by these numerical observations, we now state constraints on the couplings that imply both conditions are satisfied.
\begin{theorem}
    Assume that the couplings $t_\setsize$ for $\setsize = 0, \dots, \alpha-1$ are a smooth, weakly concave function $t(x)$ of $x = \frac{\setsize}{\alpha}$, that  $t(1) \to 0$ as $\frac{1}{n^{\gamma}}$ for some $\gamma>0$, and $\int_0^1 t(x)^{-1/2}dx$ is at most polynomially large in the system size $n$. 
    Furthermore, assume $\delta_\star$ is sufficiently large such that $V(x) = -\delta x-2 t(x)+\frac{1}{\alpha^2} \partial_x^2 t(x)$ is locally minimized for $1-\frac{1}{\alpha} < x < 1$. 
    Then both conditions (1) and (2) hold. 
\end{theorem}

\noindent\textit{Proof.} We appeal to the continuum limit of $\hbulk$, taken as the system size $n \to \infty$. 
This is equivalent to taking the largest independent set size $\alpha \to \infty$, since for the unit-disk graphs embedded on a two-dimensional lattice with constant filling fraction, $\alpha$ is proportional to $n$.
We can take $\hbulk $ to the continuum limit because we assumed that the couplings $t_\setsize$ are a smooth function of the site $x$. 
The new continuum, time-independent Schrodinger equation for the eigenstates $\psi(x)$ in the bulk is, for arbitrary $\delta$,
\begin{equation}
    \left(-\frac{1}{\alpha^2} \partial_x t(x) \partial_x +V(x)\right) \psi(x) = \varepsilon \psi(x),
\label{eq:continuum_si}
\end{equation}
where $V(x) = -\delta x-2 t(x)+\frac{1}{\alpha^2} \partial_x^2 t(x)$ and $\varepsilon$ is the energy density (energy normalized by $\alpha$). 
Note that the site-dependent couplings have two major contributions. 
First, they induce a \textit{position-dependent mass} going as \mbox{$m(x) = \frac{\alpha^2}{t(x)}$}, which imposes a metric on the chain. 
Second, the couplings induce a potential energy given by $-2 t(x)$. 
The term that goes as the second derivative in the couplings is kept as it may grow with $n$ when $t(x)$ goes to zero at the boundaries $x = 0, 1$. 
Away from the boundaries of the bulk, this second derivative term is negligible as $\alpha \to \infty$. 
In Fig.~\ref{fig:tb_gap_analysis}(c), we visualize $t(x)$ and $m(x)$ for an example unit-disk graph instance with $n=720$ vertices. 
We plot the corresponding potential $V(x)$, neglecting the second derivative term, in Fig.~\ref{fig:tb_gap_analysis}(d) for $\delta = 0$ and $\delta = \delta_\star.$

We can arrive at a more conventional position-independent problem by performing two similarity transforms. 
The first is a point canonical transformation, 
\begin{equation}
    u(x) = \int_{0}^{x} \frac{1}{\sqrt{t(y)}} dy,
\end{equation}
which transforms the position dependent mass term into a position independent kinetic term: $\partial_x t(x) \partial_x \to \partial_u^2$. 
We then employ the standard integrating factor to remove terms that are first order in the spatial derivative, leading to a typical Schrodinger problem: $\big(-\frac{1}{\alpha^2} \partial_u^2+U(u)\big)\psi(u) = \varepsilon \psi(u)$, for a doubly-transformed effective potential $U(x)$. 
The similarity transformations do not alter the smoothness nor the convexity of the original potential. 
Moreover, the contributions from $\frac{1}{\alpha^2}\partial_x^2 t(x)$ do not change the convexity of the potential. 
As such, the effective potential $U$ meets the weak convexity criterion stipulated in Andrews and Clutterbucks' proof of the fundamental gap conjecture for one-dimensional systems \cite{AndrewsGap}. 
Therefore, we can apply the fundamental gap conjecture to bound $\gbulk$ for all $\delta$ as 
\begin{equation}
    \gbulk \geq \frac{3 \pi^2}{(\alpha u(1))^2}.
\end{equation}
Thus, as long as the couplings can be well-described by a smooth $t(x)$ and $u(1)$ grows at most polynomially  in $n$, $\gbulk$ is polynomially small in $1/n$. This proves our first condition.

To validate our second condition -- that the ground state of the bulk is localized around the penultimate site on the chain -- we provide a semi-classical argument. 
Note that the semi-classical approximation is well-justified in the limit of large system sizes $\alpha \to \infty$ as the effective mass $m(x) = \frac{\alpha^2}{t(x)}$, diverges at the edges (equivalently, $t(x) \to 0$). 
The resonance condition implies that the ground state energy density of the bulk, $\varepsilon$, at the crossing is $-\delta_\star$. 
The classical minimum of the potential approaches the edge of the chain in the regime of $\delta=\delta_\star$. 
As shown in Fig.~\ref{fig:tb_gap_analysis}(d), for $\delta=\delta_\star$, in order to minimize energy, the particle seeks to lower its potential due to the electric field gradient versus the potential due to tunneling. 
As the semi-classical expectation becomes exact in the limit of an infinite mass, the particle localizes near the edge where the classical minimum of the potential lies.  
Thus, within an asymptotically exact semi-classical argument, the ground-state of the bulk should localize near the site corresponding to $\barket{\alpha-1}$. 
This validates the second condition of our argument. 
As a result, we have shown that in the $\lambda\to\infty$ limit, $\gqaa^{-1}$ is quadratically smaller than the classical Markov chain runtime lower bounds.

\subsubsection{Numerical justification}

By examining $1000$ unit-disk Maximum Independent Set problem instances for each system size between $460$ and $720$ vertices, in Fig. 
\ref{fig:tb_numerics}(a) we find a simple qualitative model for $t(x)$ is given by $t(x) = A x^{\frac{1}{2}} (1-x)^c$. 
The factor of $\frac{1}{2}$ encodes the exact scaling as $x \to 0$ and the fit parameter $c$ accounts for instance-to-instance variations as $x\to 1$. 
We find that the fitted values of $c$  fall between $c \in \{1.03, 1.09\}$ across different instances, leading to appropriate conditions for $t(x)$ such that $u(1)$ is only polynomially large in $n$. 
We confirm numerically that $u(1)$ grows approximately linearly with $n$ in Fig. \ref{fig:tb_numerics}(b). 
Thus, our numerical results confirm that $\gbulk$ is at worst polynomially small in $1/n$,  under the assumption of the validity of our continuum analysis. 
Therefore, it is well-justified to focus on the resonant level crossing between the last site and the ground state of the bulk only to determine $\gqaa$. 

This numerical evidence also supports our assumption about $V(x)$ being locally minimized for $1-\frac{1}{\alpha} < x < 1$. 
One might worry that the diverging mass near the edges of the tight-binding model causes the bulk ground state wavefunction to be classically forbidden from penetrating the region of the penultimate site on the chain. 
Indeed, by solving Eq.~\eqref{eq:continuum_si} for the ground state with energy density $-\delta$ and using the WKB approximation, one notices that there are two classically forbidden regimes: from smaller $x$ due to the an increase in the potential, and at $x\to 1$ due to terms originating from the diverging mass (e.g. terms proportional to $[\partial_x^2t(x)]/t(x)$). 
However, the latter classically forbidden regime, following the qualitative model for $t(x) = A x^{1/2} (1-x)^c$, occurs for $x > 1-\frac{c}{2 \alpha}$, where $c$ is numerically fitted to be within $1.03$ and $1.09$. 
This suggests that the classically forbidden regime occurs within the penultimate site, which occupies $1-\frac{1}{\alpha} < x<1$ on the continuum, which can be seen by simply inverting the mapping from the continuum back to discrete sites on a chain.
Thus our numerics also strongly suggest that the wavefunction is localized around $\barket{\alpha-1}$, such that $\braket{\psi_{0}|S_{\alpha-1}}$ is sufficiently large and $\gqaa = \bottleneck$, up to polynomial factors in $n$.

\subsubsection{Optimizing the modified QAA \label{subsubsec:optimization}}
For QAA to achieve a runtime which scales as $\gqaa^{-1}$ in practice, the algorithm schedule $(\Omega(t),\delta(t))$ must be optimized so that its parameters change slowly near the location of the avoided level crossing (see Fig.~\ref{fig:speedup_3}(a), main text). 
In particular, by choosing $|dH/dt|\propto \gqaa^2$ within a $\mathcal{O}(\gqaa)$ interval around the location of the minimum gap, the total QAA runtime is $\mathcal{O}(\gqaa^{-1})$~\cite{Roland_2002}.
It is therefore useful to be able to estimate the location of the avoided crossing, so that the algorithm schedule can be optimized.
Techniques to optimize QAA are a subject of active research, and recent work indicates that it is possible to optimize QAA on a wide class of disordered cost Hamiltonians when using the reflection about the uniform superposition state, $\mathds{1}-\frac{1}{2^n}\sum_{z, z'}\ket{z}\bra{z'}$, to drive the evolution, instead of $\hdrive$~\cite{jarret_2019}. 
Here we describe a simple way to optimize the modified QAA when $\lambda\to\infty$, which achieves a total runtime, including optimization, that is asymptotically smaller than the SA runtime $\mathcal{O}(\gqaa^{-2})$. 
The full quadratic speedup, with runtime $\mathcal{O}(\gqaa^{-1})$, is recovered if quantum phase estimation is used as a subroutine in optimizing the QAA. A partial speedup, with runtime $\mathcal{O}(\gqaa^{-10/7})$, is obtained if only projective measurements in the $\sigma_z$ and $\sigma_x$ bases are used.
We leave the optimization of the modified QAA at arbtitrary $\lambda$ as a subject of future research, possibly by generalizing the results of Ref.~\cite{jarret_2019}. 

Our arguments follow from Appendix~\ref{subsec:infinite_lambda}, whose results we summarize here. 
When $\lambda\to\infty$, the QAA system Hamiltonian simplifies to an effective one-dimensional tight-binding Hamiltonian $\htb$ [Eq.~\eqref{eq:tight_binding_hamiltonian}]. 
The sites of $\htb$ correspond to the uniform superpositions of each independent set size, $\{\barket{\setsize}\}$ ($\setsize=0, 1, \dots, \alpha$). 
Suppose the final coupling between $\barket{\alpha-1}$ and $\barket{\alpha}$, equal to $\Omega\alpha \sqrt{D_{\alpha}/D_{\alpha-1}}$, is small compared to all other couplings. 
Then, this coupling can be treated perturbatively, and the system is described by a Bixon-Jortner model~\cite{Heller}.
Let us take $\Omega = 1$ and consider the system ground state as a function of $\delta$, as visualized in Fig~\ref{fig:tb_gap_analysis}(b). 
At $\delta < \delta_\star$, where $\delta_\star$ denotes the location of the avoided crossing, the system ground state is the ground state $\ket{\psi_0}$ of a restricted Hamiltonian $\hbulk$, which includes all sites up to $\barket{\alpha-1}$. 
The avoided level crossing occurs when the energy $E_0(\delta)$ of $\ket{\psi_0}$ is resonant with the energy $-\delta\alpha$ of last site of the chain, $\ket{S_\alpha}$. 
Thus, the avoided crossing occurs when $E_0(\delta_\star)=-\delta_\star\alpha$ to high $\mathcal{O}(\gqaa^{2})$ accuracy by the arguments of Appendix~\ref{subsec:infinite_lambda}. 
For $\delta > \delta_\star$, $\ket{\psi_0}$ is the first excited state of $\htb$.

Therefore, if one can estimate ground state energy $E_0$ of $\hbulk$, and compare its value to the resonance condition $E_0(\delta_\star) = -\delta_\star\alpha$, one can estimate $\delta_\star$. 
Because QAA can prepare $\ket{\psi_0}$ efficiently, finding $E_0$ is computationally simple.
Note that this is \textit{distinct} from generically finding the ground state of the system Hamiltonian $\htb$.
In particular, suppose we run QAA for time $T$ with a linear schedule for $\delta(t)$, stopping the evolution at the desired value of $\delta$.
The precise choice of $T$ is algorithm-dependent and discussed below, but we always choose $1/T$ to be much less than $1/\Delta_{\rm{bulk}}^2$, where $\Delta_{\rm{bulk}}$ is the minimum gap of $\hbulk$, but larger than $\gqaa$.
Because $T^{-1/2}$ is small compared to the energy difference between the first and second excited state, this schedule should remain adiabatic with respect to all but the smallest gap $\gqaa$.
This schedule is still highly diabatic with respect to $\gqaa$, however, for which an instantaneous ramp speed $\propto \gqaa^2$ is needed to maintain adiabaticity.
When the QAA evolution is stopped at $\delta<\delta_\star$, the state thus prepared by QAA will therefore have high overlap with the ground state of $\htb$ (and thus $\hbulk$), while for $\delta>\delta_\star$ it will have high overlap with the first excited state of $\htb$ (thus the ground state of $\hbulk$).
In particular, for all values of $\delta$, because of the chosen ramp time, the prepared wavefunction will have unit amplitude in $\ket{\psi_0}$, up to small corrections from all other instantaneous eigenstates of $\htb$, which contribute small errors to the energy of the state.

Therefore $\ket{\psi_0}$ can be prepared using QAA, and one can measure its energy with error $\varepsilon$ in time $\varepsilon^{-\gamma}$. 
The value of $\gamma$ depends on the method used to compute the energy: $\gamma = 1$ using quantum phase estimation~\cite{kitaev1995quantum}, and $\gamma=5/2$ using projective measurements in the $\sigma_z$ and $\sigma_x$ bases.
The value of $5/2$ is the combined result of shot noise and the time required for a single QAA run. In particular, with a $N$ runs of time $T$ each, we expect shot noise at the level of $\mathcal{O}(N^{-1/2})$ and nonadiabatic corrections to $E_0$ at the level of $\mathcal{O}(T^{-2})$~\cite{De_Grandi_2010}.
To make both of these $\mathcal{O}(\varepsilon)$ one can choose $T=\varepsilon^{-1/2}, N=\varepsilon^{-2}$, for a total time of $\varepsilon^{-5/2}$.
Because quantum phase estimation does not require repeated runs, we can simply choose $T=\gqaa^{-1}$ so that nonadiabatic corrections of order $T^{-2}=\mathcal{O}(\gqaa^2)$ are negligible, and still retain $\gamma=1$.

\begin{algorithm}[ht!]
\caption{Optimizing the modified QAA}\label{alg:qaa_optimization}
\KwData{A subroutine that estimates $E_0(\delta)$ with error $\varepsilon$ in time $\varepsilon^{-\gamma}$, for $\delta < \delta_\star.$ 
An initial guess $r$ for the ratio $D_{\alpha-1}/D_\alpha,$ and a scale factor $k>1$ with which we will increase $r$ by during each iteration of the optimization.}
\While{An independent set of size $\alpha$ has not been found using the modified QAA.}{
$r \gets k r$\\
Use subroutine to constrain $\delta_\star$ to an $\mathcal{O}(r^{-1/(1+\gamma)})$ interval in time $\mathcal{O}(r^{\gamma/(1+\gamma)})$.
Draw $r^{1/2-1/(1+\gamma)}$ regularly spaced guesses for $\delta_\star$ from this interval.\\
\For{each guess for $\delta_\star$ in the $\mathcal{O}(r^{-1/(1+\gamma)})$ interval, in $\mathcal{O}(r^{-1/2})$ increments}
{Run the modified QAA in time $\mathcal{O}(r^{1/2})$ using a schedule that slows down at the current guess of $\delta_\star$, such that $|dH/dt|\propto 1/r$ in an $\mathcal{O}(r^{-1/2})$ range of the guess for $\delta_\star$.}}
\end{algorithm}

Our procedure in Algorithm~\ref{alg:qaa_optimization} thus  uses binary search to efficiently find  $\delta_\star$ by minimizing the absolute value of the prepared energy of $\ket{\psi_0}$ minus $-\delta \alpha$. 
In particular, we use this procedure to estimate $\delta_\star$ to some finite, high accuracy depending on $\gamma$.
We find that it is optimal to estimate $\delta_\star$ to $\mathcal{O}(\gqaa^{2/(1+\gamma)})$ accuracy.
We then run the modified QAA using an optimized schedule with runtime $\mathcal{O}(\gqaa^{-1})$, as in Ref.~\cite{Roland_2002}, for candidate guesses of $\delta_\star$ within this range of possible values. 
By using a grid search for $\delta_\star$, the optimal solution can be found with a speedup for any $\gamma>0$. 
The total runtime of Algorithm~\ref{alg:qaa_optimization} is $\mathcal{O}(\gqaa^{-2\gamma/(1+\gamma)})$, which results in a speedup over SA for any $\gamma>0$, because the SA runtime is $\mathcal{O}(\gqaa^{-2})$.
This runtime is the result of a compromise between measurement time, which improves the precision with which $\delta_\star$ is known, and the time spent grid searching for $\delta_\star$ using QAA with an optimized schedule.
In particular, if a time $\gqaa^{-2\gamma/(1+\gamma)}$ is spent measuring $\delta_\star$ to accuracy $\mathcal{O}(\gqaa^{2/(1+\gamma)})$, one wins a factor of $\mathcal{O}(\gqaa^{2/(1+\gamma)})$ in runtime relative to the $\mathcal{O}(\gqaa^{-2})$ time that is required when grid searching for $\delta_\star$ with \textit{no} knowledge of $\delta_\star$.
As a result, a total time of $\mathcal{O}(\gqaa^{-2\gamma/(1+\gamma)})$ is also spent grid searching for $\delta_\star$.

Note that in practice, one does not know $\gqaa$ \textit{a priori}, and must therefore search for this as well.
This is done efficiently in Algorithm~\ref{alg:qaa_optimization} through a grid search on an exponentially spaced grid.
Finally, we note that the two methods discussed above (phase estimation and projective measurements) are only suggestions for the subroutine required by Algorithm~\ref{alg:qaa_optimization}.
Any method (quantum or classical) which can estimate $E_0(\delta)$ to error $\varepsilon$ in time $\varepsilon^{-\gamma}$ would suffice.

\subsection{Finite $\lambda$ case \label{subsec:finite_lambda}}
\subsubsection{Numerical observations}
Here we extend the arguments for a quadratic speedup in Appendix~\ref{subsec:infinite_lambda} to the case where $\lambda$ is finite. 
We first numerically compare the runtime of the modified QAA at finite $\lambda$ and the unmodified QAA ($\lambda = 0$) for the top $1\%$ hardest instances of each system size, up to $n=80$. 
To compute $\gqaa$ for each instance and setting of $\lambda$, we use the ITensor implementation~\cite{Fishman_2022} of DMRG to find matrix product state representations of the ground and first excited state with bond dimension of up to 1500. 
We consider the system converged to its true ground state $\ket{\psi_0}$ once the truncation error falls below a threshold value of $10^{-8}$. 
In practice, this criterion is typically satisfied after $\mathcal{O}(10^2)$ sweeps. 
Once $\rvert \psi_0 \rangle$ is obtained, we compute the first excited state by repeating this procedure but with the Hamiltonian $H' = H + V \rvert \psi_0 \rangle \langle \psi_0 \lvert$, where $V=10$ is an energy penalty that ensures that the ground state of $H'$ has negligible overlap with $\ket{\psi_0}$. 
We then minimize the corresponding energy gap between the ground and first excited state over $\Omega/\delta$ to obtain $\gqaa$, using a large energy penalty $U = 100$ on independent set violations [see Eq.~\eqref{eq:hcost}].

\begin{figure*}[th!]
    \centering    \includegraphics{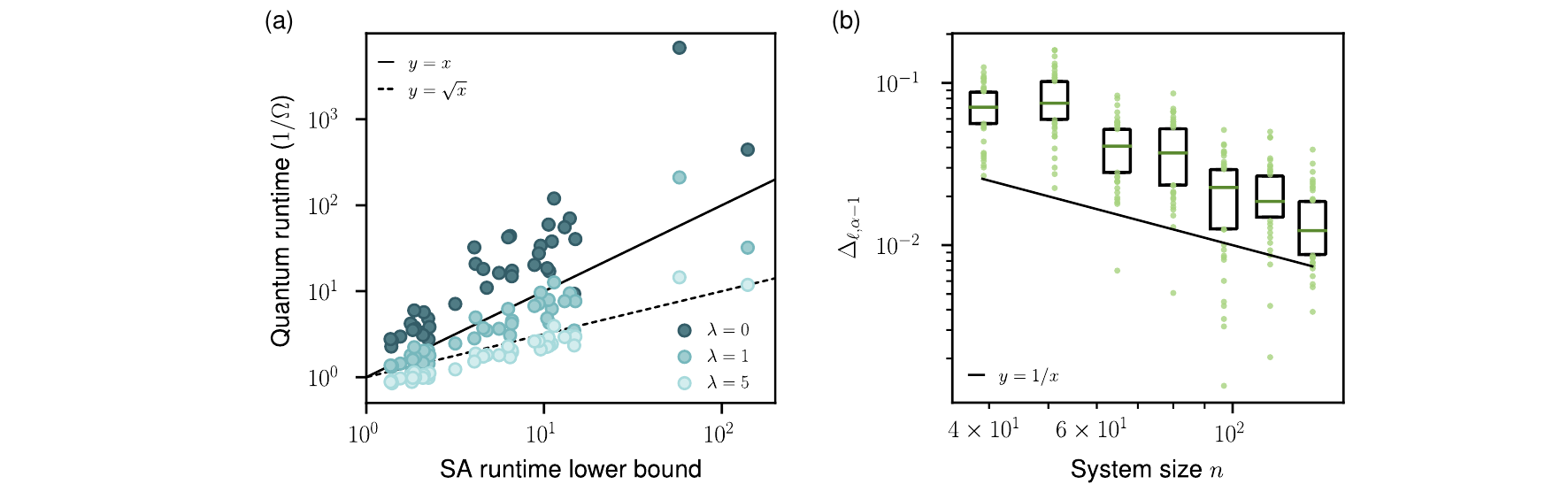}
    \caption{Modified QAA runtime at finite $\lambda$. 
    (a) The modified QAA runtime for  $\lambda = 5$ scales as the square root of the SA runtime for the top $1\%$ hardest instances of each system size up to $n=80$ (light blue points). 
    The speedup is also observed for the vast majority of instances for $\lambda = 1$ (medium blue points). 
    In both cases, the modified QAA significantly outperforms the unmodified QAA ($\lambda = 0$, dark blue points). 
    (b) For the top $5\%$ hardest instances of each system size up to $n=135$, $\glaplace{\alpha-1}$ is generally larger than $1/n$. 
    The box endpoints mark the 25th and 75th percentiles, and the box midpoint marks the 50th percentile. 
    We omit data for eight instances for which computing $\glaplace{\alpha-1}$ was too computationally expensive.
    } \label{fig:speedup_finite_lambda}
\end{figure*}

We display the numerical results in  Fig.~\ref{fig:speedup_finite_lambda}(a). We observe that for $\lambda = 5$, $\gqaa^{-1}$ is proportional to the square root of the SA runtime lower bound (the light blue data points are parallel to the line $y=\sqrt{x}$).
Furthermore, setting $\lambda = 1$  is sufficient to obtain a speedup on the vast majority of instances (medium blue data points). 
In both cases, the modified QAA vastly outperforms the unmodified QAA ($\lambda = 0$, dark blue points). 
The fact that the unmodified QAA does not frequently outperform SA suggests that typical instances of the unmodified QAA do not have favorable localization or delocalization in the ground and first excited eigenstates at the avoided level crossing, which would ensure a speedup over SA. 
Thus, the modification to QAA appears crucial to obtain a speedup over SA on these instances. 

To support these numerical observations, in the following section we further obtain analytic conditions that are sufficient, albeit not necessary, to guarantee the quadratic speedup, up to subleading polynomial factors in $n$. 
As in Appendix~\ref{subsec:infinite_lambda}, we focus on instances where the smallest coupling is between the largest independent sets of size $\alpha$ and suboptimal independent sets of size $\alpha-1$, which was overwhelmingly the most common case for the instances studied in  Appendix~\ref{sec:runtime_system_size}. We then show that when $\gqaa\simeq \Omega\alpha \sqrt{D_\alpha / D_{\alpha-1}}$ in the $\lambda \to \infty$ case, the same conditions hold for for finite $\lambda/\Omega, \lambda/\delta \gtrsim \glaplace{\alpha}^{-1},\glaplace{\alpha-1}^{-1}$, where $\glaplace{b}$ is the spectral gap of the Laplacian Hamiltonian $\hlaplace$ when restricted to independent sets of size $b$. 
To obtain the speedup in practice, it is necessary that the scaling advantage is maintained when the Hamiltonian energy scales are normalized in units of $\lambda$. By dividing the energy scales of the Hamiltonian by $\lambda$, one can see that this is equivalent to the condition that $\lambda \gqaa^{-1}$ is quadratically smaller than the classical Markov chain runtime lower bounds, up to subleading polynomial factors in $n$, where $\gqaa$ is the minimum gap in units of $\Omega$. Thus, the speedup is obtained when $\glaplace{\alpha}^{-1},\glaplace{\alpha-1}^{-1}$ grow at most polynomially in $n$. In practice, we find that $\glaplace{\alpha-1}^{-1}\geq \glaplace{\alpha}^{-1}$, so $\glaplace{\alpha-1}$ determines the strength of $\lambda$ sufficient for ensuring delocalization. 

Figure~\ref{fig:speedup_finite_lambda}(b) shows the scaling of $\glaplace{\alpha-1}$ as a function of $n$ for the top 5\% hardest instances up to $n=135.$
We observe that $\glaplace{\alpha-1}\gtrsim 1/n$ for the vast majority of instances, consistent with polynomial scaling in $n$. 
A minority of instances ($24\%$) have $\glaplace{\alpha-1} = 0$ due to a very small fraction  (median $0.2\%$) of configurations disconnected by spin exchanges, leading to degenerate ground states of $\hlaplace$ in the manifold of independent sets of size $\alpha-1$. 
For these instances, we plot the spectral gap of $\hlaplace$ in the same manifold, restricted to the largest set of configurations connected under spin exchanges. 
One can see using perturbation theory that the smaller set(s) of disconnected configurations do not change the dynamics significantly, and the larger set  determines the minimum gap. 
At small $\Omega/\delta$, the QAA Hamiltonian will energetically favor the connected subspace with the smaller expectation value of $-(\Omega^2/\delta)\hse$ under perturbation theory. 
This corresponds to the larger connected subspace, because the number of disconnected configurations in practice is very small (and thus, so is its expectation in $\hse$, which is upper-bounded by the maximum degree of a vertex in the configuration graph).
We emphasize that the numerical results in Fig.~\ref{fig:speedup_finite_lambda}(a) show that in practice, much smaller values of $\lambda$ may be necessary to obtain the speedup, depending on the graph instance. 
All instances obtain a speedup for either $\lambda = 1$ or $\lambda = 5$, which is smaller than $\glaplace{\alpha-1}^{-1}.$ This shows that while our condition is sufficient to obtain the speedup, it is not necessary in general. 

Finally, it is interesting to note the connection between the gap $\glaplace{b-1}$ and the time needed for SA to sample from the equilibrium Gibbs distribution, restricted to a manifold of independent sets of the same size. 
This corresponds to SA sampling uniformly among independent sets of the same size. 
Consider an SA algorithm that only uses spin-exchange updates to explore independent sets of the same size (of course, this SA algorithm is only ergodic among independent sets of the same size, assuming all configurations can be connected via spin exchanges).  
One can check that $\hlaplace$ is identical to the transition matrix used by SA, up to an overall rescaling and multiple of the identity. 
Thus, $\glaplace{\setsize}$ sets the mixing time for SA to sample from the uniform distribution in that manifold. 
This idea can be generalized: consider an SA algorithm now using both spin-exchange and spin-flip updates. 
Again, the matrices within a manifold are identical up to rescaling when restricted to maximal independent sets (independent sets to which no vertices can be added without removing an existing vertex). 
The fraction of maximal independent sets is approximated by the quantity $1-\frac{nD_{\setsize}}{D_{\setsize-1}}$, which is close to one on instances with a large SA runtime lower bound. 
Thus, $\hlaplace$ and the SA transition matrix are near-identical, up to rescaling.  
As a result, $\glaplace{\setsize}^{-1}$ sets the equilibration time to uniformly sample independent sets within that manifold. 
SA will thus need $\mathcal{O}( \glaplace{\alpha-1}^{-1})$ updates to converge to uniformly sample independent sets for the $\alpha-1$ manifold. 
Because this  quantity is polynomial in $n$, SA rapidly mixes within the $\alpha-1$ manifold. 
The same is true of the manifold of independent sets of size $\alpha$, because $\glaplace{\alpha-1}^{-1}\geq \glaplace{\alpha}^{-1}$. 
Thus, we expect the SA runtime $\tsa(\varepsilon)$ is set by the time to find an optimal solution, which is exponential in $\sqrt{n}$, rather than the time to equilibrate within a manifold of independent sets of the same size. 
This is consistent with the numerical results in Fig.~\ref{fig:speedup_2}, which put together, suggests that $\tsa(\varepsilon)$ is a good proxy for the SA time to find an optimal solution.

\subsubsection{Sufficient analytic conditions for the speedup}

We now show that $\gqaa\simeq \Omega\alpha \sqrt{D_\alpha / D_{\alpha-1}}$ for finite $\lambda/\delta,\lambda/\Omega \gtrsim \glaplace{\alpha}^{-1},\glaplace{\alpha-1}^{-1}$.
To this end, we will use the resolvent formalism developed in Appendix~\ref{subsec:resolvent_derivation}, and let $\ketG=\barket{\alpha}, \ketE=\barket{\alpha-1}$. 
When $\lambda/\delta,\lambda/\Omega \gtrsim \glaplace{\alpha}^{-1},\glaplace{\alpha-1}^{-1}$, we expect these states to have significant overlap with the  ground and first-excited state of $\hqaa$ at $\loc$. 
As a result,
\begin{align}\label{eq:finite_lambda_gap}
    \gqaanaive &= 2|\braE \heff(\ecrit)\ketG| \\
    &=2\left|-\braE\hdrive\ketG+\braE\hdrive Q\frac{Q}{\ecrit-QH Q}Q\hdrive\ketG\right| \nonumber
\end{align}
is a good estimator of $\gqaa$ (see Appendix~\ref{subsec:resolvent_validity}), where $H=\hcost-\hdrive+\lambda\hlaplace$ is the modified QAA Hamiltonian. 
The first term of this expression is the  coupling $-\Omega\alpha\sqrt{D_{\alpha}/D_{\alpha-1}}$ from the $\lambda\to\infty$ limit, which is responsible for the quadratic speedup. 
To argue that the speedup is maintained at finite $\lambda$, it remains to argue that the second term does not cancel with the first to reduce the gap. 
We do this by analyzing the dependence of this second term on $\lambda$.

We first simplify the second term. 
Let $\kettildeG=\frac{1}{\alpha\Omega}\hdrive\ketG$, where the $\alpha\Omega$ factor is used to make $\braket{\tilde{\mathcal{G}}|\tilde{\mathcal{G}}}\simeq 1$. 
Note that $\kettildeG$ has support only on independent sets of size $\alpha-1$. 
We now write the second term from Eq.~\eqref{eq:finite_lambda_gap} as
\begin{widetext}
\begin{align}
    \braE\hdrive Q\frac{Q}{\ecrit-QHQ}Q\hdrive\ketG&=\alpha\Omega\braE\hdrive Q\frac{Q}{\ecrit-QHQ}Q\kettildeG \nonumber\\
    &=\alpha\Omega\braE\hdrive Q\frac{Q}{\ecrit-QHQ}Q\hdrive Q\frac{Q}{\ecrit-Q(\hcost+\hlaplace)Q}Q\kettildeG \nonumber\\
    &=\alpha\Omega\braE\hdrive Q\frac{Q}{\ecrit-QHQ}Q\hdrive Q\frac{Q}{\ecrit+\delta(\alpha-1)-Q\hlaplace Q}Q\kettildeG
\end{align}
\end{widetext}
where in the second line we used the Woodbury matrix identity, and dropped a term which is unable to connect $\kettildeG$ to $\ketE$.
Now, because we have taken $\lambda\glaplace{\alpha-1}\gg \Omega, \delta$, we may make the approximation
\begin{align}
\frac{Q}{\ecrit+\delta(\alpha-1)+Q\hlaplace Q}Q\kettildeG \approx \hlaplace^{+}\kettildeG,
\end{align}
where $\hlaplace^+$ denotes the Moore-Penrose pseudoinverse of $\hlaplace$, restricted to the space of sets of size $\alpha-1$.
Here we rely on the fact that $\ecrit+\delta(\alpha-1)=\mathcal{O}(\delta)$, as argued in Appendix~\ref{subsec:late_crossing} because the avoided level crossing happens at $\loc \ll 1$. 
If the perturbative avoided level crossing condition is not met, then the same conclusion holds if we take $\lambda/\delta,\lambda/\Omega \gtrsim n\glaplace{\alpha}^{-1},n\glaplace{\alpha-1}^{-1}$, which introduces a subleading factor of $n$ to the runtime.
We note that this approximation neglects a term that is $\mathcal{O}(\Omega/[\lambda\glaplace{\alpha-1}], \delta/[\lambda\glaplace{\alpha-1}])$, which we will argue below is subleading. 
The second term of Eq.~\eqref{eq:finite_lambda_gap} thus reduces to
\begin{align}\label{eq:lambda_scaling_gap}
    &\braE\hdrive Q\frac{Q}{\ecrit-QHQ}Q\hdrive\ketG \nonumber \\
    &\qquad\quad =\alpha\Omega\braE\hdrive Q\frac{Q}{\ecrit-QHQ}Q\hdrive\hlaplace^+\kettildeG.
\end{align}

The central point of our argument is that the factor of $\hlaplace^+$, which scales as  $\mathcal{O}(1/\lambda)$, ensures that the leading $\lambda$-dependence of this expression is $\mathcal{O}(1/\lambda)$. 
This will make it impossible for the second term of Eq.~\eqref{eq:finite_lambda_gap} to always cancel exponentially with the first term. 
To see this, suppose for the sake of contradiction that for a specific value of $\lambda$, the second term was equal to $\alpha\Omega\sqrt{D_{\alpha}/D_{\alpha-1}}(1+\varepsilon)$, for some exponentially small $\varepsilon$ (leading to a suppressed gap in Eq.~\eqref{eq:finite_lambda_gap} of order $\alpha\Omega\sqrt{D_{\alpha}/D_{\alpha-1}}\varepsilon$). 
Then, if Eq.~\eqref{eq:lambda_scaling_gap} is $\mathcal{O}( 1/\lambda)$, doubling $\lambda$ will yield a gap from Eq.~\eqref{eq:finite_lambda_gap} equal to $\alpha\Omega\sqrt{D_{\alpha}/D_{\alpha-1}}(1/2-\varepsilon/2)$, which still achieves the quadratic speedup, losing only a factor of two.

It therefore remains to argue that Eq.~\eqref{eq:lambda_scaling_gap} decreases with $\lambda$ as $1/\lambda$ or faster. 
Since $\hlaplace^+$ scales as $1/\lambda$, the only way this could not be the case is if the $\lambda$-dependence of the denominator $\ecrit-QHQ$ changes the scaling to be slower than $1/\lambda$.
This would occur if there were a leading order $\mathcal{O}(1/\lambda)$ term in $\ecrit-QHQ$.
However, since we have chosen $\ketG,\ketE$ to satisfy the overlap condition of Theorem~\ref{thm:resolvent_condition} in Appendix~\ref{sec:resolvent}, we know that the smallest eigenvalue of $QHQ-\ecrit$ is at least $(E_2-\ecrit)$, up to polynomial factors in $1/n$, where $E_2$ is the energy of the second excited state of $H$ at the gap closing.
In the limit we are considering, by standard perturbation theory in $\Omega,\delta/\lambda$, the leading (in particular, zeroth-order) contribution to $E_2-\ecrit$ will be independent of $\lambda$. 
As a result, the leading contribution to Eq.~\eqref{eq:lambda_scaling_gap} will be $\mathcal{O}(1/\lambda)$.

It is now also clear why the $\mathcal{O}(\Omega/[\lambda\glaplace{\alpha-1}], \delta/[\lambda\glaplace{\alpha-1}])$ term we dropped is unimportant. 
By nearly identical arguments to the above, this term will have a leading $\mathcal{O}(1/\lambda^2)$ scaling, which will not modify the overall argument that the second term in Eq.~\eqref{eq:finite_lambda_gap} cannot cancel with the first term for generic values of $\lambda$ (due to the $\lambda$-dependence of the second term).

\let\oldaddcontentsline\addcontentsline 
\renewcommand{\addcontentsline}[3]{} 
\bibliographystyle{apsrev4-2}
\bibliography{ref.bib}
\let\addcontentsline\oldaddcontentsline 

\end{document}